\documentclass[numberedappendix]{emulateapj}
\usepackage{graphicx}
\slugcomment{Draft --- \today}

\newcommand{\be}{\begin{equation}}
\newcommand{\bel}[1]{\be\label{eq:#1}}
\newcommand{\ee}{\end{equation}}
\newcommand{\ba}{\begin{eqnarray}}
\newcommand{\ea}{\end{eqnarray}}
\newcommand\bp{\begin{figure}}
\newcommand\ep{\end{figure}}
\newcommand\bpm{\begin{figure*}}
\newcommand\epm{\end{figure*}}
\newcommand{\btab}{\begin{tabular}}
\newcommand{\etab}{\end{tabular}}
\newcommand{\bt}{\begin{table}}
\newcommand{\et}{\end{table}}
\newcommand{\ben}{\begin{enumerate}}
\newcommand{\een}{\end{enumerate}}
\newcommand\reffig[1]{Figure \ref{fig:#1}}
\newcommand\refeq[1]{Equation \ref{eq:#1}}
\newcommand\refsec[1]{\S \ref{sec:#1}}
\newcommand\reftbl[1]{Table \ref{tbl:#1}}
\newcommand\refapp[1]{Appendix \ref{app:#1}}
\newcommand{\cm}{\rm{cm}}

\newcommand\mean[1]{\langle #1 \rangle}

\newcommand{\bcn}{\begin{center}}
\newcommand{\ecn}{\end{center}}

\newcommand{\nup}{\nu_{\rm p}}

\newcommand{\rmbf}[1]{{\rm\bf #1}}
\newcommand{\EM}{{\rm EM}}
\newcommand{\planck}{plc}
\newcommand{\Tgas}{T_{\rm gas}}

\newcommand{\nH}{n_{\rm H}}
\newcommand{\mmu}{\mu_{0}}
\newcommand{\degree}{^\circ}
\newcommand{\COBE}{{\it COBE}}
\newcommand{\IRAS}{{\it IRAS}}

\begin{document}

\title{Constraining Spinning Dust Parameters with the WMAP Five-Year Data}

\author{Gregory Dobler,\altaffilmark{1,3}
  Bruce Draine\altaffilmark{2}
  \& Douglas P. Finkbeiner\altaffilmark{1}}

\altaffiltext{1}{
  Institute for Theory and Computation, 
  Harvard-Smithsonian Center for Astrophysics, 60 Garden Street, MS-51,
  Cambridge, MA 02138 USA
}
\altaffiltext{2}{Department of Astrophysical Sciences, Princeton
  University, Princeton, NJ 08544 USA
}
\altaffiltext{3}{gdobler@cfa.harvard.edu}

\begin{abstract}

  We characterize spinning dust emission in the warm ionized medium by
  comparing templates of Galactic dust and H$\alpha$ with the 5-year
  maps from the \emph{Wilkinson Microwave Anisotropy Probe}.  The
  H$\alpha$-correlated microwave emission deviates from the thermal
  bremsstrahlung (free-free) spectrum expected for ionized gas,
  exhibiting an additional broad bump peaked at $\sim40$ GHz which
  provides $\sim20$\% of the peak intensity.  We confirm that the
  bump is consistent with a modified Draine \& Lazarian (1998)
  spinning dust model, though the peak frequency of the emission is
  somewhat lower than the 50 GHz previously claimed.  This frequency
  shift results from systematic errors in the large-scale modes of the
  3-year WMAP data which have been corrected in the 5-year 
  data release.  We show that the bump is not the result of 
  errors in the H$\alpha$ template by analyzing regions of 
  high free-free intensity, where the WMAP K-band map may be 
  used as the free-free template.  We rule out a pure 
  free-free spectrum for the H$\alpha$-correlated emission 
  at high confidence: $\sim27\sigma$ for the nearly full-sky 
  fit, even after marginalizing over the CMB 
  cross-correlation bias.  We also extend the previous 
  analysis by searching the parameter space of the Draine \& 
  Lazarian model but letting the amplitude float.  The best 
  fit for reasonable values of the characteristic electric 
  dipole moment and density requires an amplitude factor of 
  $\sim0.3$.  This suggests that small PAHs in the warm 
  ionized medium are depleted by a factor of $\sim 3$.

\end{abstract}
\keywords{ 
diffuse radiation ---
dust, extinction --- 
ISM: clouds --- 
radiation mechanisms: non-thermal --- 
radio continuum: ISM 
}

\section{Introduction}

Following formulation of a physical model \citep[][hereafter 
DL98]{DL98b}, spinning dust has been invoked to explain the 
dust-correlated microwave emission seen by the 
\emph{Wilkinson Microwave Anisotropy Probe} (WMAP) from 
23-41 GHz \citep{deO04,finkbeiner04,gb04,boughn07,DF08b}.  
Given the WMAP data at 94 GHz, the dust-correlated emission 
at 23-41 GHz was anomalously high to be explained by a 
purely thermal emission mechanism.

\citet{bennett03} attempted to obviate the inconsistency by 
suggesting that the first year data from WMAP did not 
require a spinning dust component to construct an internally 
consistent foregrounds model, so long as the dust-correlated 
emission was actually dust-correlated synchrotron at low 
frequencies.  However, when the 3-year WMAP data (23-94 GHz) 
were combined with lower frequency data from Green Bank 
\citep[8,14 GHz;][]{gb04}, Tenerife \citep[10,15 
GHz;][]{deO04}, and other telescopes \citep[19 
GHz;][]{boughn07}, it became clear that the dust-correlated 
emission at $\sim20$ GHz did not behave spectrally like 
synchrotron emission, and was instead consistent with a 
spinning dust spectrum.

The principal source of ambiguity in interpreting the low 
frequency WMAP data as evidence for spinning dust was that 
the spectrum of the dust-correlated emission falls from 
23-41 GHz.  The implication is that either the peak in the 
spectrum is near or below the lowest WMAP band or that the 
emission is synchrotron, and it is only with data at lower 
frequency that the spinning dust spectrum is recovered.  In 
the DL98 spinning dust models however, the peak frequency of 
the spinning dust emission varies depending on model 
parameters (e.g., the grain size, geometry, ambient density, 
etc.), and at higher temperatures with smaller grain size is 
expected to be $\sim40$ GHz.  This regime is termed the 
``warm ionized medium'' (WIM, $T \sim 8000$ K) in DL98 
compared to the ``cold neutral medium'' (CNM, $T \sim 100$ 
K) emission that peaks at $\sim$ 20-30 GHz and is thought to 
be the source of the dust-correlated emission at low 
frequencies in WMAP.

Dobler \& Finkbeiner (2008b, hereafter DF08b) showed that 
the ambiguity is eliminated and that the turnover in the 
spinning dust spectrum is recoverable \emph{within} the WMAP 
frequency range using H$\alpha$-correlated emission as a 
tracer of WIM spinning dust emission.  Typically, H$\alpha$ 
is used as a tracer of free-free (thermal bremsstrahlung) 
emission from $\sim 10^4$K gas in regions where extinction 
from dust is minimal.  However, because a map of H$\alpha$ 
represents an emission measure (density squared integrated 
along the line of sight) and because spinning dust emission 
is generated via dust grains that are spun up by collisions 
with ions (which should also be roughly proportional to 
density squared integrated along the line of sight), the 
H$\alpha$ emission should also spatially trace WIM spinning 
dust emission.  In the analysis of DF08b, the peak frequency 
of the emission was found to be $\sim 50$ GHz, which was 
slightly higher than the $\sim 40$ GHz predicted by DL98.

In this paper, we show that the evidence for spinning dust 
in the H$\alpha$-correlated emission has strengthened with 
the release of the 5-year data, due to both decreased noise 
as well as corrections made to large-scale fluctuations that 
were present in the 3-year data.

\section{Template Fits}
\label{sec:template_fits}

Microwave emission mechanisms in the WMAP frequency range 
can be separated into four broad categories: free-free, 
thermal dust, spinning dust, and synchrotron 
emission.\footnote{\citet{DL99} point out that if 
interstellar dust contains an appreciable fraction of 
ferrimagnetic or ferromagnetic materials, thermal 
fluctuations in the magnetization would generate appreciable 
amounts of magnetic dipole emission at frequencies 
$\nu<100$GHz, which would be a fifth type of CMB foreground.  
However \citet{cassasus08} find that this type of emission 
is inconsistent with the morphology of cm emission from the 
$\rho$ Oph cloud, implying that magnetic materials are not 
abundant in interstellar dust. Thus, we do not consider it 
further in this paper; we will use the term ``thermal'' 
emission to refer specifically to emission from thermal 
fluctuations in the electric dipole moments of grains.}  
Free-free emission is generated by electron-ion collisions 
which produce thermal bremsstrahlung in ionized gas; thermal 
dust refers to electric dipole emission from thermal 
fluctuations (e.g., lattice vibrations) in the electric 
charge distribution in the grain; and spinning dust is 
electric dipole radiation from the smallest dust grains 
which are excited into rotational modes through a variety of 
collisional mechanisms.  Synchrotron emission consists of a 
soft component ($T \propto \nu^{-\beta_S}$ with $\beta_S 
\approx 3$) originating from supernova shock accelerated 
electrons which spiral around the Galactic magnetic field 
and a hard component ($\beta \approx 2.5$) centered on the 
Galactic center and extending roughly 20 degrees.  This hard 
component has been termed the ``haze'' and its origin 
remains uncertain \citep{finkbeiner04,DF08a}.

Each of these emission mechanisms is approximately traced by 
maps of the sky at other frequencies, described below.  This 
external information about the spatial structure of the 
foregrounds makes possible a multi-linear fit of the 
spectrum of each foreground.  This multi-linear regression 
may be performed over the whole sky, or in selected regions 
to study the spectral variation of each component. We now 
briefly describe the templates used, and the fitting method 
\citep[see][for more details]{DF08a}.

\subsection{The templates}
\label{sec:templates}

\begin{deluxetable*}{|c|l|ccccc|}
  \tablehead{CMB type & \multicolumn{1}{c|}{description} & 
    \multicolumn{5}{c|}{ILC coefficients} \\
    & & K & Ka & Q & V & W}
  \startdata
    1 & published WMAP 5yr ILC\footnote{available at 
      http://lambda.gsfc.nasa.gov/} & & & N/A & & \\
    2 & ILC using WMAP 5yr Kp2 coefficients & 
      0.134 & -0.646 & -0.377 & 2.294 & -0.405 \\
    3 & minimum variance ILC over unmasked, fit pixels  & 
      0.048 & -0.603 & 0.342 & 0.530 & 0.683 \\
    4 & WMAP W band minus (thermal dust + free-free model)  & 
      & & N/A & & \\
    5 & minimum variance ILC with thermal dust model presubtracted  & 
      0.182 & -0.638 & 0.044 & 0.325 & 1.086 \\
    6 & TOH 5yr map\footnote{Tegmark et al.\, priv communication}  & 
      & & N/A & &
  \enddata
\tablecomments{
The different types of CMB estimators used in the fits.  
Because of contamination of these estimators by foregrounds, 
the inferred foreground spectra vary depending on the 
estimator used.  Our preferred estimator is CMB5, which 
consists of an ILC with a minimum variance (over unmasked 
pixels) weighting of the thermal dust model presubtracted 
WMAP data.  One of the attractive features of this map is 
that it has well understood noise properties.
}\label{tbl:cmbtbl}
\end{deluxetable*}

For fitting purposes, we distinguish between the soft synchrotron
component originating from supernova-shock accelerated electrons, and
a harder synchrotron component in the inner Galaxy, possibly with a
different physical origin.  Soft synchrotron emission is well traced
by the \citet{haslam82} 408 MHz map.  Though the spectral index of
this emission is expected to vary slightly from place to place in the
sky, \citet{laporta08} showed that from 408 to 1420 MHz, $T \propto
\nu^{-\beta_S}$ with $\beta_S \approx 3$.  A similar spectral index
was derived by \citet{page07} using polarization in the WMAP data.

The WMAP data, even in the first year, contained evidence of a harder
synchrotron component in the inner Galaxy, which has become known as
the ``haze'' because of its relatively featureless morphology
\citep{finkbeiner04}.  Dobler \& Finkbeiner (2008a, hereafter DF08a)
interpret the haze as a separate physical component in the template
fit described in DF08a and below, and the spectrum of this emission
was found to be $T \propto \nu^{-\beta_H}$ with $\beta_H \approx 2.5$.
The template we use for this component is $T \propto 1/r_{\rm gal}$
where $r_{\rm gal}$ is the distance from the galactic center in
degrees.  Though systematics related to contamination of any CMB
template by residual foregrounds leads to uncertainty in the derived
soft and hard synchrotron spectra, we note that neither the presence
nor the spectrum of the haze has any significant effect on the
conclusions in this paper.  For a detailed study of microwave
synchrotron in the Galaxy see Dobler \& Finkbeiner (2008c, in
preparation).

Our thermal dust template is the Finkbeiner, Davis, \& Schlegel (1999,
hereafter FDS99) two-component model of thermal dust, evaluated at 94
GHz.  The FDS99 map uses column densities and temperatures from
IRAS/ISSA \citep{issa94} and COBE/DIRBE \citep{dirbesupp95} given by Schlegel,
Finkbeiner, \& Davis (1998; SFD), constrained to fit COBE/FIRAS data in the
microwave and sub-mm \citep{firas_supp}.  The preferred FDS99 model (model 8)
gives $I_\nu \propto \nu^{1.7}B_\nu(T_{\rm dust})$.  Converting this
to antenna temperature, $T\propto\nu^{\beta_D}$ with
$\beta_D\approx1.55$ in the WMAP bands for typical $T_{dust}$ values.
We use the same template for the diffuse, cold neutral medium (CNM)
spinning dust which has a peak frequency $\sim 20$ GHz.

For a free-free and WIM spinning dust template we use the H$\alpha$ map
assembled by \citet{finkbeiner03} using data from three surveys: VTSS
\citep{dennison98}, SHASSA \citep{gaustad01}, and WHAM
\citep{haffner03}.  The template is corrected for dust extinction
assuming the dust and ionized gas are uniformly mixed along the line of
sight.  This approximation fails in regions of very high dust column
density, so we mask out regions in our fit where the SFD extinction at
H$\alpha$ is $A(\mbox{H}\alpha) \equiv 2.65 E(B-V) > 1$ mag.
Additionally, we mask out all point sources listed in the WMAP5 point
source list as well as the LMC, SMC, M31, Orion-Barnard's Loop, NGC
5090, and the HII region around $\zeta$ Oph.  This mask covers 22.2\% of
the sky and is shown in \reffig{diff53} below.  Roughly 1/4 of the
masked pixels are point sources.

\subsection{CMB Estimators}
\label{sec:cmb_estimator}

Finally, we also need an estimator for the CMB since it 
contributes a large variance at WMAP frequencies (and in 
fact constitutes the \emph{biggest} source of noise when 
deriving foreground properties).  We use the six CMB 
estimators described in DF08a and summarize their features 
in \reftbl{cmbtbl}.  As noted in DF08a \citep[see 
also][]{hinshaw07}, every CMB estimator is contaminated to 
varying degrees by foregrounds which leads to a systematic 
bias in the inferred foreground spectra.  In the case of 
``internal linear combination'' (ILC) type estimators, in 
which the WMAP maps in each band $b$ are weighted and summed 
with weight coefficients $\zeta_b$ that are chosen to 
approximately cancel the foregrounds while preserving unity 
response to the CMB, the bias is proportional to the chance 
spatial correlation of the true foregrounds (as opposed to 
the templates) with the CMB (see DF08a).

The spectral cross correlation coefficients given in e.g. 
Fig. \ref{fig:halpha-twopanel} depend on the CMB estimator 
used, but because we fit out a CMB spectrum in the 
interpretation of the correlation spectra (see 
\refsec{interpretation}), \emph{the final results for the 
H$\alpha$-correlated spinning dust do not depend on our 
choice of CMB estimator}.

\subsection{Template fit procedure}

To infer the spectra of the individual foreground 
components, we use the multi-linear regression template fit 
described in DF08a.  We construct the template matrix $P$, 
whose columns consist of the templates described above, and 
derive the spectrum of the template correlated emission, 
$a^i_b$, where $\rmbf{a}^i$ denotes the spectrum of 
foreground $i$ and $\rmbf{a}_b$ is the vector of correlation 
coefficients for band $b$. For each band, we solve the 
matrix equation,
\bel{mateq}
  P \rmbf{a}_b = \rmbf{w}_b,
\ee
for $\rmbf{a}_b$, where $\rmbf{w}_b$ is the WMAP map for 
band $b$.  To determine the best fit $\rmbf{a}_b$ in 
\refeq{mateq} we evaluate the ``pseudoinverse'' $P^+$ (see 
DF08a) and note that $\rmbf{a}_b = P^+ \rmbf{w}_b$ minimizes 
the quantity $\Delta^2 \equiv \sum_p \left| 
P\rmbf{a}_b-\rmbf{w}_b \right|^2$ where the sum is over 
unmasked pixels. Dividing $P$ and $\rmbf{w}_b$ by $\sigma_b$ 
\citep[the mean noise in each band\footnote{ The WMAP scan 
strategy visits some pixels, e.g. the ecliptic poles, far 
more than average.  Because we are in the 
systematics-dominated limit, we use the average noise in 
each band to avoid over-weighting certain parts of the sky.
}, as in][]{bennett03,hinshaw07}, this solution for $\rmbf{a}_b$
minimizes,
\bel{chisq}
  \left\|\frac{P}{\sigma} \ \rmbf{a}_b -
  \frac{\rmbf{w}_b}{\sigma_b}\right\|^2 = 
  \frac{\|P\rmbf{a}_b-\rmbf{w}_b\|^2}{\sigma_b^2} \equiv \chi_b^2.
\ee
The total $\chi^2 = \sum_b \chi_b^2$ for the fit is given by 
the sum over the 5 WMAP bands.  We also define the residual 
maps $\rmbf{r}_b \equiv \rmbf{w}_b - P\rmbf{a}_b$.

With our template fits, the spectrum of each foreground 
emission component $\rmbf{a}^i$ is left completely 
unconstrained, though it is assumed to be constant across 
the sky.  To appropriately account for the CMB, we also 
force $\rmbf{a}^{\rm CMB} =1$ in units of thermodynamic 
temperature.

\section{Foreground Cross-Correlation Spectra}

Our foreground spectrum fits are characterized by the cross-correlation 
coefficient $\rmbf{a}$ as well as the formal error in the fit in each band.  
These errors take into account both the noise in each WMAP band
and morphological correlations between the different templates.  By 
our definition of $\rmbf{a}$ above, the derived (cross-correlation) spectra are 
in $I_{\nu}$ units of kJy/sr per template unit --- Rayleighs for 
H$\alpha$-correlated emission and mK for FDS99-correlated emission.  Both the 
data and templates are mean-subtracted over unmasked pixels to remove 
sensitivity to zero-point levels in the maps.

\subsection{H$\alpha$ and Dust Correlated Emission}
\label{sec:corremission}
\bpm
\centerline{
  \includegraphics[width=0.47\textwidth]{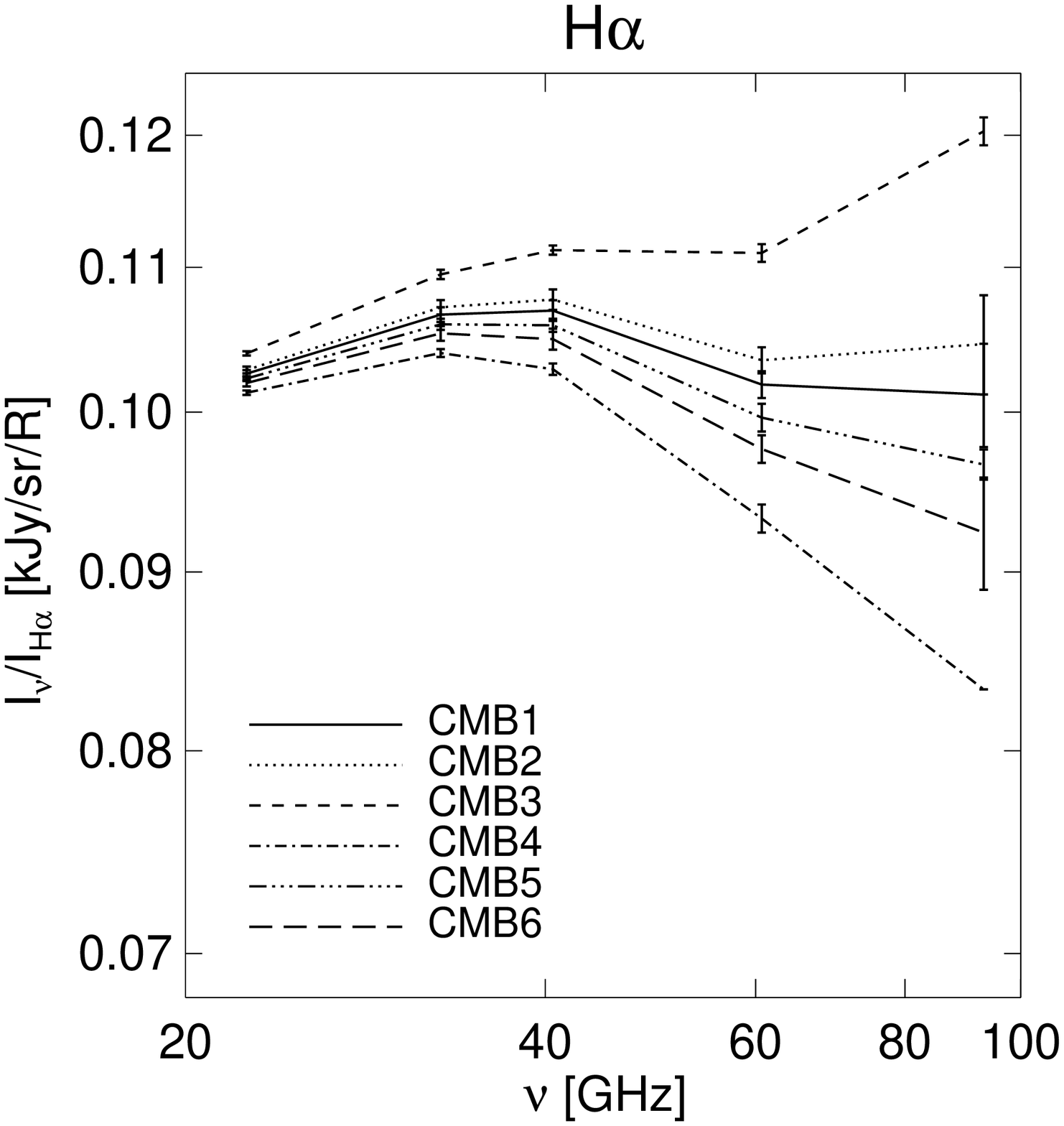}
  \includegraphics[width=0.47\textwidth]{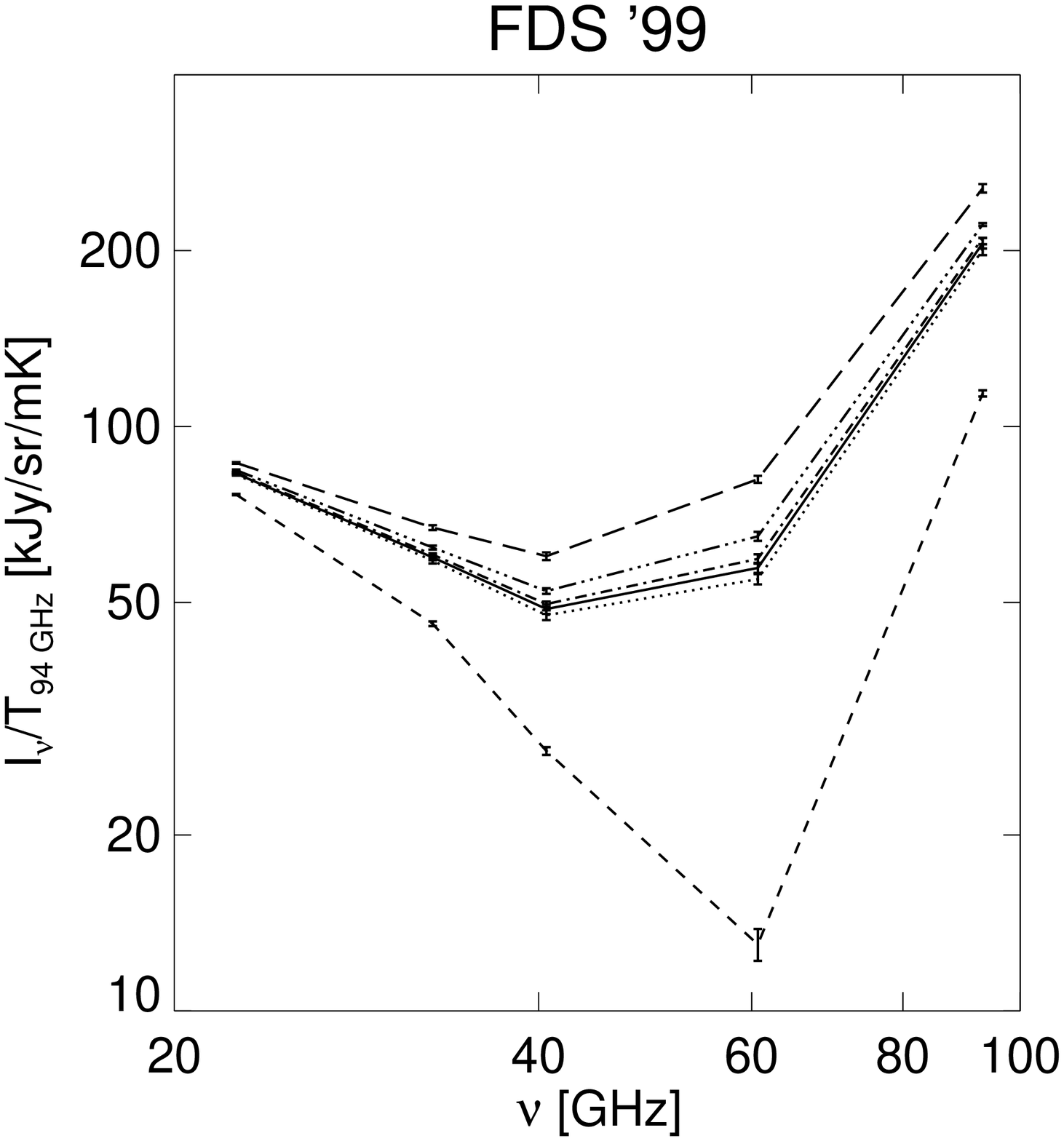}
}
\caption{
Cross-correlation spectra for the H$\alpha$ (\emph{left}), and FDS99
dust (\emph{right}) templates using various estimators of the CMB
anisotropy (see \refsec{cmb_estimator}).  Although the formal error
in the fit is small, in both cases the unknown degree of CMB
contamination introduces a highly covariant uncertainty.  In each case,
the spectra are the same up to addition of a CMB spectrum of unknown
amplitude.  }
\label{fig:halpha-dust-spec}
\epm

\reffig{halpha-dust-spec} shows our derived cross-correlation spectra for the 
H$\alpha$- and FDS99-correlated microwave emission from 23 to 94 GHz for our six 
CMB estimators.  The dust-correlated spectrum exhibits the familiar thermal 
tail from 94 to 61 GHz and then the rise from $\sim$50 to 23 GHz from anomalous 
emission.  While the statistical error bars on the spectra are very small due 
to both the high sensitivity of WMAP as well as the large sky coverage ($\sim$ 
150,000 pixels for HEALPix $N_{\rm side} = 128$), the systematic effects from 
the contamination of the CMB estimators by foregrounds are significant.

As identified in DF08a and DF08b, the spectrum of H$\alpha$-correlated emission 
does \emph{not} follow the free-free power law as expected.  Instead, there is a 
bump in the spectrum with a peak frequency $\nup \approx 40$ GHz.  This bump is 
present for all CMB estimators.  We note that it cannot be generated by a 
contamination of the CMB estimator by foregrounds (since this bias has the 
spectrum of the CMB) and DF08b argued that this bump is most easily explained by 
a WIM spinning dust component that is traced by the H$\alpha$ map.  We address 
both of these considerations in detail below.

\subsection{Changes from 3-year to 5-year data}

\bpm
\centerline{
  \includegraphics[width=0.85\textwidth]{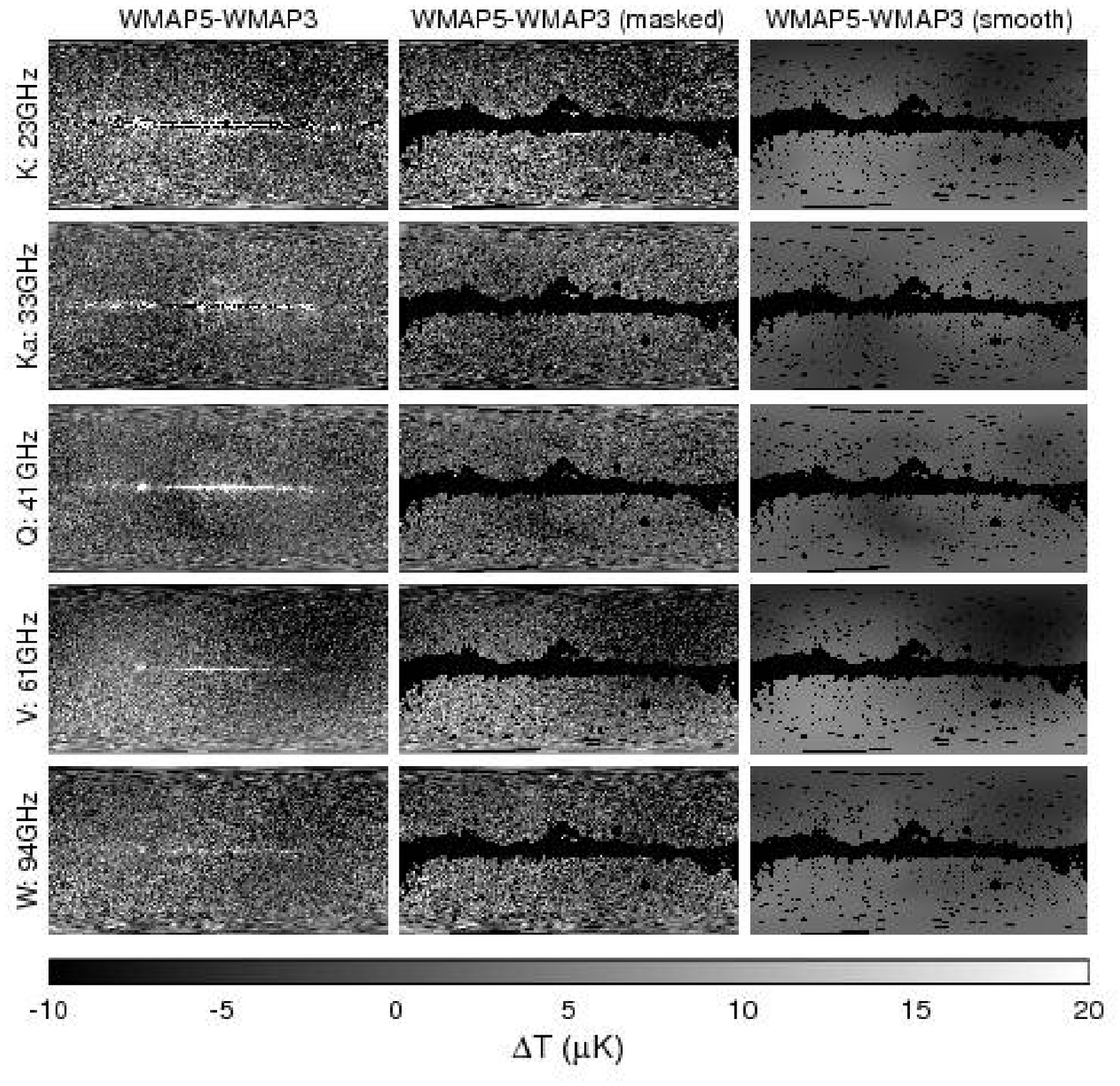}
}
\caption{
The difference between the WMAP 5-year and 3-year data in 
the 5 bands. \emph{Left:} WMAP5$-$WMAP3.  Updated beams and 
gains lead to differences in the plane, however large scale 
gradients (and in some cases dipoles) are seen in each band.  
\emph{Center:} WMAP5$-$WMAP3 masked with the pixels that are 
used in the multi-linear regression fit (see 
\refsec{template_fits}).  \emph{Right:} WMAP5$-$WMAP3 
smoothed to $30\degree$ FWHM and masked.  Here the 
large-scale fluctuations are clearly evident.  In 
particular, the peaks range from $\sim -8$ $\mu$K to $\sim$ 
15 $\mu$K.  All maps were mean subtracted over unmasked 
pixels prior to subtraction and smoothing.
} \label{fig:diff53}
\epm

There are two notable differences when comparing our results 
using the 3-year versus 5-year WMAP data.  First, the 
inferred peak frequency $\nup \approx 40$ GHz is 
significantly lower than the 50 GHz seen in the 3-year WMAP 
data (see DF08b).  This shift in peak frequency is primarily 
due to spurious large-scale power that was present in the 
3-year data but has been corrected in the 5-year data 
release.  \reffig{diff53} shows that the amplitude of these 
fluctuations can be quite large.  In particular, the V band 
dipole is clearly apparent and has peak values from $\sim -7 
\mu$K to $\sim 10 \mu$K.

As described above, we have mean subtracted the maps over 
unmasked pixels to perform the template fits, and so our 
results are not sensitive to zero point offsets.  However if 
there is spurious low $\ell$ power, as in the 3-year data, 
our fit results will be affected.  \citet{hinshaw08} point 
out that these large-scale fluctuations in the 3-year data 
were due to imperfect characterization of the instrument 
gain.  While the amplitudes of the fluctuations are not 
large enough to have a significant impact on CMB analyses, 
H$\alpha$-correlated emission mechanisms (free-free and WIM 
spinning dust) are sufficiently subdominant at WMAP 
frequencies that $\sim 10 \mu$K offsets become important.  
Because the 61 GHz band (V band) was the most affected, and 
because the WIM spinning dust peak frequency is in the range 
30-50 GHz, it is not surprising that our best fit peak 
frequency has shifted.  Had the spurious low-$\ell$ power 
been morphologically similar in all bands, the peak 
frequency would have been similar between the 3- and 5-year 
data.  The problem was exacerbated by the CMB estimators, 
which necessarily contained some complicated combination of 
these fluctuations.

The second difference between the 3- and 5-year analyses is 
that, for a given CMB estimator, the ILC coefficients 
(defined in \refeq{ilc} below) have changed.  Although this 
does not affect our interpretation of the fit results (see 
\refsec{interpretation}), the spectrum of 
H$\alpha$-correlated emission has also changed by more than 
the formal fit uncertainties.  Since this is due primarily 
to changes in the coefficients used when constructing a CMB 
estimator, the extent of the contamination of these 
estimators by foregrounds has changed from year three to 
year five.  For example, in the case of a simple ILC which 
minimizes the variance over unmasked pixels (CMB3) with 
respect to the ILC coefficients $\zeta_b$,
\be
  \frac{\partial \mean{T^2_{\rm ILC}}}{\partial 
  \zeta'_b}\Bigr\vert_{\zeta'_b=\zeta_b} = 0,
\ee
where
\be
\label{eq:ilc}
  T_{\rm ILC} = \sum_b \zeta_b \rmbf{w}_b
\ee
is the ILC map in thermodynamic mK and the ILC coefficients 
$\zeta_b$ are constrained to sum to unity to preserve 
response to the CMB.  These coefficients are sensitive not 
only to the large-scale fluctuations mentioned above, but 
also to the \emph{noise} in the measurements themselves 
since $\rmbf{w_b} = \rmbf{c} + \rmbf{f}_b + \rmbf{n}_b$, 
where $\rmbf{c}$, $\rmbf{f}_b$, and $\rmbf{n}_b$ are the 
CMB, total foregrounds in each band, and noise in each band 
respectively.  Thus the signal variance in the noise also 
contributes to the determination of $\zeta_b$.  This is 
illustrated by the limit of large measurement noise, in 
which case the minimum variance linear combination would 
have $\zeta_b=1/\sigma_b^2$.

\subsection{H$\alpha$ as a tracer of spinning dust}

For the 3-year WMAP data, DF08b presented an argument for why the
H$\alpha$ map should trace spinning dust emission at WMAP frequencies.
Briefly, the H$\alpha$ intensity is proportional to the \emph{emission
measure} (EM), $\int n_e^2 d\ell \equiv$ EM, where $n_e$ is the electron
density and the integral is along the line of sight.  We will show below
that evaluating the DL98 spectrum with parameters appropriate for the
WIM, the total emission is indeed roughly proportional to $n_e^2$.  That is,
including other (de-)excitation mechanisms such as plasma drag,
far-infrared emission from the grains, etc., the behavior of the total
emission with density implies that the H$\alpha$ EM map should trace a
component of the spinning dust emission in the Galaxy.

\section{Spinning Dust Model}
\label{sec:spinmodel}

\begin{deluxetable*}{|c|c|c|c|c|c|c|c|c|}
  \tablehead{ Environment & $\nH (\cm^{-3})$ & $\mmu$ (debye) & $\Tgas$ (K) & 
              $T_{\rm dust}$ (K) & $x_{\rm H}$ & $x_{\rm M}$ & $\chi$ & $y$}
  \startdata
    WIM & 0.01-0.6 & 0.5-12 & 3000 & 20 & 0.99   & 0.001  & 1.0 & 0.0 \\
    \hline
    CNM & 5.0-40.0 & 0.5-12 & 100  & 20 & 0.0012 & 0.0003 & 1.0 & 0.0
  \enddata
\tablecomments{
Spinning dust parameters for the ``warm ionized medium'' (WIM) and ``cold 
neutral medium'' (CNM) environments.  For a detailed definition of the 
parameters shown here see DL98.
}\label{tbl:spindustparams}
\end{deluxetable*}
                      
\bp
\centerline{
  \includegraphics[width=0.47\textwidth]{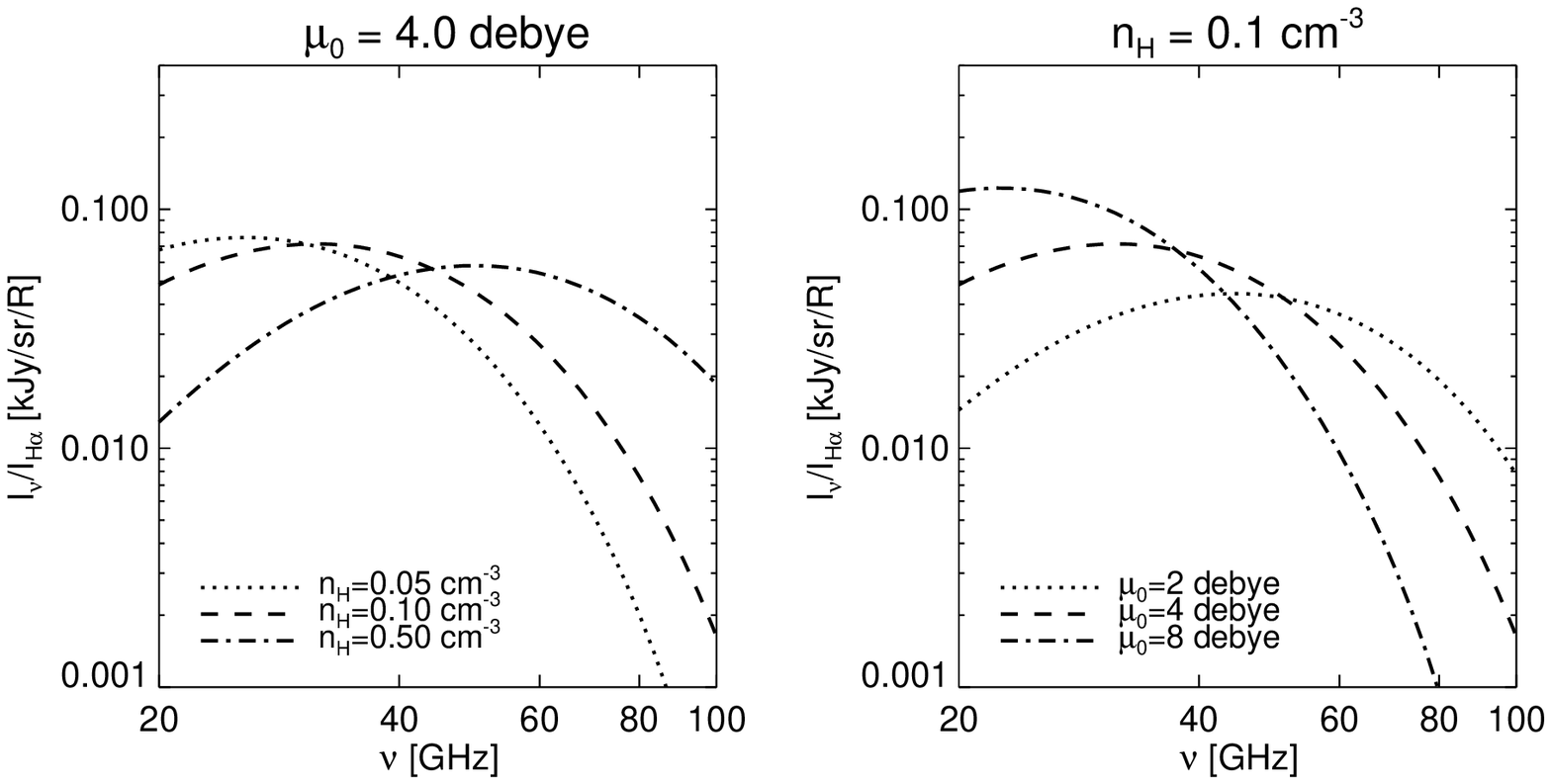}
}
\centerline{
  \includegraphics[width=0.47\textwidth]{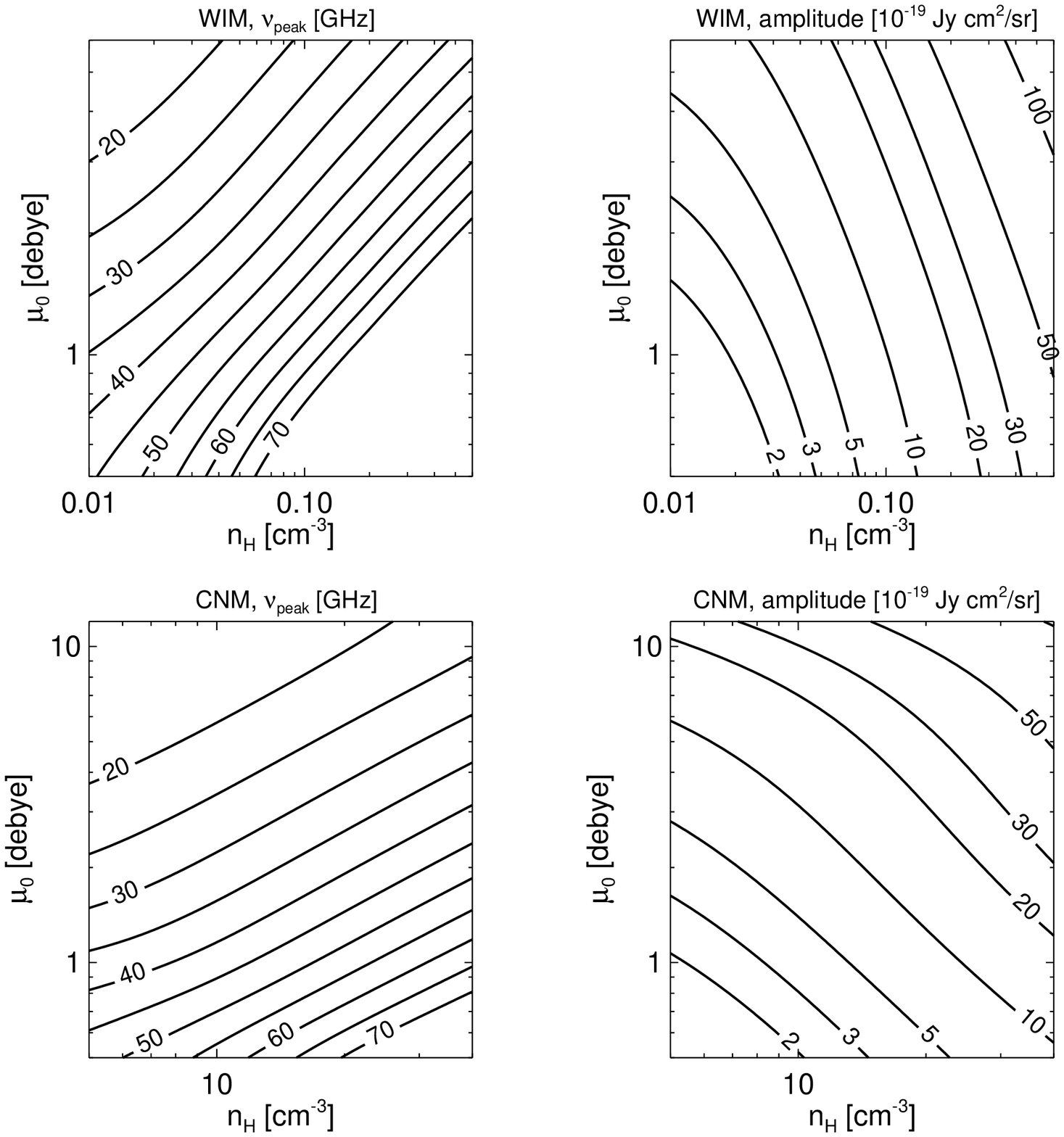}
}
\caption{
\emph{Upper left:} Spinning dust emission per H$\alpha$ 
intensity for WIM parameters, keeping $\mmu$ constant and 
varying $\nH$.  Increasing $\nH$ increases the peak 
frequency but changes the total power per H$\alpha$ little. 
\emph{Upper right:} Same, but for constant $\nH$ and varying 
$\mmu$.  Decreasing $\mmu$ also increases the peak frequency 
but decreases the total power per H$\alpha$.  Note that in 
this part of $(\nH,\mmu)$ parameter space, the shape of the 
spectrum is nearly unchanged, but shifts in frequency and 
amplitude. \emph{Middle panels:} Contours of the peak 
frequency (\emph{left}) and peak amplitude (\emph{right}) in 
the $(\nH,\mmu)$ plane. \emph{Lower panels:} Same, but for 
CNM parameters.  See the discussion in \refsec{spinmodel} 
for details.
}
\label{fig:spin_spec_ex}
\ep

\bpm
\centerline{
  \includegraphics[width=0.47\textwidth]{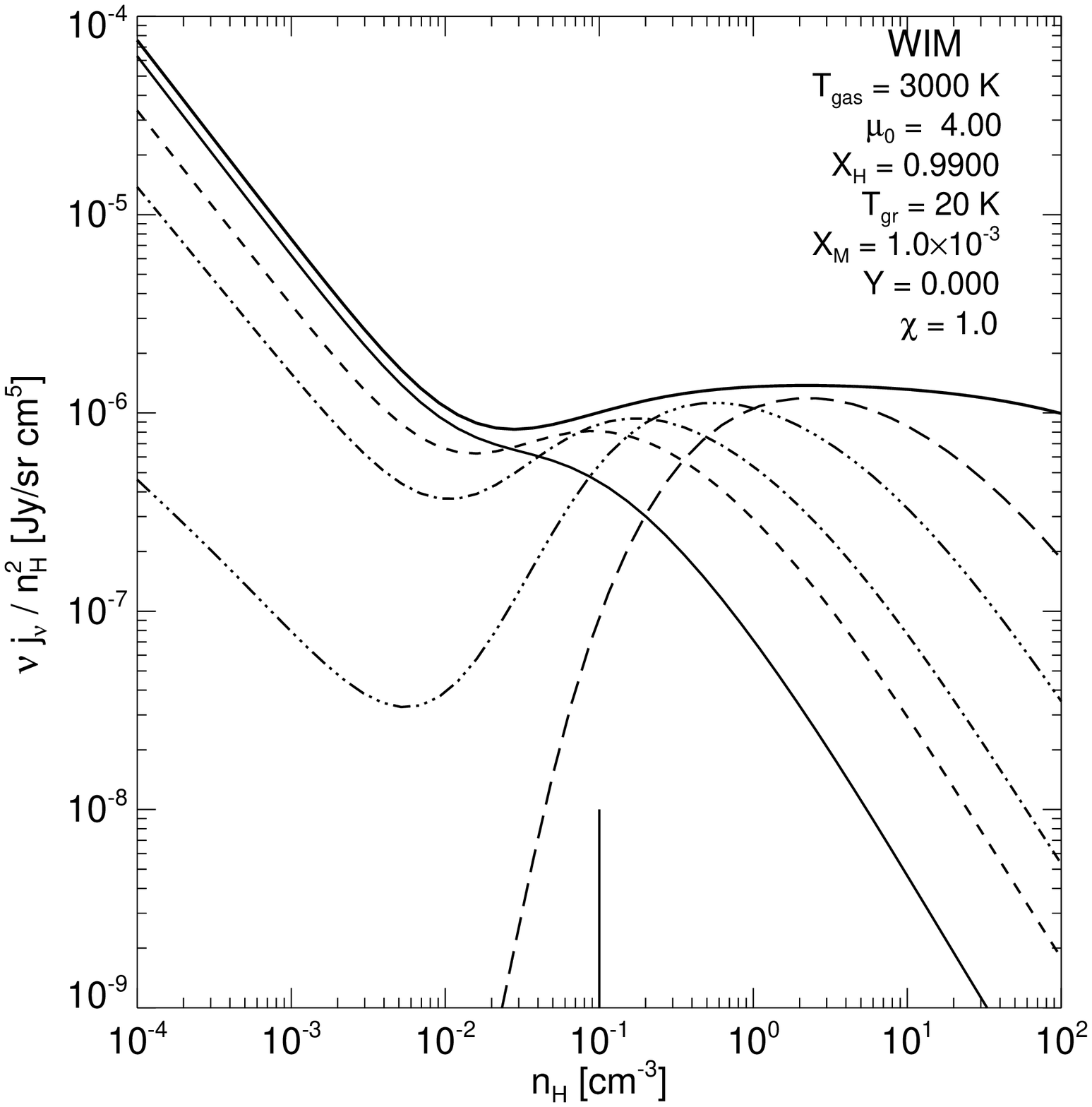}
  \includegraphics[width=0.47\textwidth]{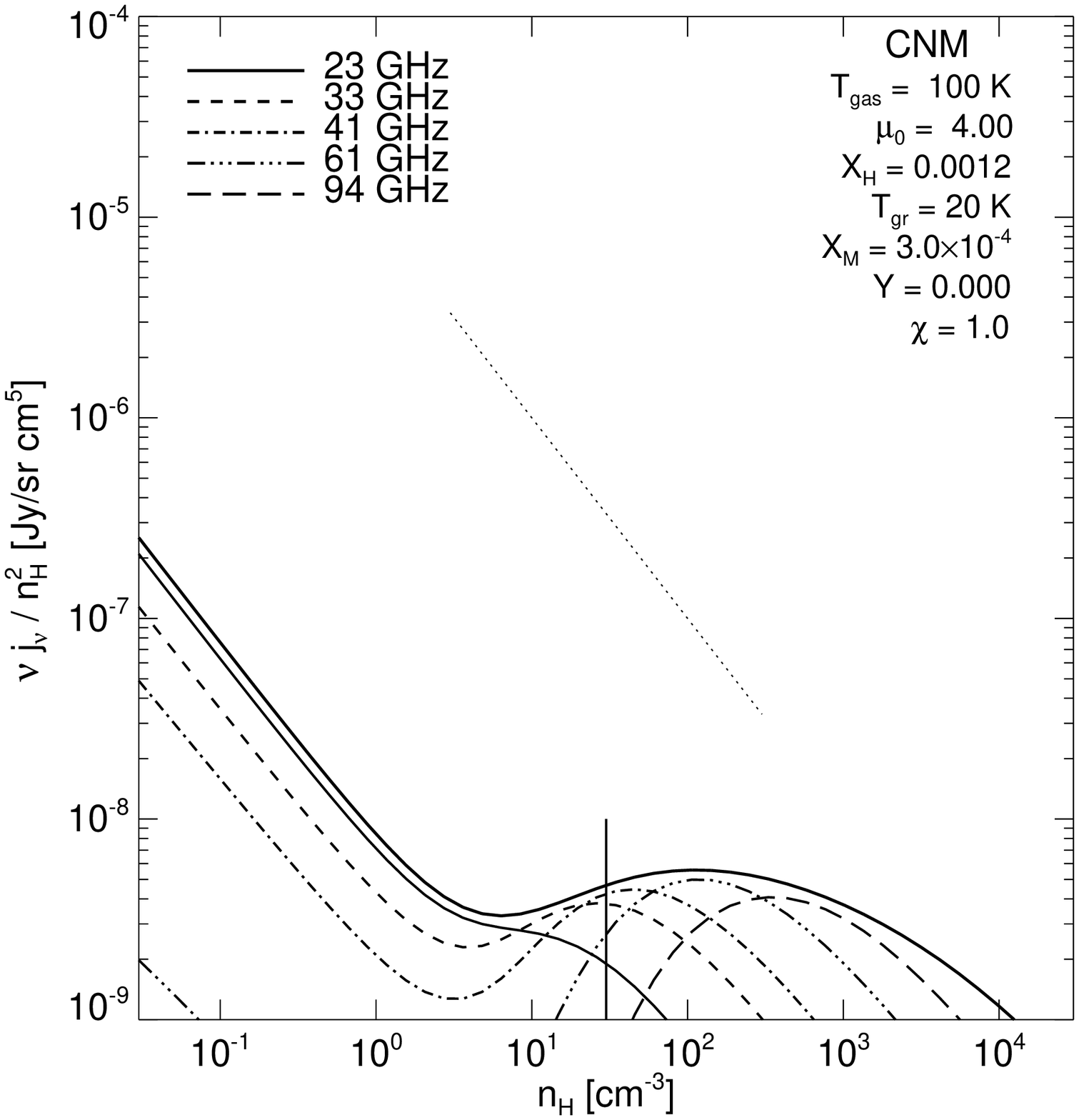}
}
\centerline{
  \includegraphics[width=0.47\textwidth]{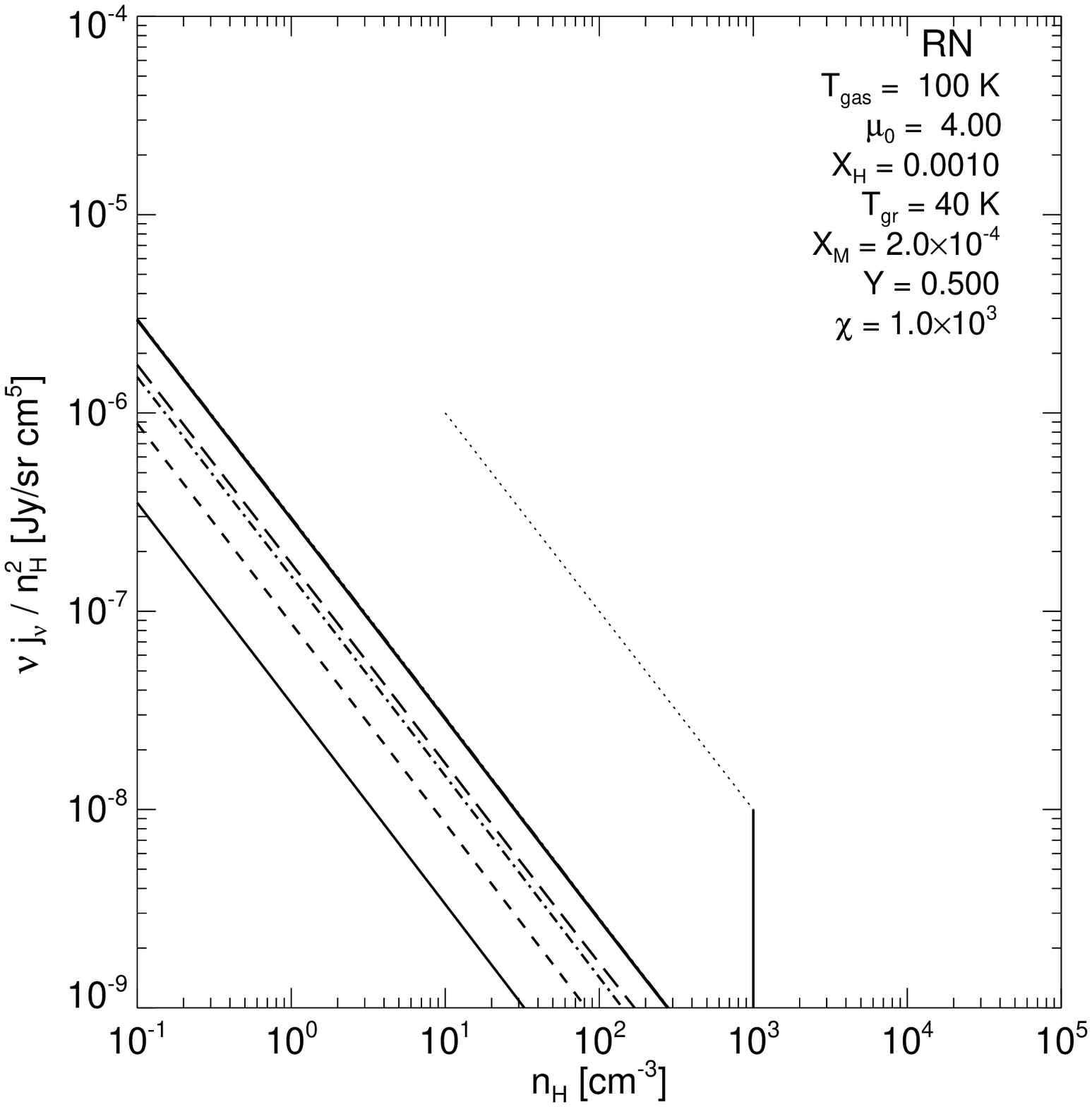}
  \includegraphics[width=0.47\textwidth]{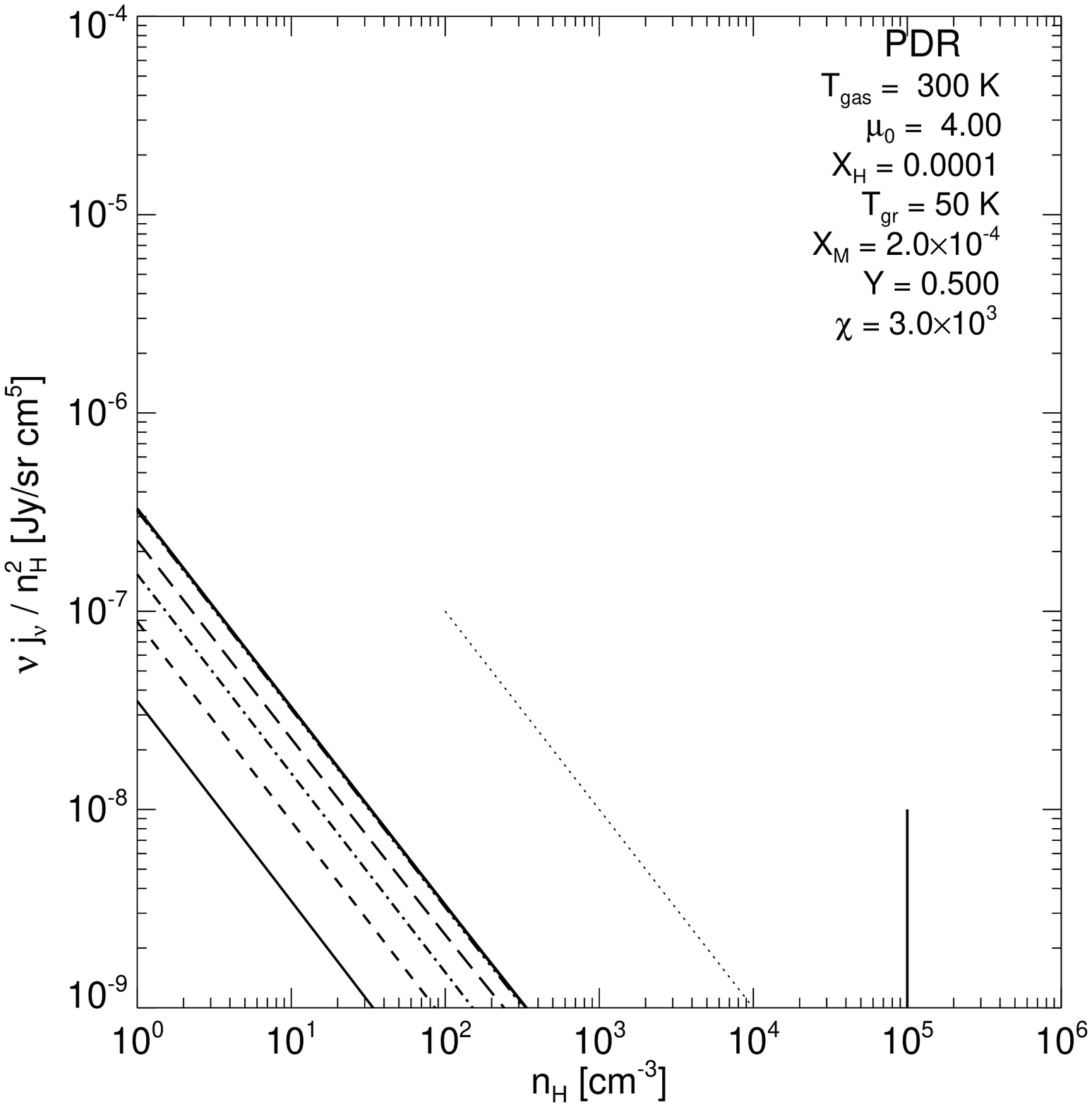}
}
\caption{
Spinning dust emission per $\nH^2$ for the WIM (\emph{upper 
left}) CNM (\emph{upper right}), RN (\emph{lower left}), and 
PDR (\emph{lower right}).  Each panel contains $\nu j_\nu / 
\nH^2$ for the 5 WMAP bands: 23 GHz (\emph{solid}), 33 GHz 
(\emph{dashed}), 41 GHz (\emph{dash dot}), 61 GHz 
(\emph{dash triple-dot}), and 94 GHz (\emph{long dash}).  
The upper solid line is the integrated emissivity, $\int 
j_\nu d\nu / \nH^2$.  Some panels include a line slope of -1 
(\emph{dotted}), corresponding to $j_\nu / \nH$ constant 
with $\nH$.  This limit is generally approached when photons 
dominate, either because of low density or intense radiation 
field.  The solid vertical line marks the value chosen by 
DL98 for each environment.
}
\label{fig:nujnu_per_em}
\epm

In this section, we summarize the spinning dust model used 
in our analysis of the 5-year H$\alpha$- and 
dust-correlated spectra below. The model for excitation 
mechanisms and dipole moment distributions is described in 
detail in DL98.  We find that the most relevant parameters 
are the density, $\nH$ and the typical electric dipole 
moment of 1 nm grains, $\mmu$.

We use the updated grain size distribution given in 
\cite{draine07}.  Our grain size distribution is a double 
lognormal distribution characterized by a mean and width for 
the small and large grain populations --- $\mean{a} =$ 4 and 
$20$\AA\ and $\sigma_a =$ 0.4 and 0.55 respectively 
\citep{draine07}.  This size distribution has a lower cutoff 
of $a_{\rm min} = 3.55$\AA\, which corresponds to $N_{\rm 
atoms} = 20$ Carbon atoms. The dipole moment distribution 
for the grains consists of three delta functions so that 
50\% of the grains have $\mu = \mmu (a/1\mbox{ nm})^{3/2}$, 
25\% have $\mu$ half as large, and 25\% have $\mu$ twice as 
large.

The spinning dust emission in the DL98 model comes from 
polycyclic aromatic hydrocarbons (PAHs) of very small size, 
$N_{\rm atom} \sim 100$ or less.  Small asymmetries in the 
grain geometry result in non-zero dipole moments that are 
taken to scale with the number of atoms as $\mu \propto 
\sqrt{N_{\rm atom}}$.  As these grains are spun up by 
various excitations processes (see DL98), they emit electric 
dipole radiation with total power $P \propto \omega^4$, 
where $\omega/2\pi$ is the angular rotation frequency and is 
tens of GHz.\footnote{ An improved treatment of the spinning 
dust problem by Ali-Hamoud \& Hirata has made a few 
refinements of the DL98 model but those modifications do not 
significantly affect the results of this paper (Ali-Hamoud 
\& Hirata priv. comm.)}

The parameters which characterize the environment and grain 
properties, with values appropriate for the WIM and CNM 
(cold neutral medium), are summarized in 
\reftbl{spindustparams}.  The top two panels of 
\reffig{spin_spec_ex} shows example spinning dust spectra 
for WIM parameters while holding the density $\nH$ fixed 
while varying the characteristic dipole moment $\mmu$ and 
vice-versa.  These spectra are in units of 
$I_{\nu}/I_{H\alpha}$, or total intensity per intensity at 
H$\alpha$ with $\Tgas =$ 3000 K (our justification for using 
this gas temperature, which is lower than the commonly used 
8000 K, is given in \refsec{hacorr}).  For fixed $\mmu$, as 
$\nH$ is increased, the spectrum shifts up in frequency with 
very little change in overall power per H$\alpha$.  Since 
$I_{H\alpha} \propto \nH^2$ in fully ionized environments, 
the total spinning dust intensity goes roughly like $\nH^2$.  
This point will be explored in more detail shortly, but it 
illustrates that an EM map like the H$\alpha$ map should 
trace spinning dust emission.

For fixed $\nH$, as $\mmu$ is increased 
\reffig{spin_spec_ex} shows that the spectrum increases in 
amplitude and decreases in peak frequency.  This can be 
understood in terms of the radiated power per grain which 
goes as $\mu^2$: grains with larger dipole moments radiate 
away power faster and are thus harder to spin up leading to 
a lower peak frequency, but they also emit more total power 
leading to an increased amplitude.

The lower four panels of \reffig{spin_spec_ex} are contour 
plots showing the effects of varying $\nH$ and $\mmu$ on the 
amplitude and peak frequency of the spinning dust spectrum.  
Contours are shown for both the WIM and CNM conditions. The 
contours of peak frequency show that, if the amount of dust 
per H is assumed unknown (i.e., the amplitude is allowed to 
vary), then there is a strong degeneracy in the peak 
frequency between $\nH$ and $\mmu$.  The implication of this 
strong degeneracy is that a measurement of the spinning dust 
spectrum with sparse frequency sampling that only gives 
information about the peak frequency (as is the case with 
the WMAP data below, see \reffig{comp-contours}) cannot 
uniquely constrain these two parameters.  On the other hand, 
if the grain abundance is assumed known, than the degeneracy 
is broken and the parameters are more tightly constrained.

\reffig{nujnu_per_em} shows the behavior with $\nH$ of the 
emissivity per EM for both the total emission, $\int j_{\nu} 
d\nu$, as well as $\nu j_{\nu}$ evaluated at the five WMAP 
frequencies.  The parameters of the WIM and CNM environments 
are also given in \reftbl{spindustparams} while those for 
the reflection nebula (RN) and photo-dissociation region 
(PDR) are shown on the plot.  In the limit of low $\nH$, all 
four environments exhibit roughly $j_{\nu} \propto \nH$ 
behavior. In this limit, the dominant excitation and 
de-excitation mechanism are absorption and emission of 
infrared photons and thus the total emission scales linearly 
with density.  This is in contrast to the argument presented 
in DF08b who argued that in the lowest density regions, the 
grains are in the episodic limit and so the total emission 
should scale as $\nH^2$.  That argument neglected the 
importance of photon interactions.

While the linear $\nH$ scaling persists through all 
densities shown here for the RN and PDR regions, the WIM and 
CNM both flatten to $j_{\nu} \propto \nH^2$ behavior at $\nH 
\sim 0.01$ and 5 cm$^{-3}$ respectively. In this regime, the 
dominant spin up mechanism is collisions with ions and the 
$\nH^2$ scaling persists for four orders of magnitude in 
$\nH$ for the WIM ($\sim 10^{-2}-10^2$ cm$^{-3}$).  As the 
density increases further in the CNM, the spectrum turns 
over and again goes roughly like $j_{\nu} \propto \nH$.  At 
these high densities, both spin up and damping are dominated 
by thermal processes (collisions with ions and atoms).  
However, it is interesting to note that over the density 
ranges shown in \reftbl{spindustparams}, both the WIM and 
CNM models scale as $\nH^2$ indicating that a map that 
scales with EM like \emph{the H$\alpha$ map should trace WIM 
and CNM spinning dust emission}.

\section{Three Component Foreground Spectra}
\label{sec:interpretation}
Given that the H$\alpha$ map should trace both the free-free and a spinning dust 
emission component, and given the fact that our foreground spectra are 
necessarily contaminated by a component with the spectrum of the CMB (flat in 
thermodynamic $\Delta T$) from the CMB cross-correlation bias described in DF08a 
and \refsec{template_fits}, we interpret the H$\alpha$-correlated 
emission in \reffig{halpha-dust-spec} as a three component spectrum.  As in 
DF08b, we fit a free-free plus CMB plus DL98 WIM spinning dust model,
\begin{eqnarray}
  I^{\rm mod}_{\nu} & = & F_0 \left( \frac{\nu}{23\mbox{ GHz}} \right)^{-0.15}
    + C_0 \left(\frac{\nu}{23\mbox{ GHz}} \right)^{2} \frac{1}{\planck(\nu)}
    \nonumber\\
  & & + D_0 \times \left(\mbox{DL98 WIM}\right),
\label{eq:threecomp}
\end{eqnarray}
to the measured cross-correlation spectrum $I_{\nu}$.  Here, $plc$ is the 
``planck correction'' factor which converts thermodynamic $\Delta T$ to antenna 
temperature at frequency $\nu$ \citep{firas_supp}.   We choose the parameters 
$\nH$ and 
$\mmu$, generate a spinning dust spectrum, and minimize $\chi^2 \equiv \sum_i 
(I_{\nu} - I^{\rm mod}_{\nu})^2/\sigma_i^2$, where $\sigma_i$ are the
errors in the fit in each band $i$, over the parameters $F_0$, $C_0$,
and $D_0$.  We allow $D_0$ to float in the fit because of uncertainty
about the PAH size distribution and abundance in the WIM.  We
concentrate on CMB5, but our results are minimally changed for the other
CMB estimators, it is only the value of $C_0$ which is significantly
affected.

\subsection{H$\alpha$-Correlated Emission}
\label{sec:hacorr}

\bp
\centerline{
  \includegraphics[width=0.47\textwidth]{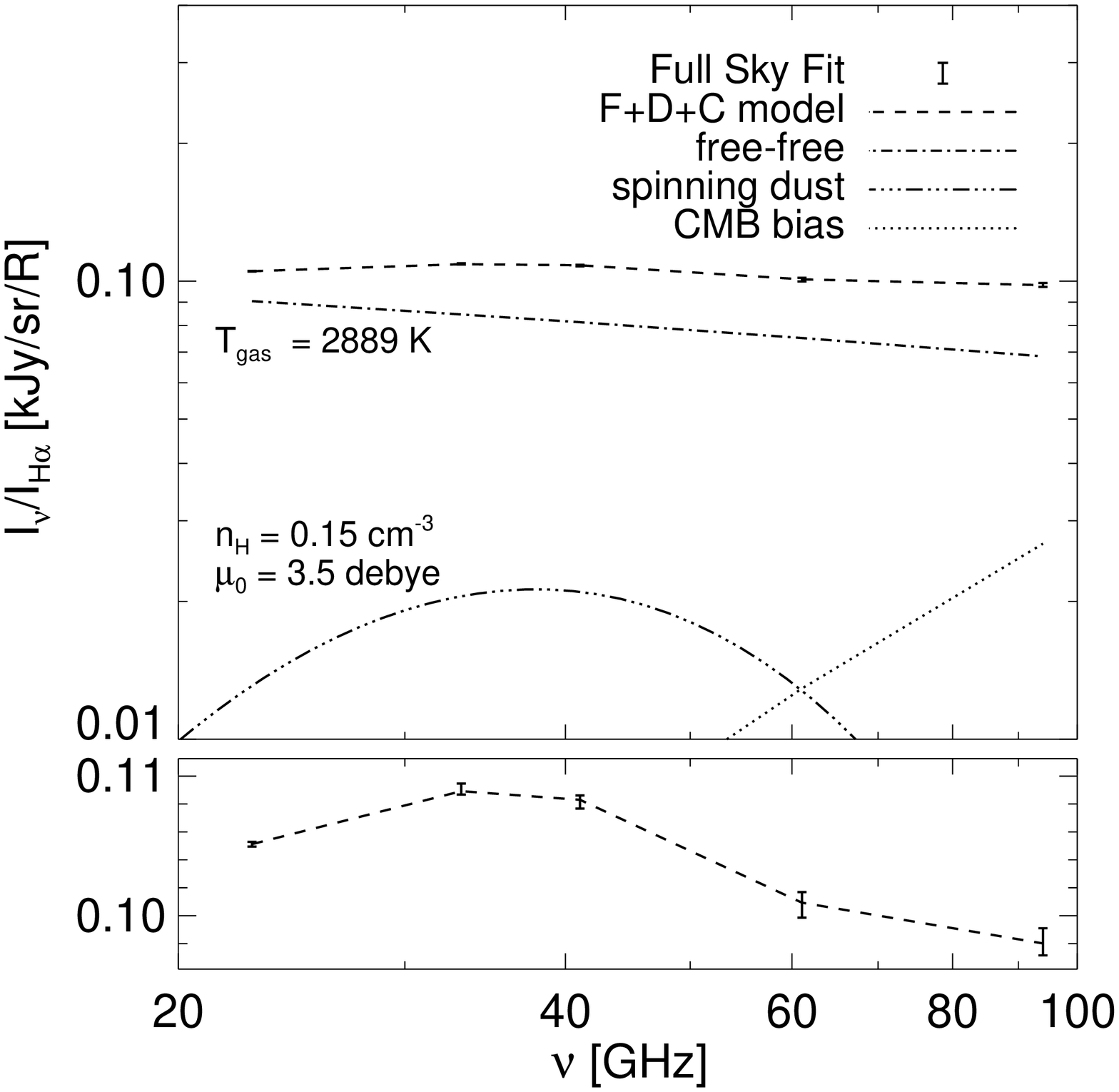}
}
\caption{
\emph{Upper:} The H$\alpha$-correlated spectrum from 23 to 
94 GHz (points) using CMB5 and the three-component fit to 
the data (dashed line).  The three components, free-free, 
spinning dust, and CMB, are shown separately.  The amplitude 
of the free-free component yields $\Tgas \approx 3000$ K, 
while the amplitude of the spinning dust component is less 
than unity indicating that either PAHs are depleted in the 
WIM or the grain size distribution is altered.  
\emph{Lower:} A zoom in of the spectrum showing that, 
although the bump is subtle, the statistical significance is 
very high.
}
\label{fig:halpha-twopanel}
\ep

\bpm
\centerline{
  \includegraphics[width=0.47\textwidth]{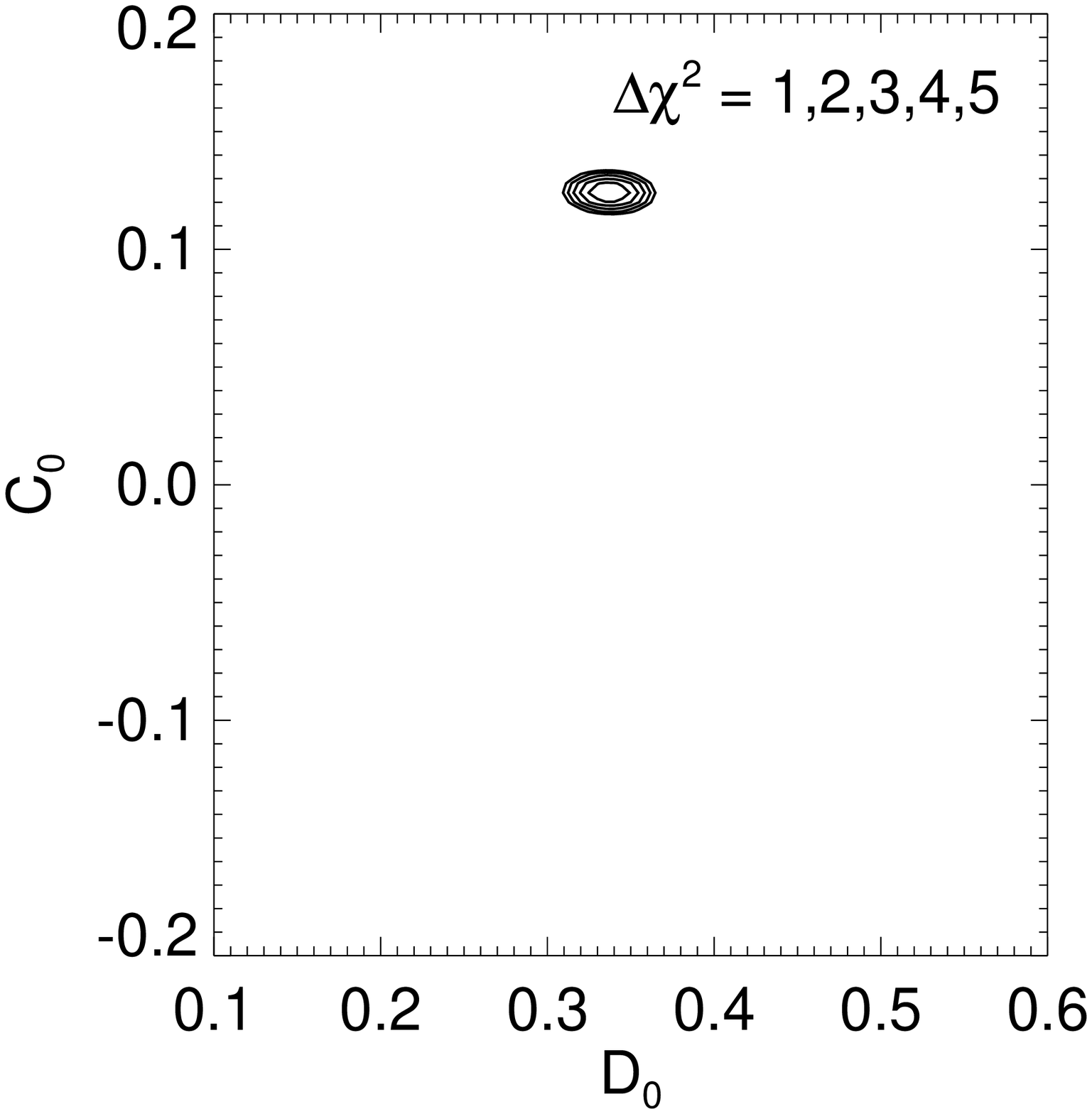}
  \includegraphics[width=0.47\textwidth]{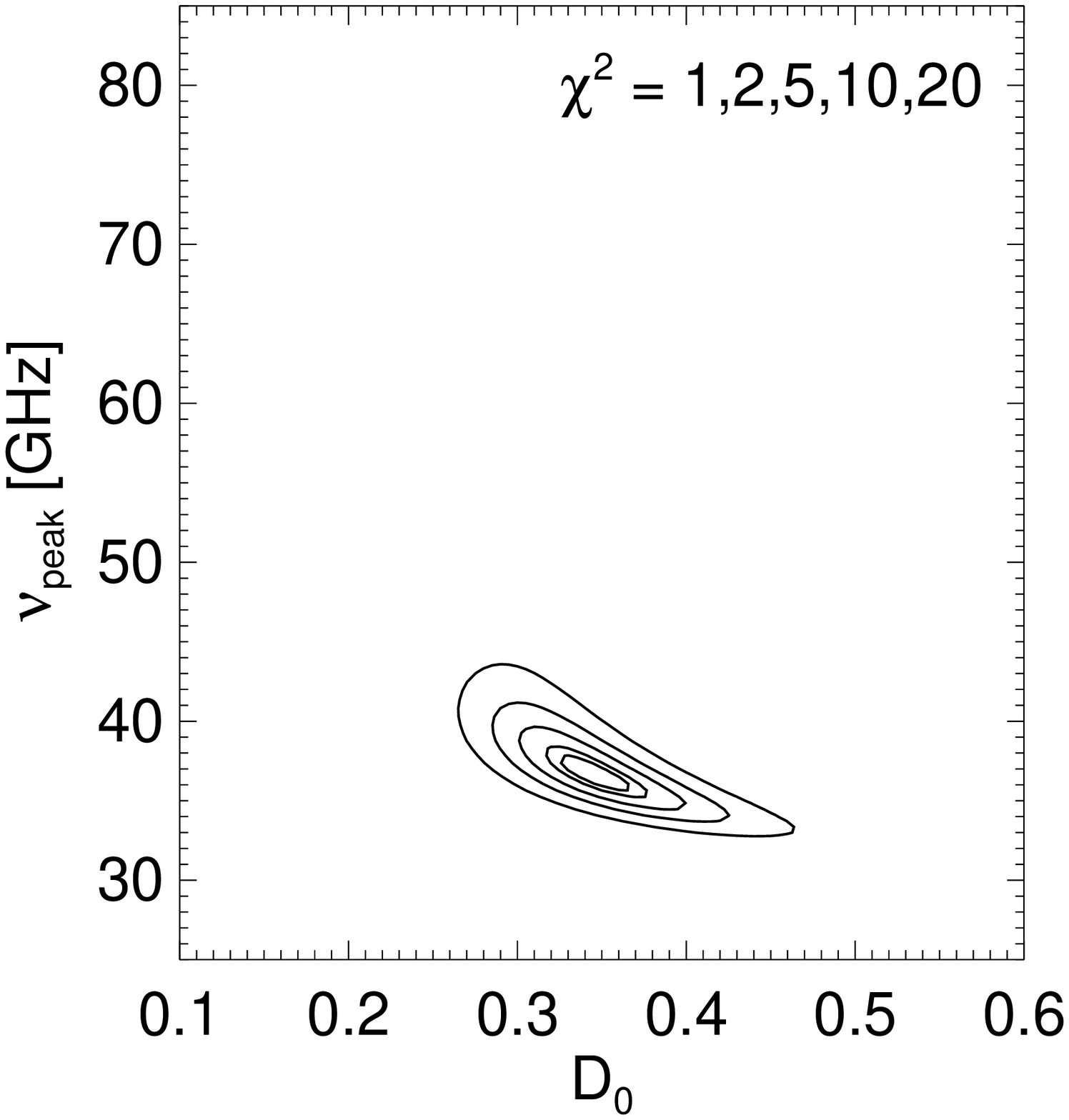}
}
\caption{
\emph{Left:} $\Delta\chi^2$ contours in the $(D_0,C_0)$ 
plane.  The null hypothesis that the H$\alpha$-correlated 
emission follows a free--free power law with a contamination 
by the CMB spectrum is ruled out to very high significance 
($\chi^2 \sim 740$ for 3 degrees of freedom at $D_0=0$).  
\emph{Right:} $\chi^2$ contours in the $(D_0,\nu_{\rm 
peak})$ plane.  Though the peak frequency is not constrained 
to high accuracy, $\nu_{\rm peak} \sim 37$ GHz is roughly 
the best fit value.
}
\label{fig:simp-contours}
\epm

\bp
\centerline{
  \includegraphics[width=0.47\textwidth]{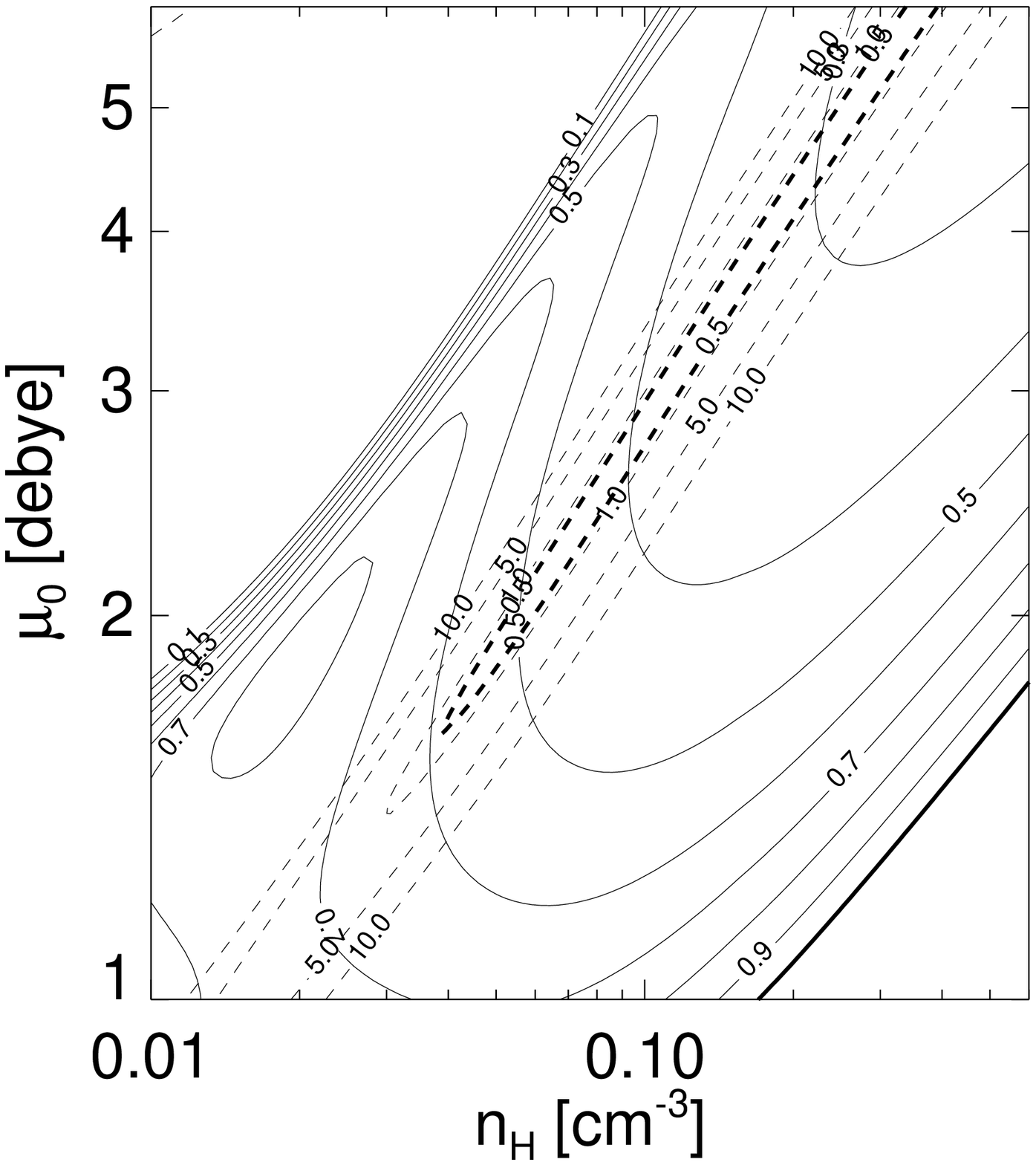}
}
\caption{
The $\chi^2$ (dashed) and $D_0$ (solid) contours in the 
$(\nH,\mmu)$ plane.  There is a deep narrow valley in the 
$\chi^2$ surface representing the near degeneracy in effects 
on the spinning dust peak frequency model with increasing 
$\nH$ and decreasing $\mmu$.  The values $\nH = 0.15 
\cm^{-3}$ and $\mmu = 3.5$ debye produce an acceptable fit 
with $D_0 \approx 0.3$.  Although $D_0$ does vary along the 
narrow valley in $\chi^2$, $D_0 = 1$ does not yield an 
acceptable fit, indicating that either the PAHs are depleted 
in the WIM or the grain size distribution is altered.
}
\label{fig:comp-contours}
\ep

\reffig{halpha-twopanel} shows the results of our three 
component fit for the H$\alpha$-correlated emission.  From 
the figure, it is clear that a spinning dust spectrum with 
$\nH$ and $\mmu$ set to $0.15 \ \cm^{-3}$ and $3.5$ debye 
respectively (and with the WIM parameters in 
\reftbl{spindustparams}) fits the data remarkably well.  
This value for $\mmu$ is comparable to the earlier estimates 
in DL98.  This spectrum has a peak frequency of $\nup = 
37.2$ GHz and $D_0$ fit coefficient of 0.34.  It is 
interesting to note that the value of $D_0$ is less than 
unity.  This may indicate either that PAHs are depleted in 
the diffuse WIM by a factor of $\sim 3$ or that the grain 
size distribution differs from that in the CNM from which 
the values of the log-normal mean and width of our grain 
size distribution were derived \citep{draine07}.

From the amplitude of the $F_0$ fit coefficient, we can 
derive a gas temperature in the WIM which we find to be 
$\Tgas \approx 3000$ K.  This is the origin of the choice of 
$\Tgas$ in the WIM spinning dust model.  This value for 
$\Tgas$ is much lower than the temperature inferred from 
emission line ratios \citep[e.g., see][]{madsen06}.  We 
stress that this low gas temperature is \emph{not} the 
result of assigning some of the intensity to a spinning dust 
component.  Ignoring the bump and merely fitting the data 
with a free-free only spectrum, still gives a gas 
temperature below 5000K.  Using a technique similar to ours, 
\citet{davies06} also report anomalously low gas 
temperatures based on the free-free to H$\alpha$ ratio in 
the WMAP 1-year data and DF08b found similar temperatures in 
the 3-year data.  The origin of this discrepancy is unclear, 
though it may result from temperature variations in the WIM 
along the line of sight as discussed in 
\citet{heiles01}.\footnote{Another possibility is that our 
template fit of the spectrum is dominated by the strongest 
H$\alpha$ features which naturally have a lower 
temperature.}  We will explore implications of a 
two-component temperature model in future work.  Lastly, we 
find that the total intensity in H$\alpha$-correlated 
spinning dust emission is roughly 25\% of that in free-free 
emission at 41 GHz.

The upper panel in \reffig{halpha-twopanel} shows that the 
bump in the spectrum is indeed subtle, but the zoom in 
beneath reveals that it is highly statistically significant.  
Since we minimize $\chi^2$ over three amplitudes ($F_0$, 
$D_0$, and $C_0$), set the values of $\nH$ and $\mmu$ by 
hand, and fit to five data points, the number of degrees of 
freedom is formally zero. However, Figures 
\ref{fig:simp-contours} and \ref{fig:comp-contours} show 
that the $\chi^2$ rises quickly as we deviate from these 
parameters.  In the left panel of \reffig{simp-contours}, 
$\Delta\chi^2$ contours are shown in the $(D_0,C_0)$ plane 
and it is clear that the null hypothesis that the 
H$\alpha$-correlated emission is simply a linear combination 
of free-free plus a CMB bias (i.e., $D_0=0$), is ruled out 
at very high confidence.  In fact we find that the $D_0 = 0$ 
case is ruled out at $\sim 27\sigma$.  The right panel shows 
$\chi^2$ contours in the $(D_0,\nup)$ plane as $\nup$ is 
varied by hand.  This parameter is less well constrained but 
is broadly consistent with $\nup \sim$ 35-40 GHz.

\reffig{comp-contours} shows $\chi^2$ and $D_0$ contours in 
the $(\nH,\mmu)$ plane for the full-sky fit.  The $\chi^2$ 
contours show that there is a long narrow valley in the 
$\chi^2$ surface reflecting the degeneracy in peak frequency 
shifts resulting from varying $\nH$ and $\mmu$ (see 
\reffig{spin_spec_ex}). Since varying $\mmu$ affects the 
overall amplitude of the spinning dust component in units of 
$I_{\nu}/I_{H\alpha}$ while $\nH$ does not (because of the 
density squared behavior of the spinning dust emission at 
these $\nH$), the value of $D_0$ changes as we move along 
the narrow valley of the $\chi^2$ surface.  It is important 
to note that small deviations from these parameters quickly 
lead to large $\chi^2$ values.  Further, given this set of 
environmental conditions and grain sizes, $D_0 = 1$ is ruled 
out; again, implying either PAH depletion or modified grain 
size distributions in the WIM.

\subsection{The Gum Nebula}

\bpm
\centerline{
  \includegraphics[width=0.23\textwidth]{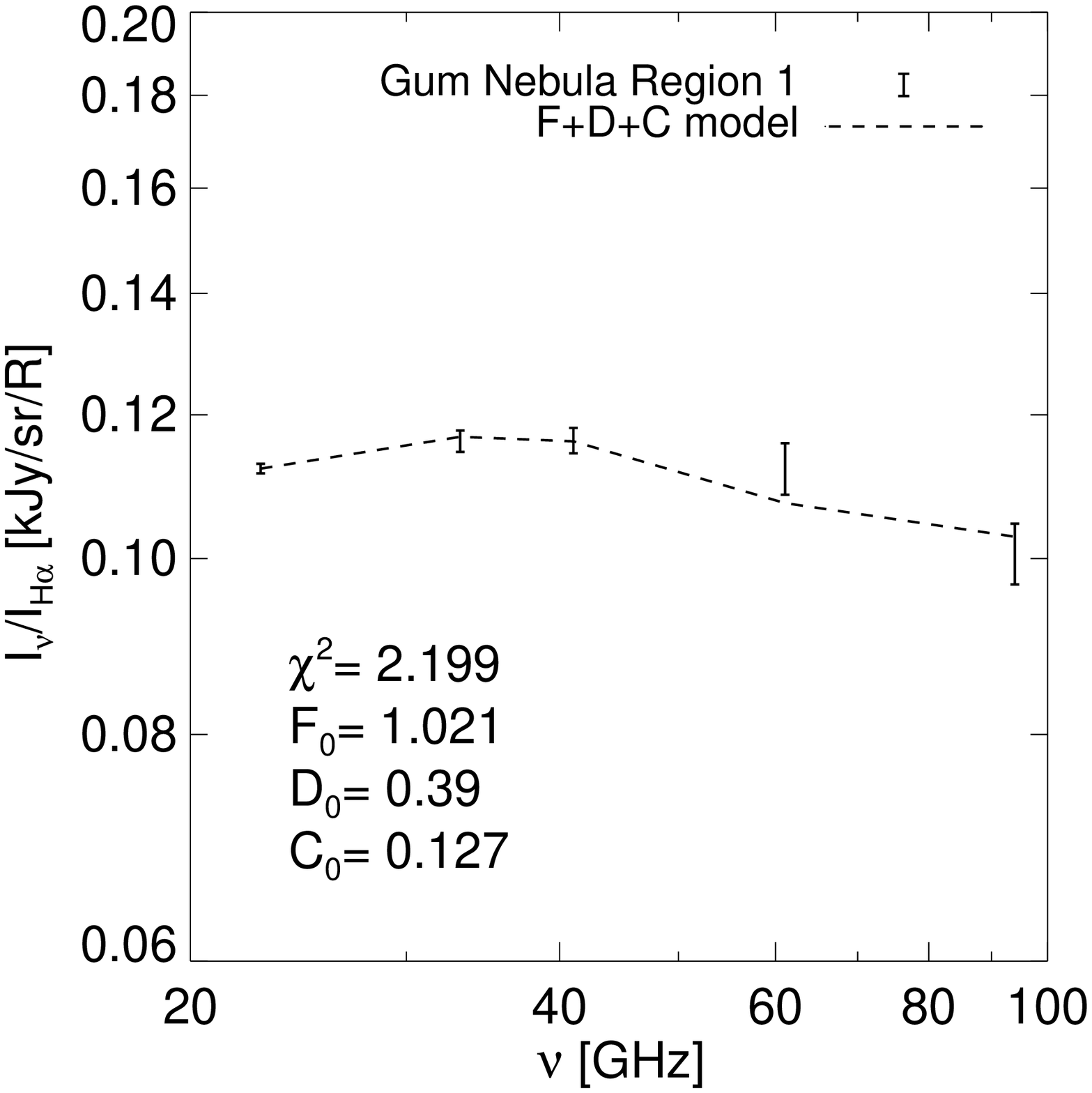}
  \includegraphics[width=0.23\textwidth]{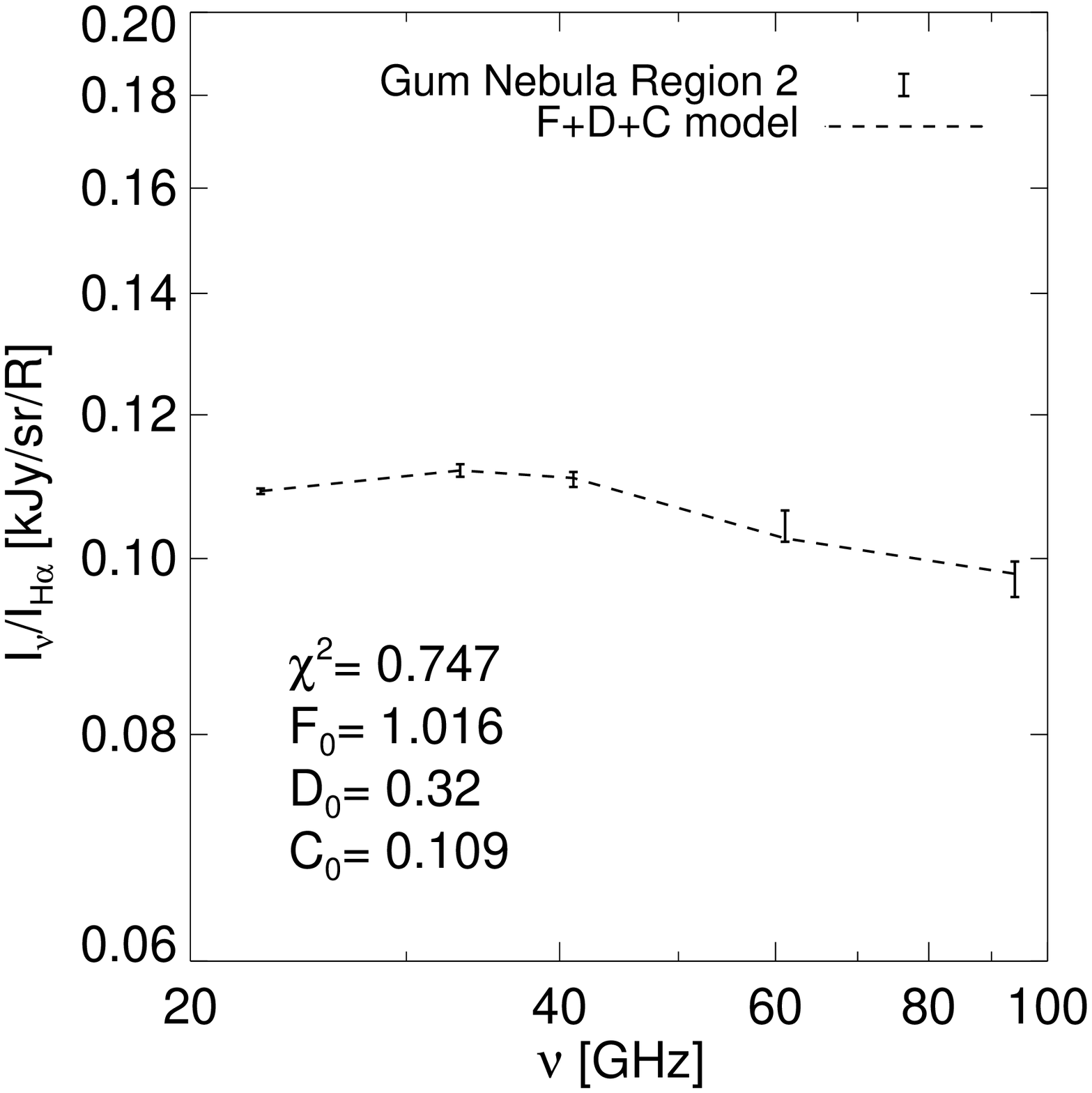}
  \includegraphics[width=0.23\textwidth]{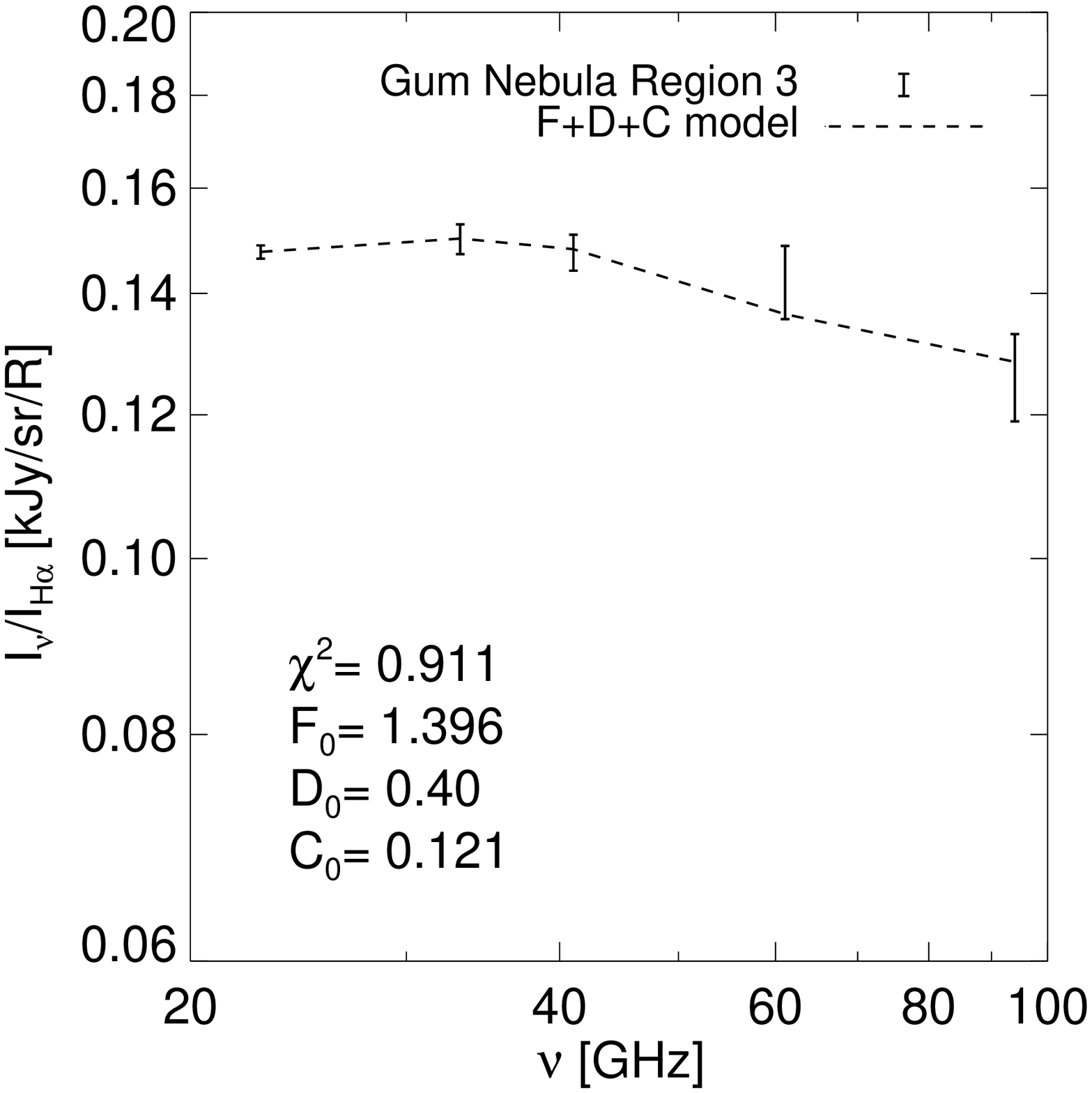}
  \includegraphics[width=0.23\textwidth]{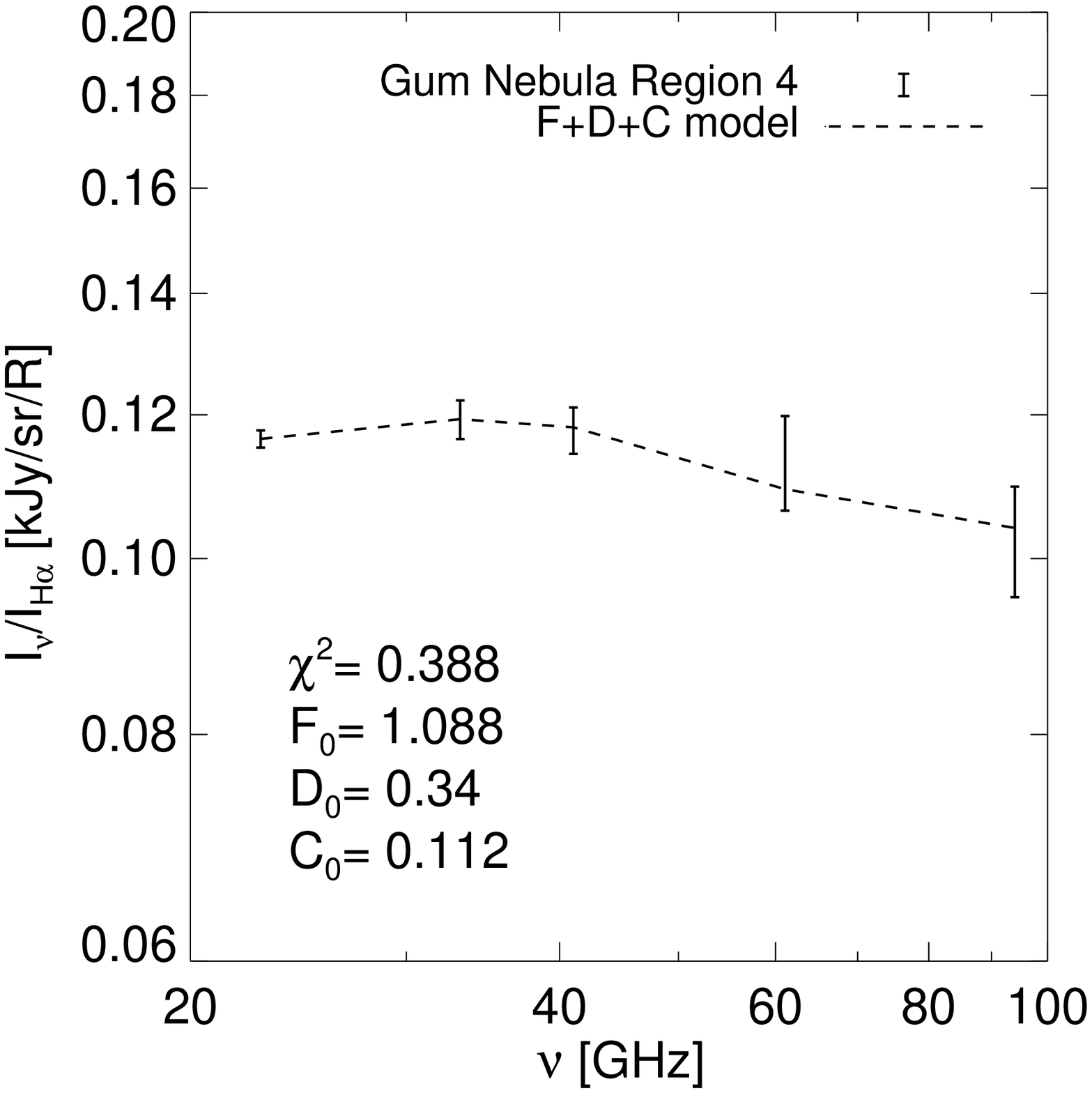}
}
\centerline{
  \includegraphics[width=0.23\textwidth]{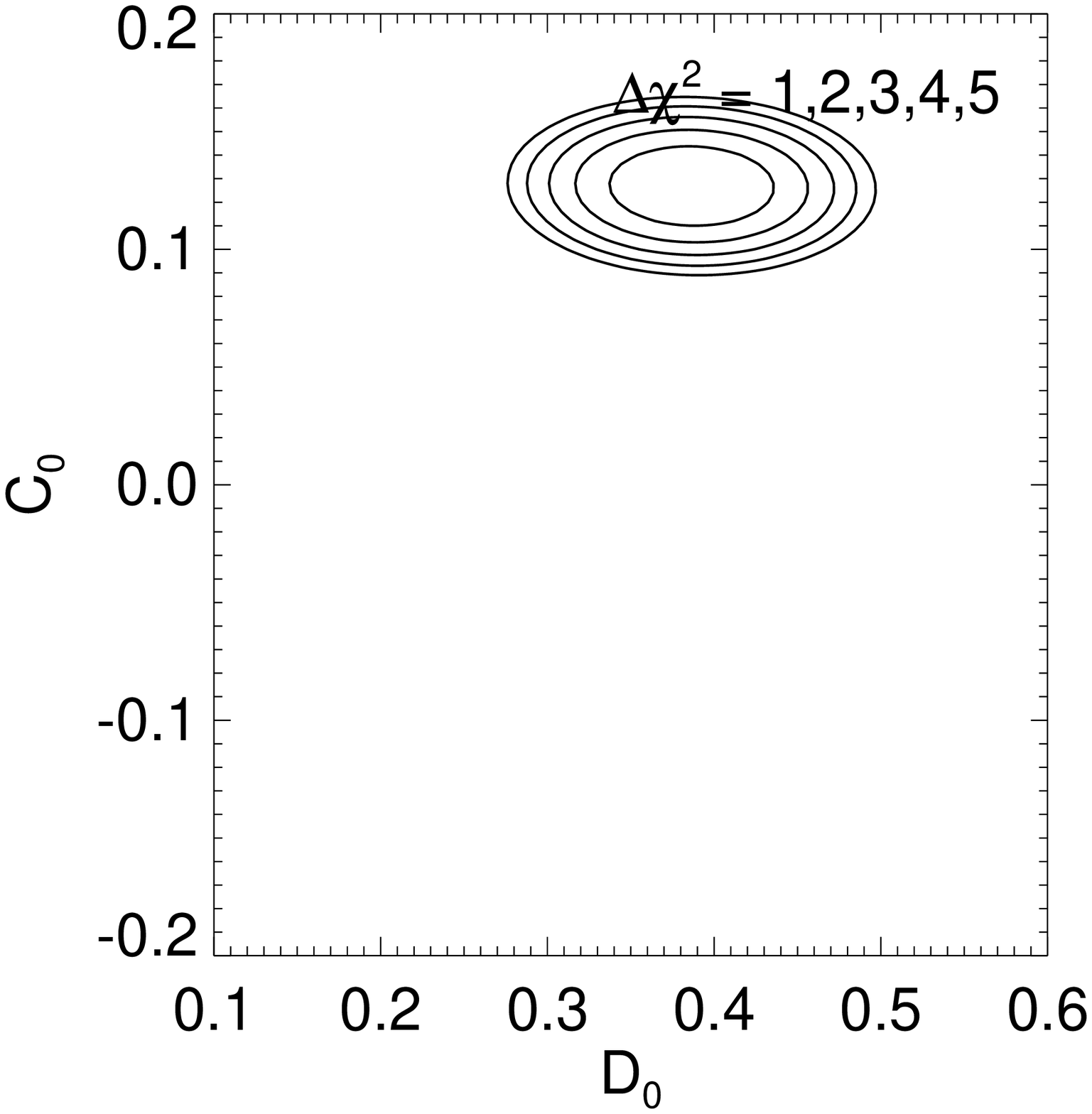}
  \includegraphics[width=0.23\textwidth]{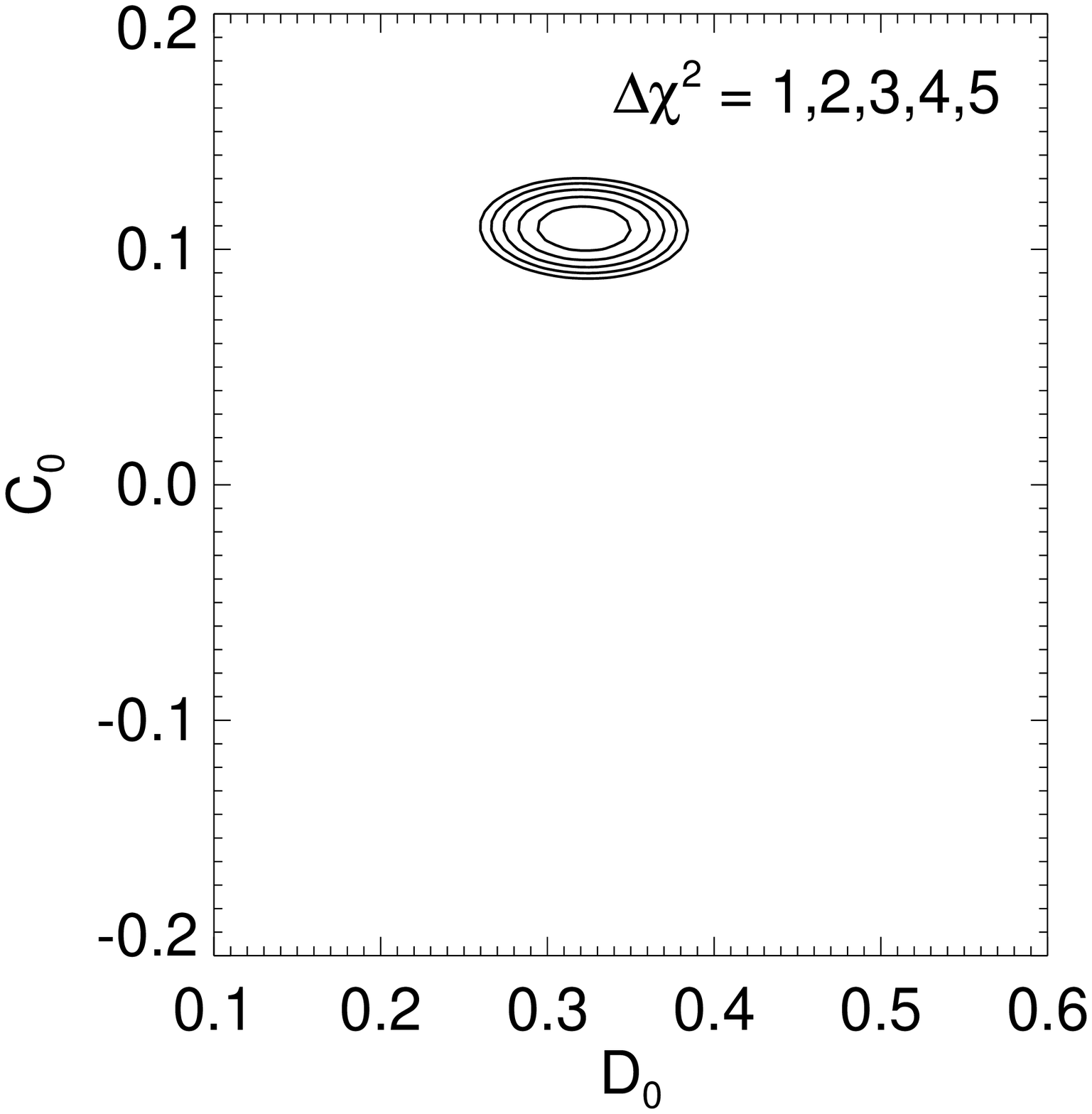}
  \includegraphics[width=0.23\textwidth]{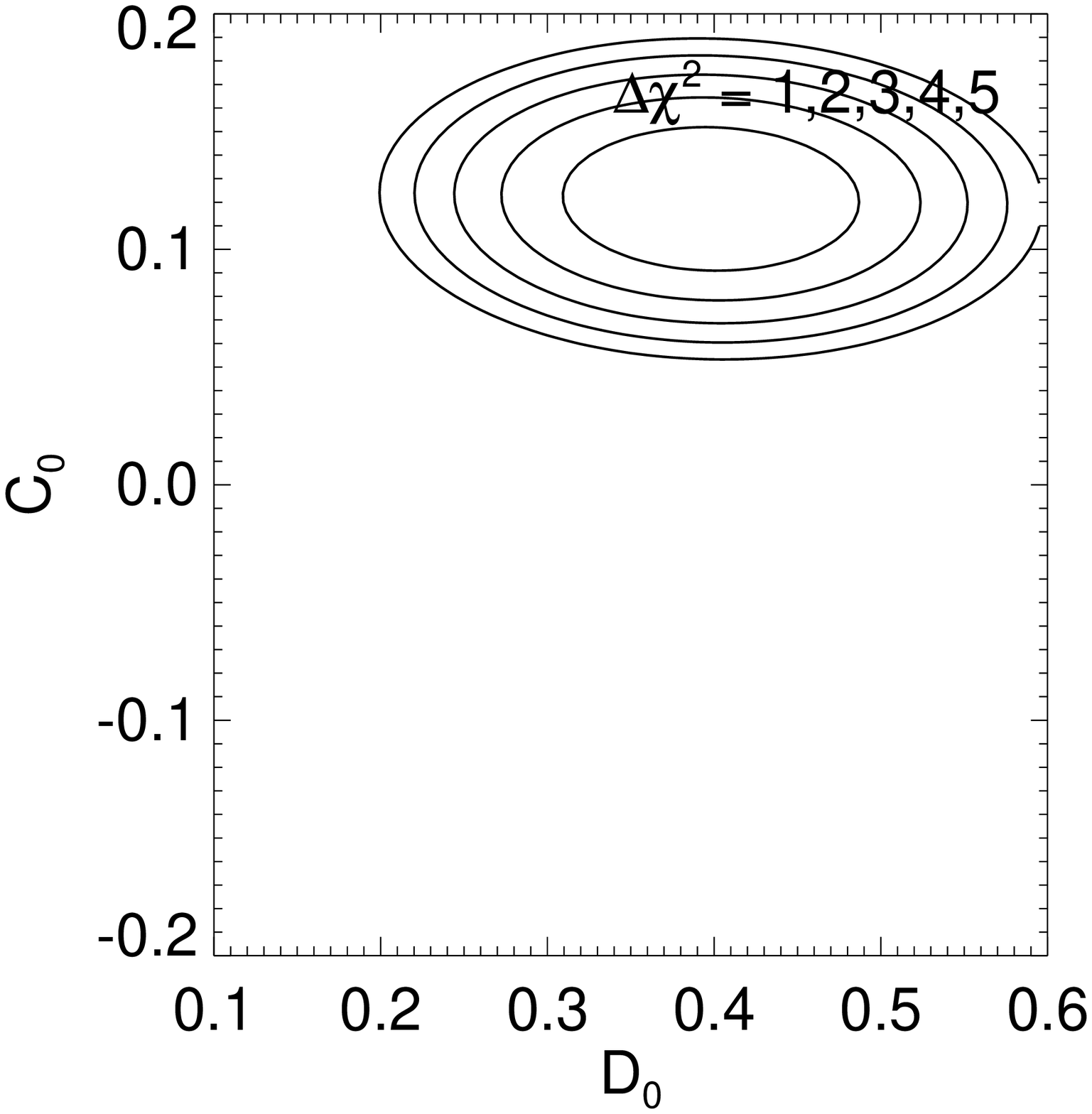}
  \includegraphics[width=0.23\textwidth]{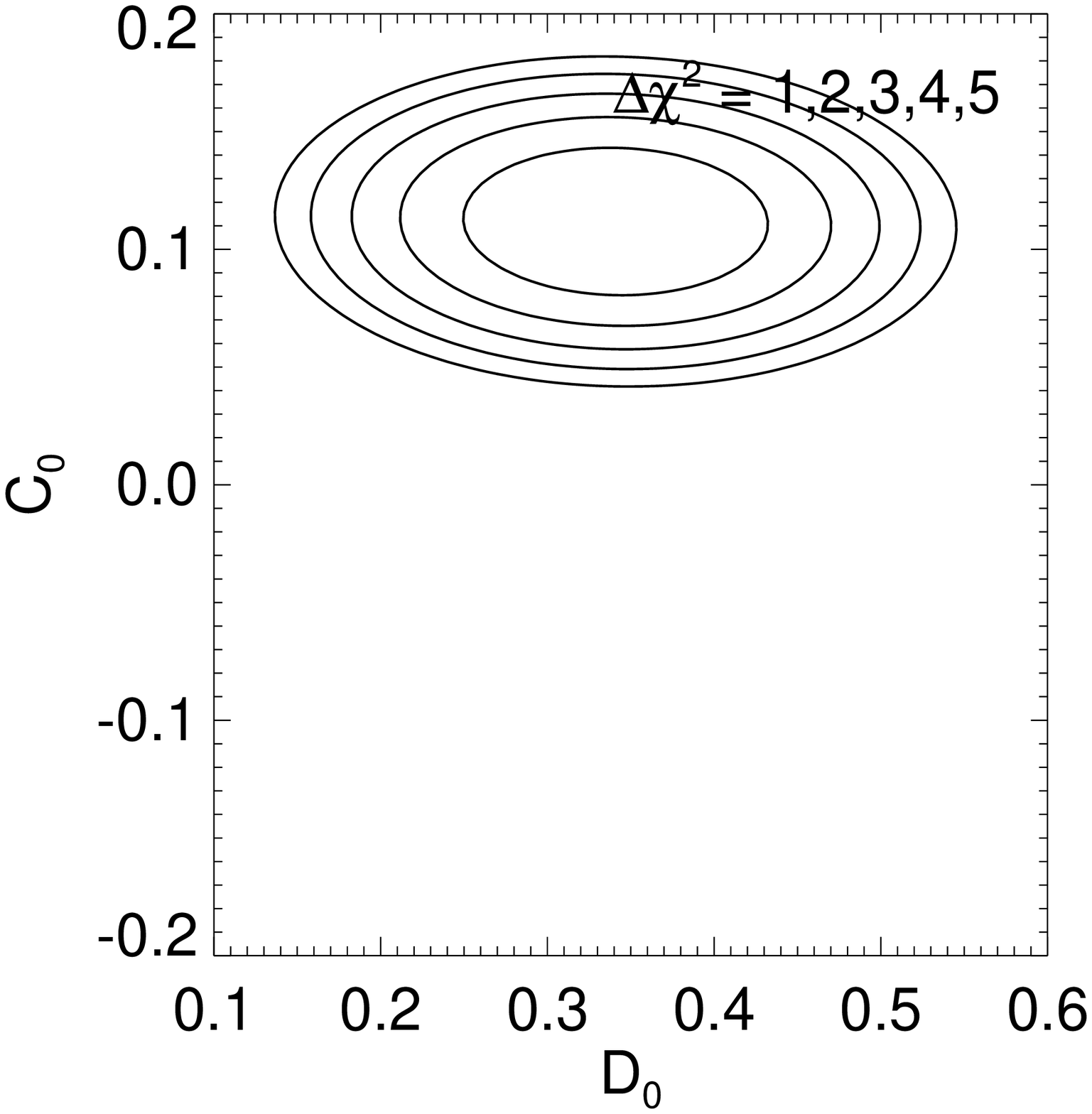}
}
\centerline{
  \includegraphics[width=0.23\textwidth]{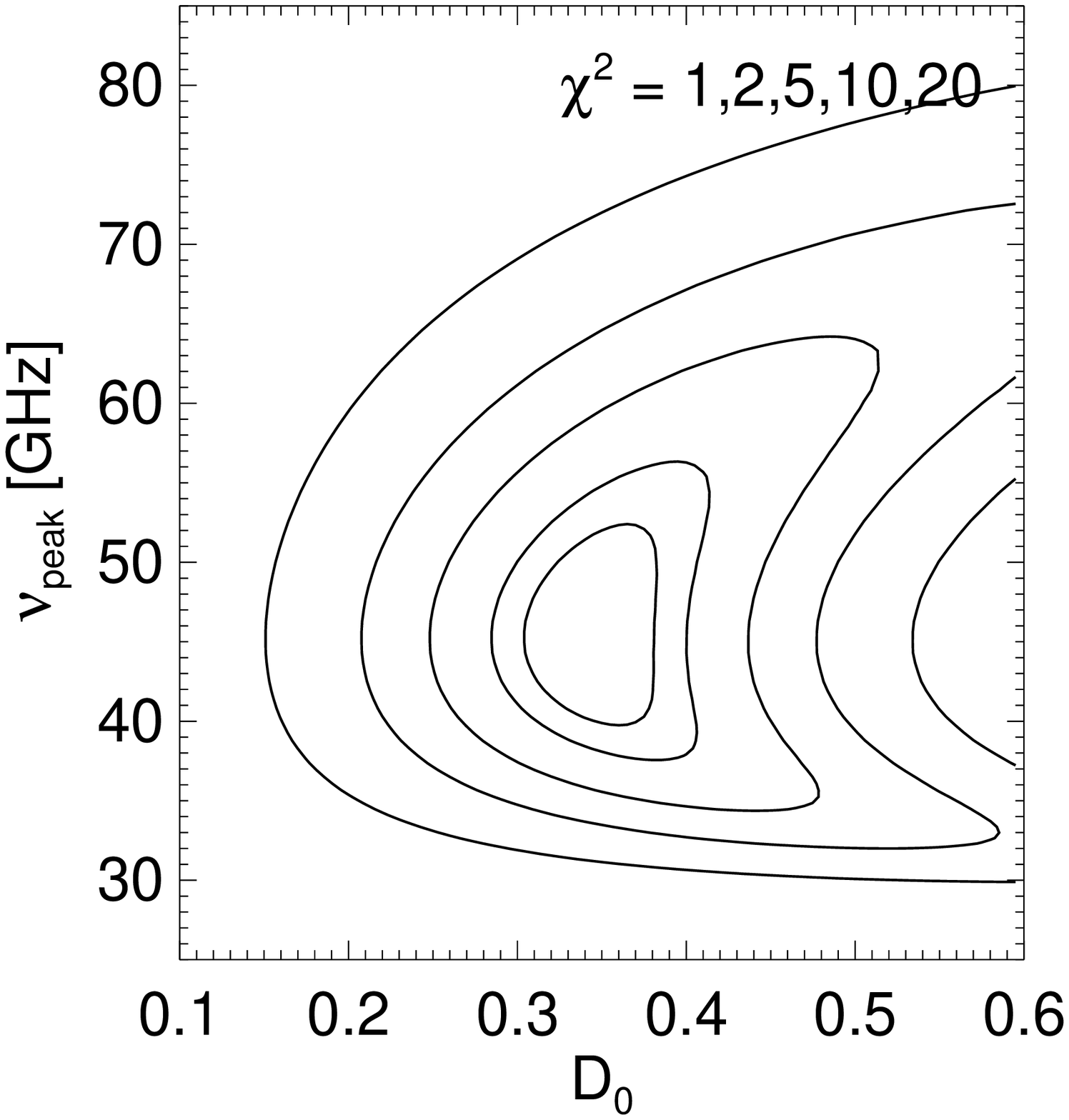}
  \includegraphics[width=0.23\textwidth]{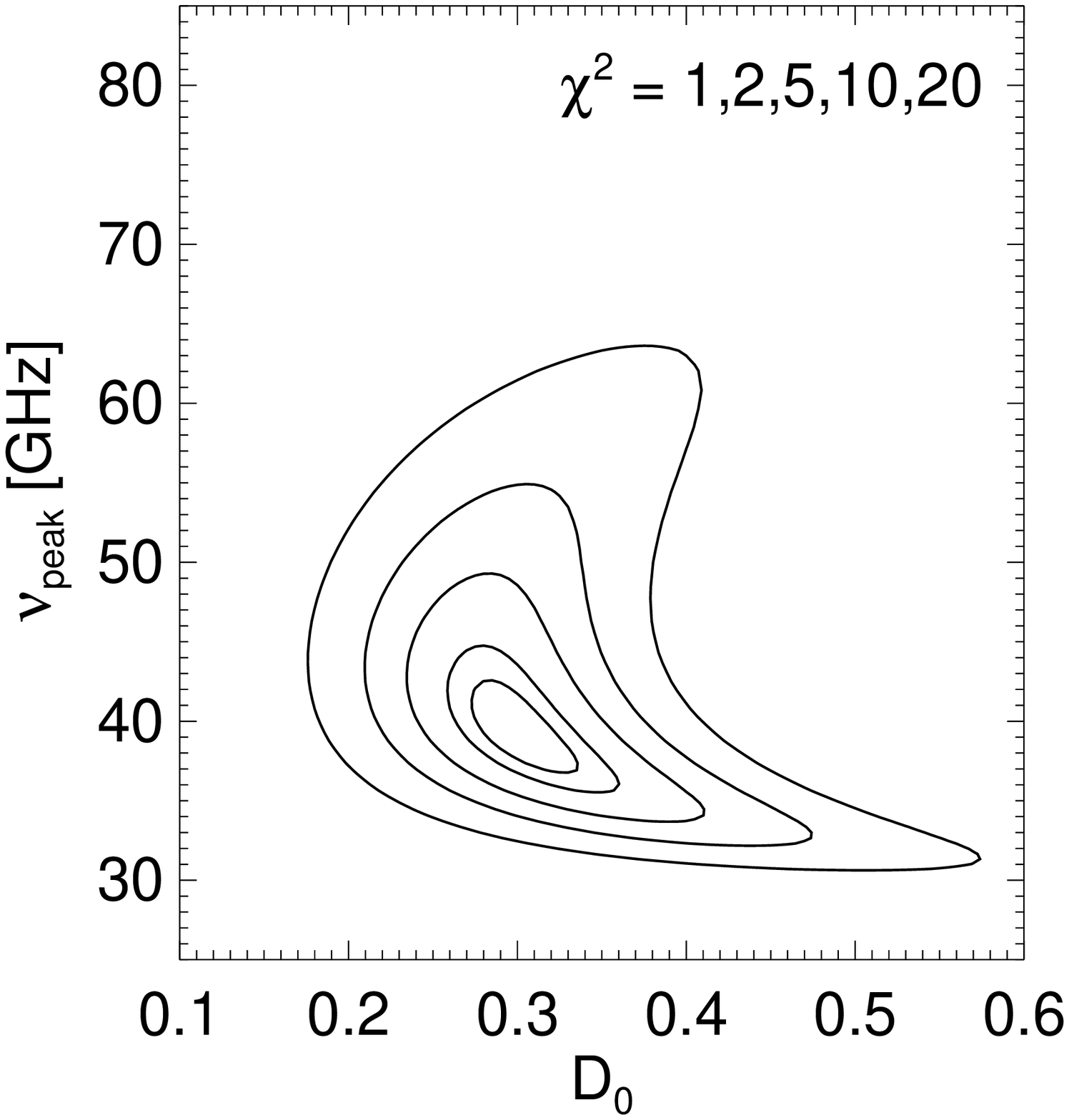}
  \includegraphics[width=0.23\textwidth]{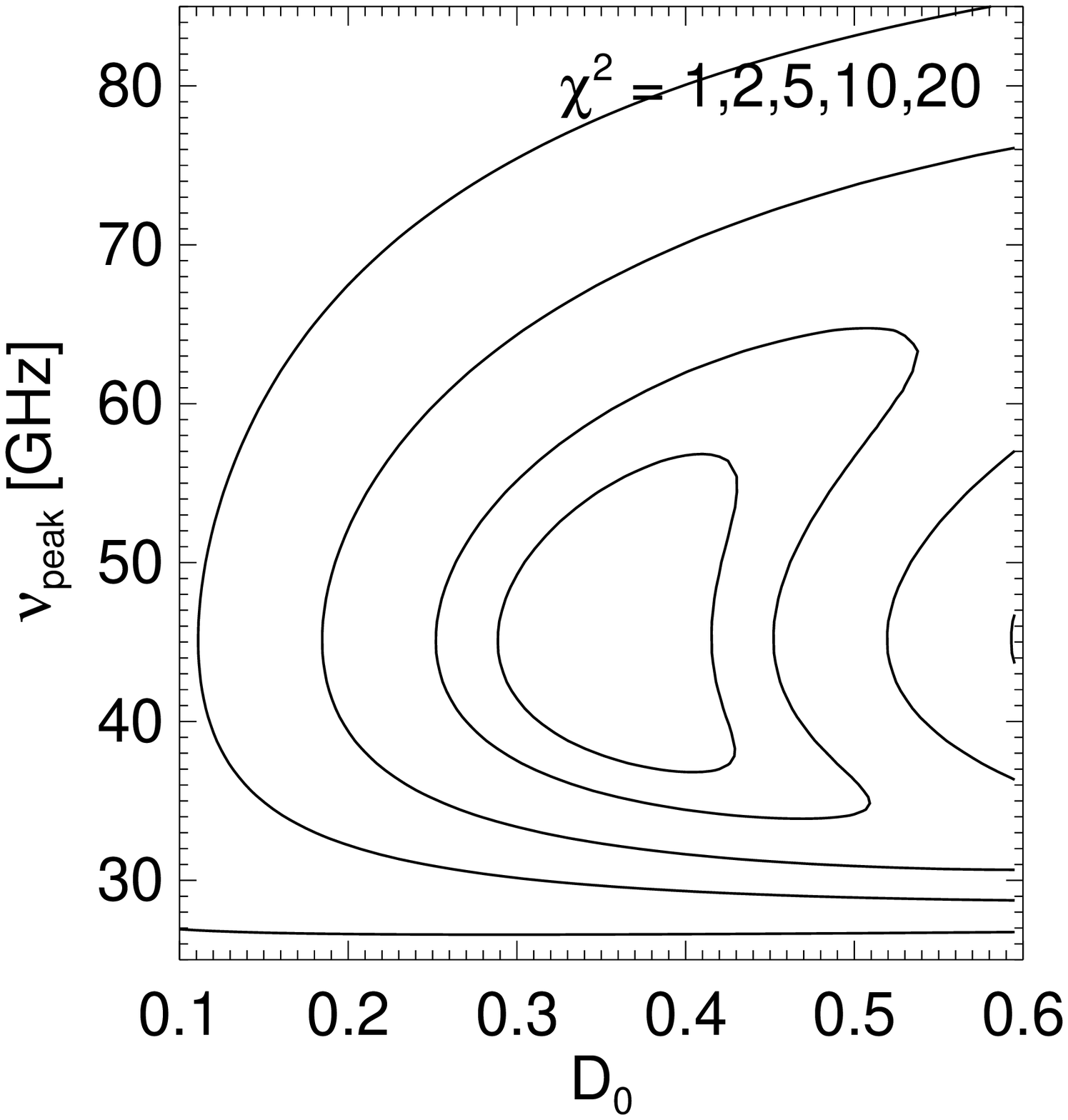}
  \includegraphics[width=0.23\textwidth]{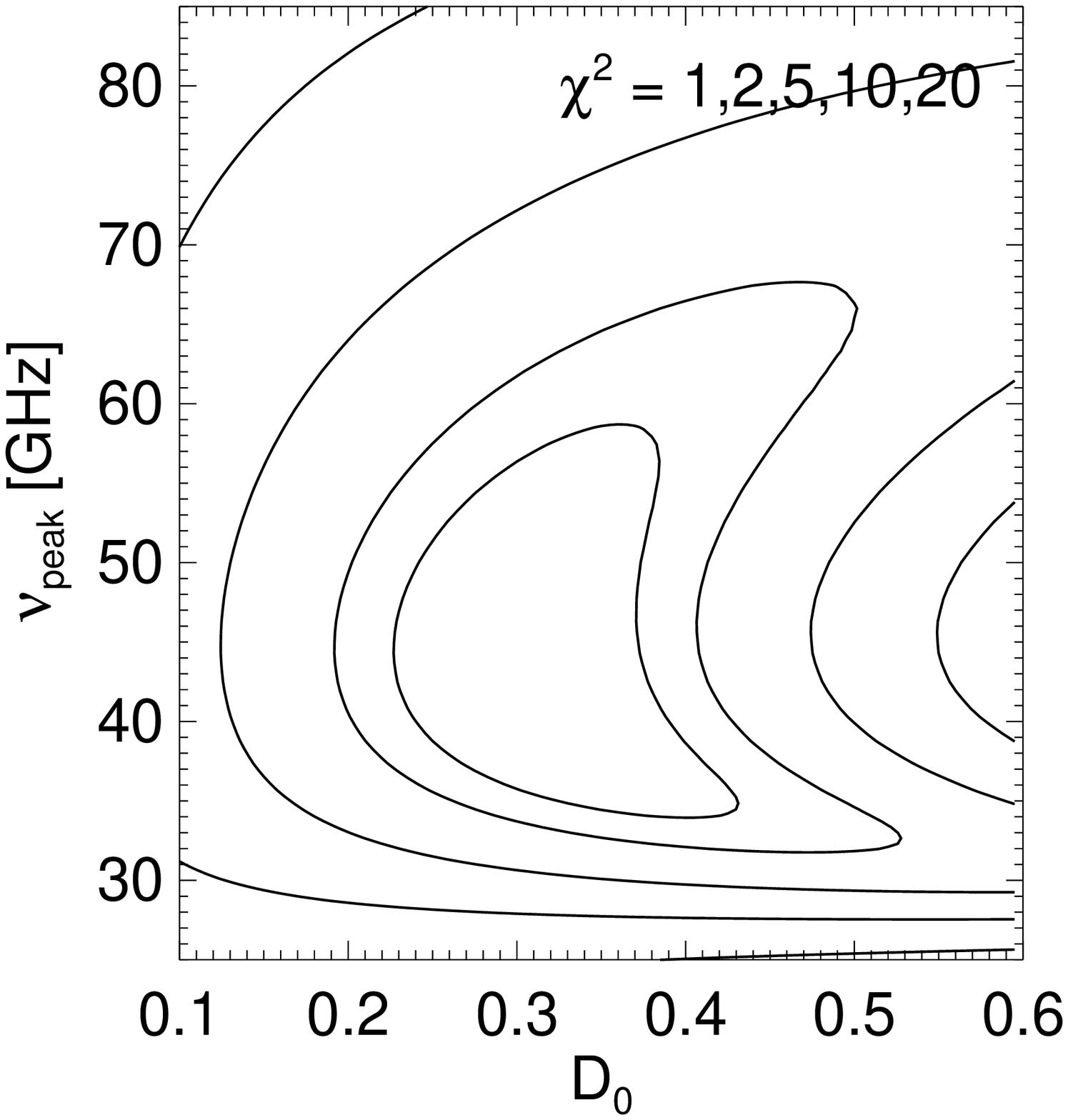}
}
\centerline{
  \includegraphics[width=0.23\textwidth]{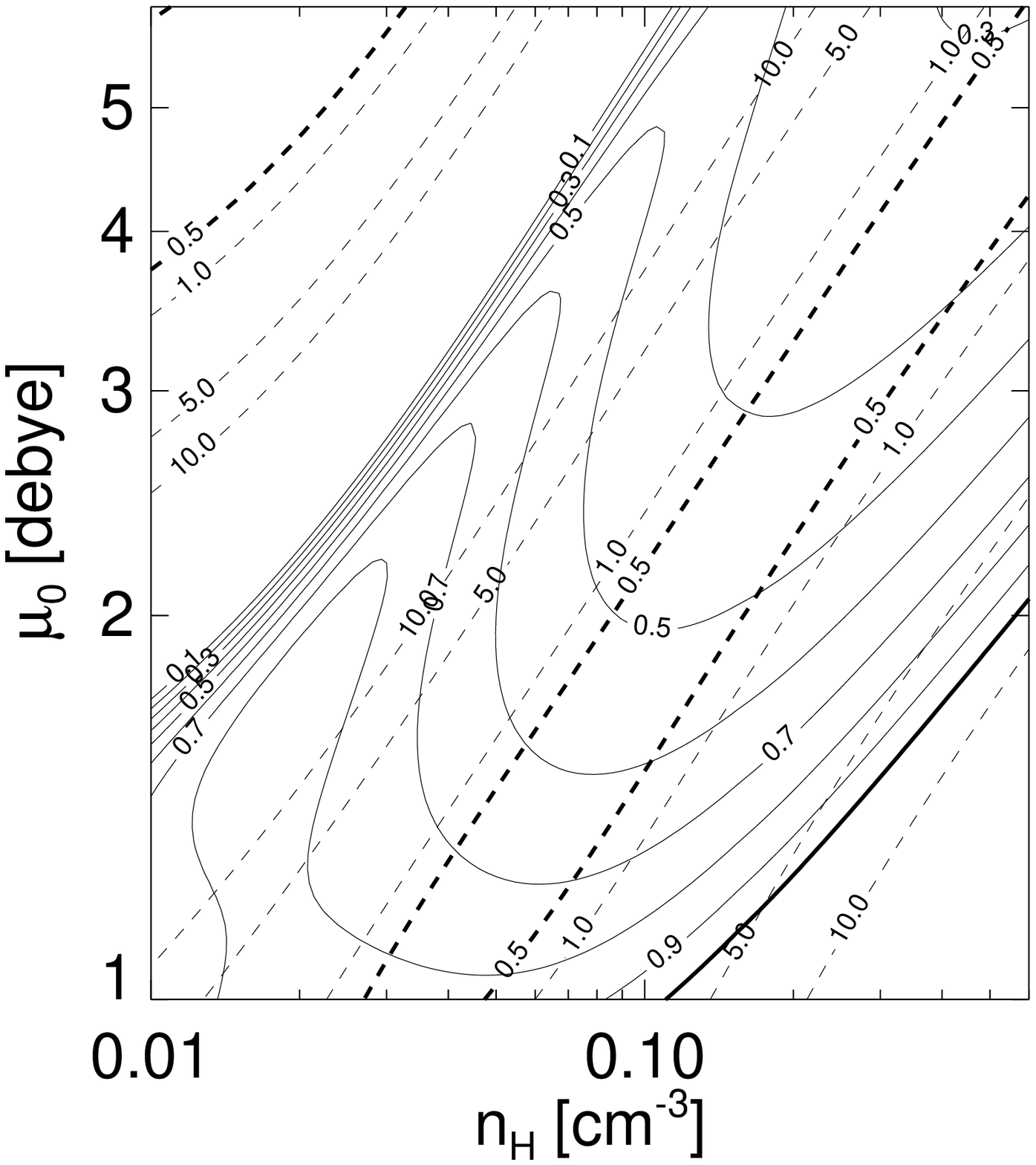}
  \includegraphics[width=0.23\textwidth]{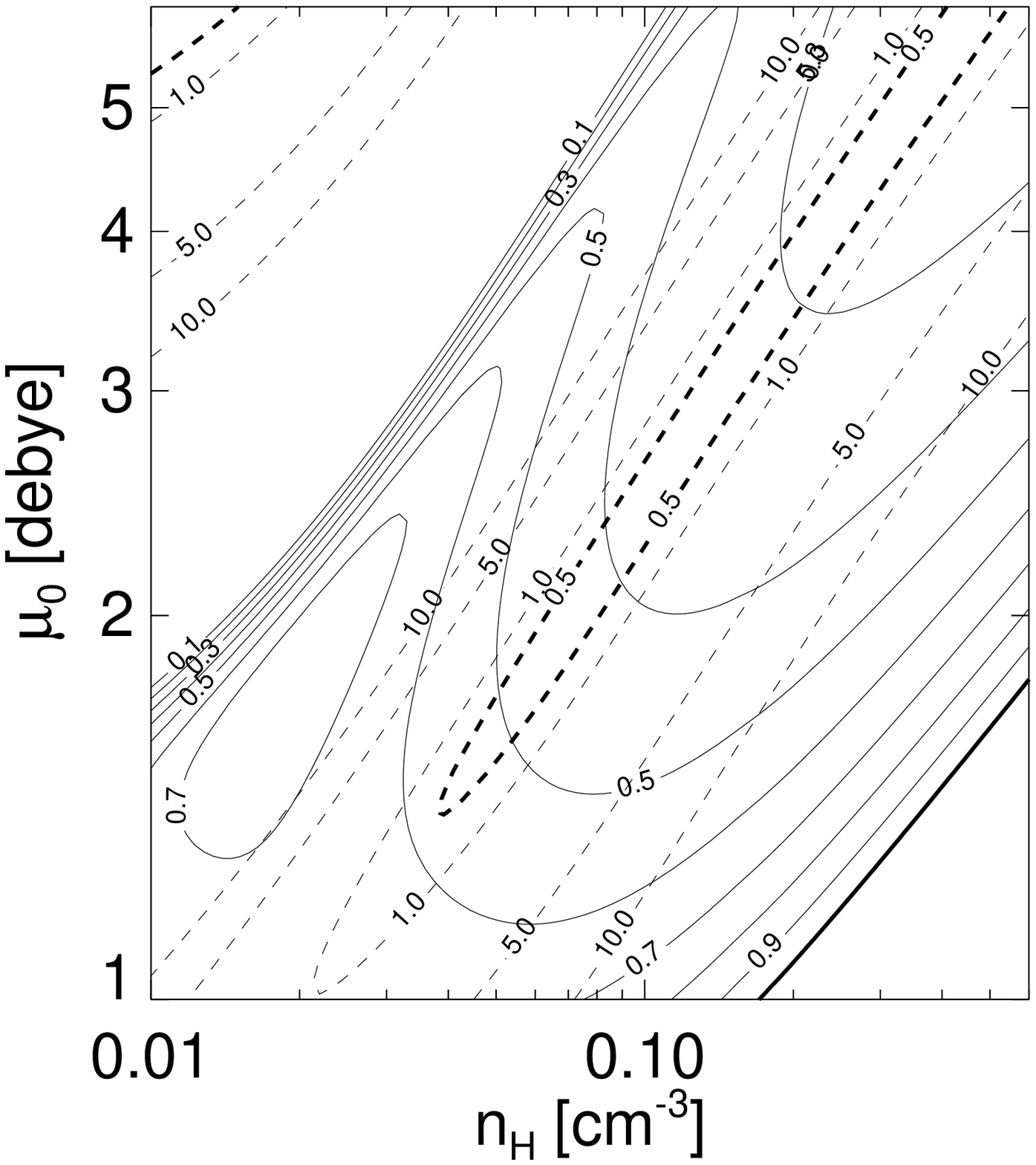}
  \includegraphics[width=0.23\textwidth]{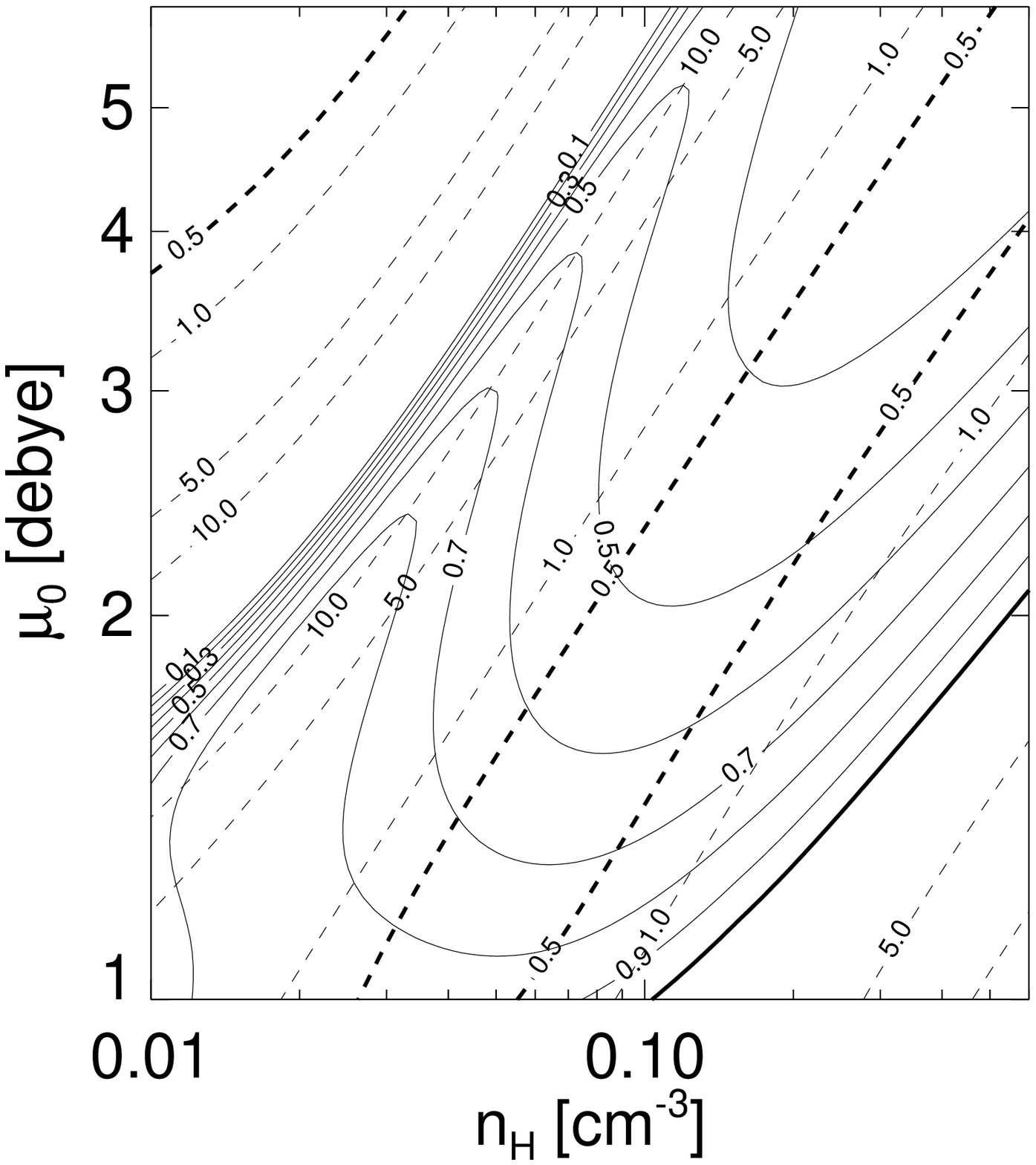}
  \includegraphics[width=0.23\textwidth]{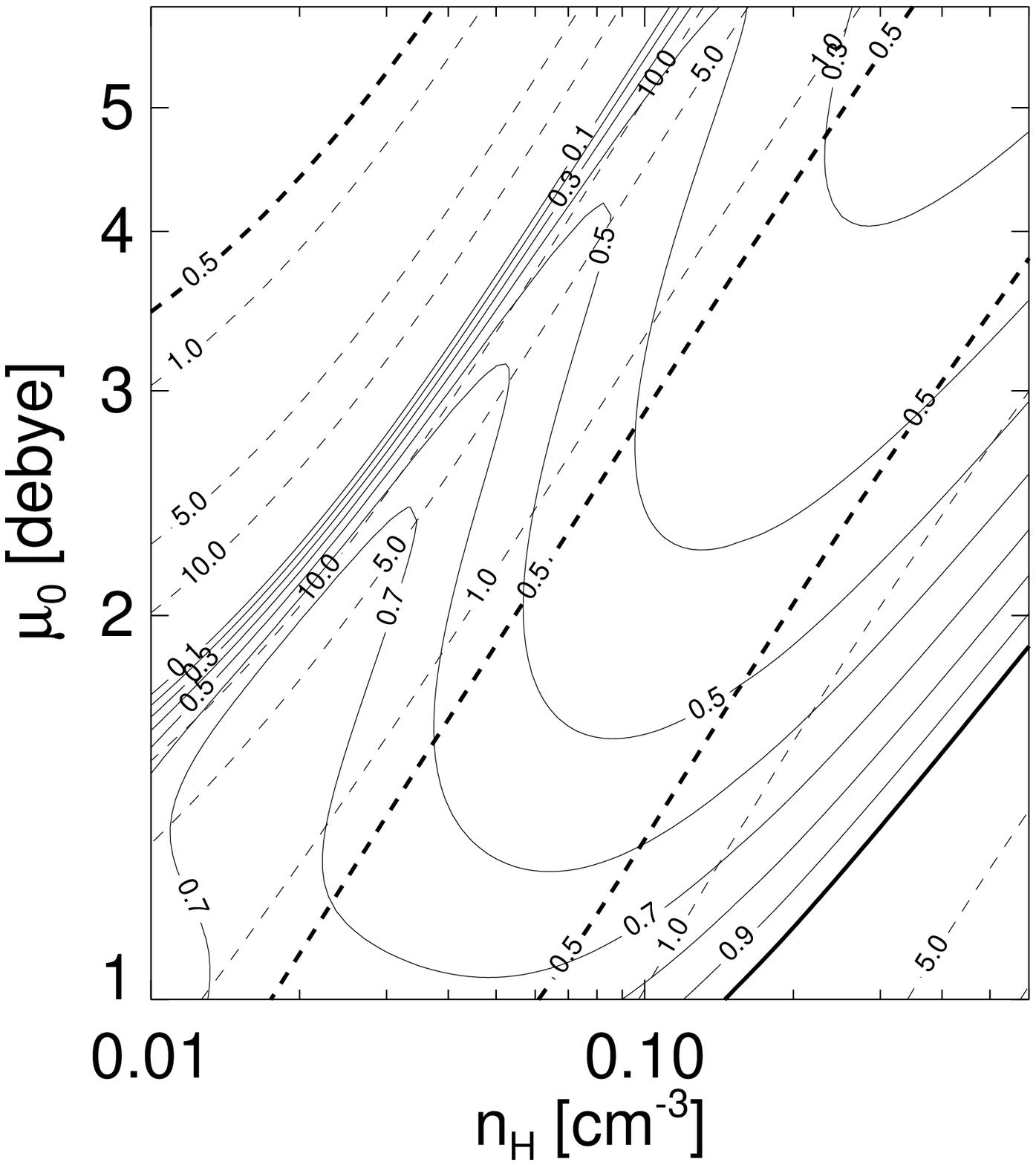}
}
\caption{
The same as Figures \ref{fig:halpha-twopanel}, 
\ref{fig:simp-contours}, and \ref{fig:comp-contours} except 
for the four regions of the Gum Nebula defined and 
illustrated in \citet{DF08b}.  The null hypothesis that the 
H$\alpha$-correlated emission is well represented by a 
free-free spectrum with a CMB contamination is ruled out to 
high significance in all regions ($\chi^2 \geq 14$ for 3 
degrees of freedom at $D_0 =0$).
}
\label{fig:gum-nebula}
\epm

In addition to our nearly full sky template fit, we have 
also carried out smaller regional fits.  Of particular 
interest are the four regions of the Gum Nebula shown in 
Figure 4 of DF08b.  This is a region that is very bright in 
H$\alpha$ but has little in the way of thermal dust emission 
or synchrotron.  The null hypothesis is that the spectrum of 
this region should correspond almost entirely to a free-free 
spectrum.

\reffig{gum-nebula} shows the three component model fit as 
well as the contour plots for the four regions of the Gum 
Nebula.  The bump in the H$\alpha$-correlated spectrum 
persists for all four regions, though the spinning dust 
amplitudes $D_0$ and CMB correlation bias amplitudes $C_0$ 
vary slightly from region to region.  There is some evidence 
that the peak frequency $\nup$ also varies slightly from 
region to region, but it is the least well constrained 
parameter and all four regions indicate roughly $\nup \sim$ 
35-45 GHz.  From the $\Delta\chi^2$ contours in the 
$(D_0,C_0)$ plane, we find that the null hypothesis of $D_0 
= 0$ is ruled out at high confidence in all regions.

Due to the larger error bars in the template fit (the 
smaller regions include fewer pixels), the $\chi^2$ valley 
in the $(\nH,\mmu)$ plane is somewhat broader than in the 
full sky fit.  Nevertheless, all of the spectra are 
consistent with a WIM spinning dust spectrum with $\nH \sim 
0.1 \ \cm^{-3}$, $\mmu \sim 2$ debye, and $D_0 \sim 0.4$.

\subsection{Thermal Dust-Correlated Emission}

\bpm
\centerline{
  \includegraphics[width=0.47\textwidth]{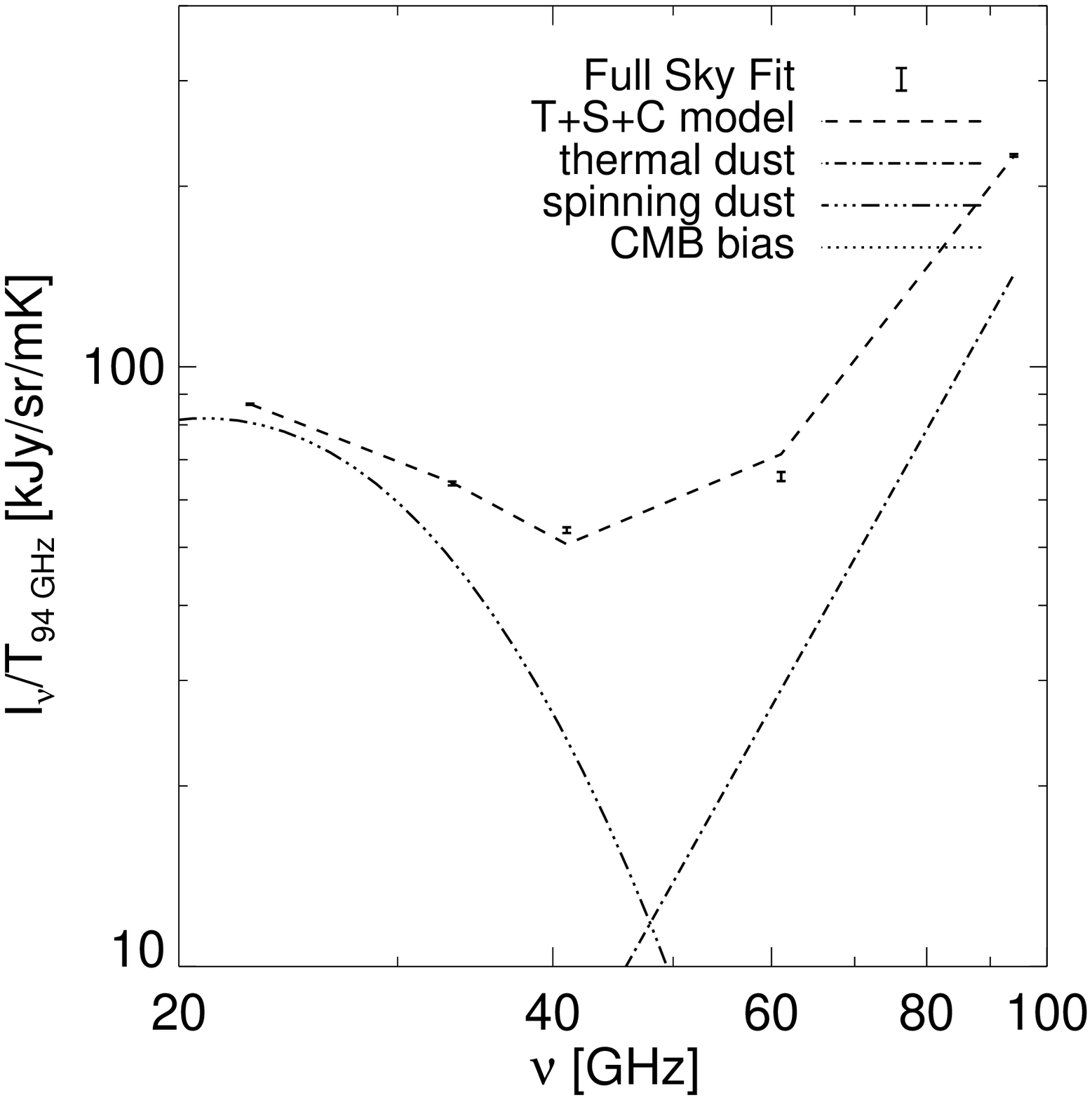}
  \includegraphics[width=0.47\textwidth]{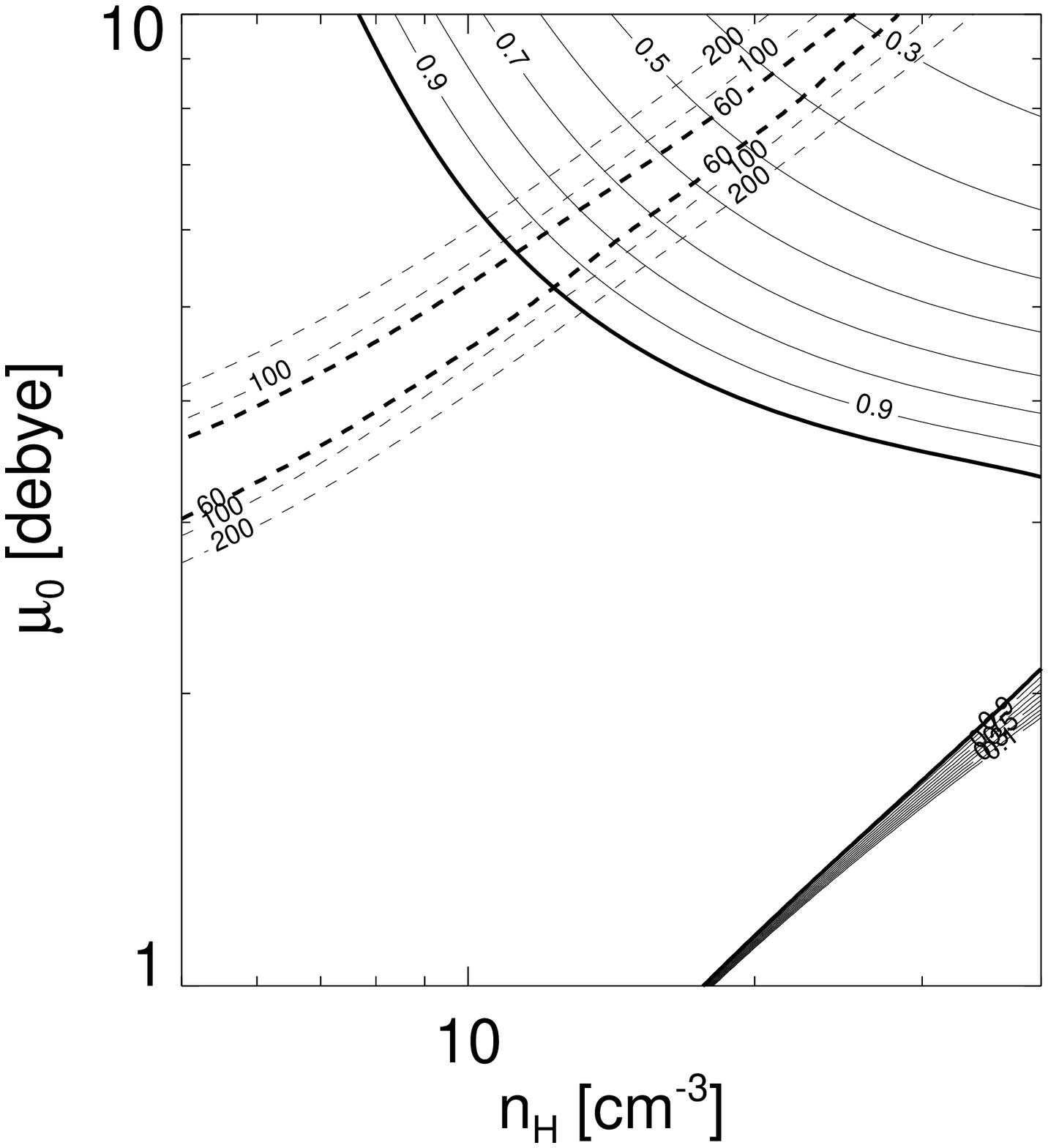}
}
\caption{
\emph{Left:} A three component model fit to the 
FDS99-correlated emission spectrum from 23 to 94 GHz.  The 
fit includes a thermal dust component ($T \propto 
\nu^{1.7}$), a CNM spinning dust component ($\nH = 11 
\cm^{-3}$ and $\mmu = 5.5$ debye; see 
\reftbl{spindustparams}), and a CMB component.  Close 
inspection reveals that the curvature in this single 
component spinning dust spectrum is incompatible with the 
data and over-predicts the amplitude at 61 GHz.  
\emph{Right:} The same as \reffig{comp-contours} except for 
CNM parameters and fitting to FDS99-correlated emission.  
Although the amplitude is roughly $D_0 = 1$, the $\chi^2$ 
values are unacceptably high. In fact, for the model in the 
left hand panel, $\chi^2 = 50.5$.  Thus, for reasonable CNM 
parameters, a single component DL98 spinning dust model is 
ruled out at high confidence.
}
\label{fig:tsc-fit}
\epm

As shown in \reffig{halpha-dust-spec}, dust-correlated emission falls from 23 
GHz to 60 GHz and it was this departure from a thermal spectrum which initially 
lead to speculation that this ``anomalous'' dust component originated from 
spinning grains.  In this section, we fit our three component model to this data 
but replace the WIM spinning dust model in \refeq{threecomp} with a CNM spinning 
dust model and the free-free component with a thermal dust spectrum $T = T_0 
\times (\nu/94 \mbox{ GHz})^{\beta_D}$.  We choose $\beta_D = 1.7$ but note that our 
results are not significantly changed for the range $1.5 < \beta_D < 2$.

The left hand panel in \reffig{tsc-fit} shows the results of fitting a CNM 
spinning model to the WMAP dust-correlated spectrum.  The spinning dust model 
has $\nH = 11 \ \cm^{-3}$ and $\mmu = 5.5$ debye.  From the figure, it is clear 
that this particular spinning dust spectrum is not a very good fit to the data 
($\chi^2 = 50.5$), especially considering that there are few free parameters.  
However, the right hand panel of \reffig{tsc-fit} indicates that, although 
the amplitude is approximately correct ($T_0 \approx 1.0$), for the CNM 
parameters there is no point in parameter space which yields a significantly 
better fit.

The poor fit is due to the fact that the 23, 33, and 41 GHz data
points exhibit roughly powerlaw behavior ($T \propto \nu^{-2.83}$) so that the 
spinning dust spectrum has too much curvature to fit the data.  This powerlaw 
behavior has lead several authors to misidentify this emission as 
``dust-correlated synchrotron'', however both WMAP polarization data 
\citep{kogut07} and data at lower frequencies over large areas 
\citep{gb04,deO04} argue strongly against the synchrotron hypothesis.
Furthermore, we note that, along any line of sight, there may very well be 
regions with different environment and grain properties leading to a 
superposition of spinning dust spectra.  This degeneracy between synchrotron 
and a superposition of spinning dust spectra can only be broken with sufficient 
frequency coverage.  In particular full-sky, high-resolution maps between 5 
and 15 GHz could eliminate the ambiguity.

Lastly, we point out that the anomalous dust-correlated 
emission in the CNM could originate from an emission 
mechanism that is neither synchrotron nor spinning dust.  
For example, \cite{DL99} suggest that emission from 
magnetized dust grain could contribute to the total emission 
in the lower frequency WMAP bands.  However, in the case of 
the WIM, the fact that the H$\alpha$ map and the DL98 
spinning dust models both scale roughly as density squared 
suggests that the bump in the \emph{H$\alpha$-correlated} 
emission originates from spinning dust and \emph{not} DL99 
magnetic dust.  Furthermore, the observations of 
\citet{cassasus08} appear to strongly limit the contribution 
of magnetic grain materials to the microwave emission from 
dust.

\section{Alternative Explanations for the H$\alpha$-Correlated Bump}

In this section we will address alternative explanations for the origin of the 
bump in the H$\alpha$-correlated emission.  The alternatives listed here are not 
meant to be an exhaustive list of possibilities but rather constitute 
potentially important features of the maps, our fitting procedure, etc., and 
thus require careful attention about whether they can artificially generate a 
bump in our spectrum.  For each of these, we argue why the bump is more likely 
to be explained by spinning dust.

\subsection{Bandpass Effects}

Although the spectra presented in previous sections consisted of five
data points, the individual channels of WMAP have broad
($\Delta\nu/\nu\approx 0.2$) bandpasses that must be taken
into account.  For a given emission mechanism, the total intensity in a WMAP 
channel is the spectrum of that emission integrated over the bandpass 
of that channel,
\be
  I_{b}^{\rm tot} = \frac{\int I_{\nu} \omega(\nu) d\nu}{\int \omega(\nu) d\nu},
\ee
where $I_{b}^{\rm tot}$ is the total intensity in band $b$ and $\omega(\nu)$ is 
the bandpass for band $b$ normalized so that $\int \omega(\nu) d\nu \equiv 1$.  
Thus, for a free-free spectrum ($I_{\nu} \propto \nu^{-0.15}$), the observed 
spectrum will be slightly modified from a straight powerlaw when plotted against 
the CMB weighted band centers (see below).  We find that the deviations are at 
the level of $\sim 1\%$.  Nevertheless, this total intensity, evaluated at each 
band, is what must be used to fit the derived cross-correlation spectrum.

Additionally, the WMAP data themselves must be converted from thermodynamic 
$\Delta T$ to $I_{\nu}$ in each band assuming a CMB weighted band 
center,\footnote{By design the WMAP receivers measure energy and so the natural 
data unit is $\int I_{\nu} d\nu$, however since they are then calibrated off of 
the CMB dipole, the data are presented in thermodynamic $\Delta T$.}
\be
  \nu_{\rm CMB} = \frac{\int \nu I_{\nu}^{\rm CMB} \omega(\nu) d\nu}
                       {\int I_{\nu}^{\rm CMB} \omega(\nu) d\nu},
\ee
where the spectrum of the CMB $I_{\nu}^{\rm CMB} \propto \nu^2 plc(\nu)$ (i.e., 
flat in thermodynamic $\Delta T$).  All of the fits presented in previous 
sections take the bandpass weighting into account, and we find that the changes 
to the results are also at the $\sim 1\%$ level.

\subsection{Dust Extinction and Masking Effects}

\bp
\centerline{
  \includegraphics[width=0.47\textwidth]{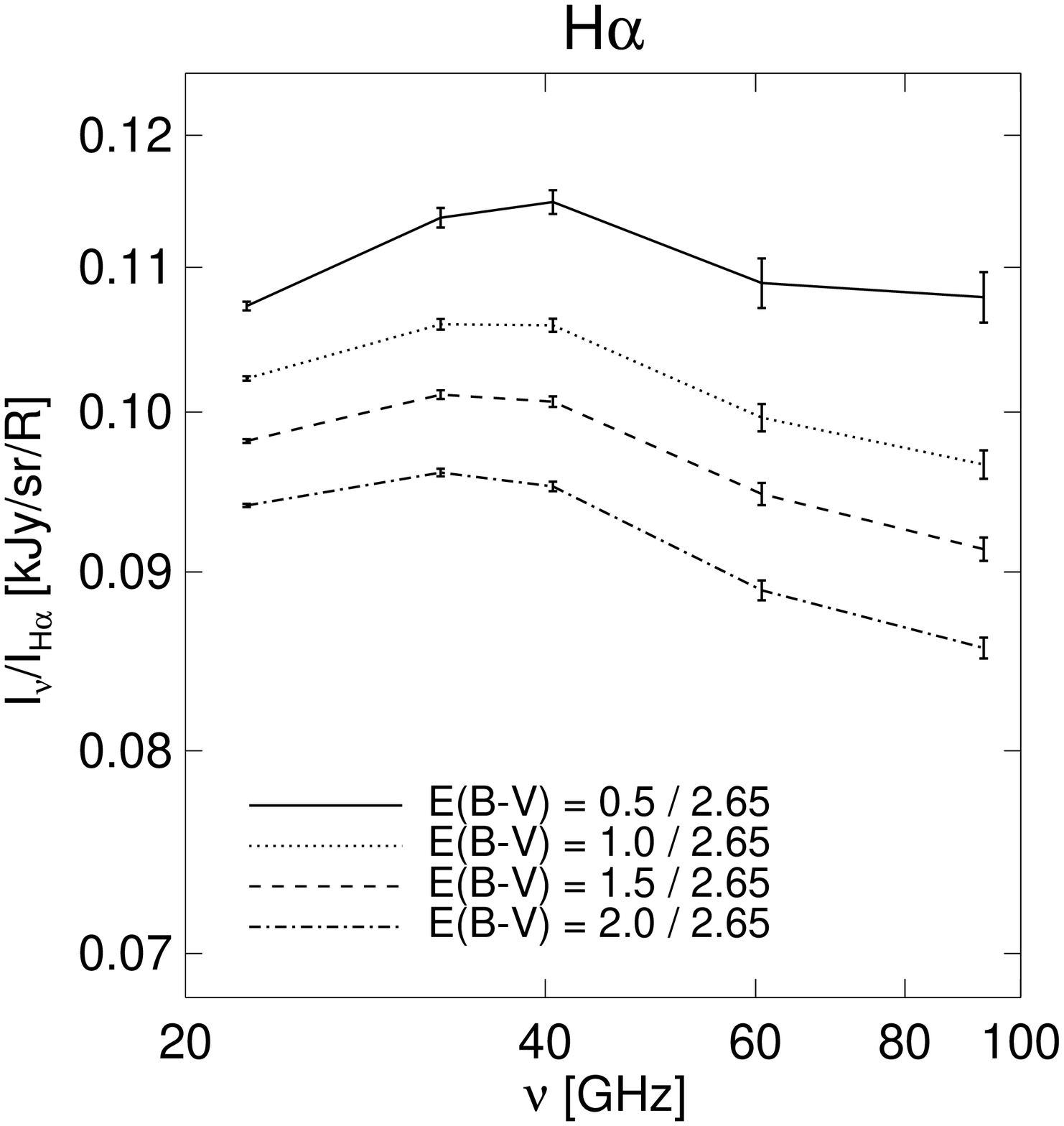}
}
\caption{
Foreground spectra derived via our multi-linear regression template fit as a 
function of the extinction cut used for the mask.  The bump in the 
H$\alpha$-correlated emission persists for all mask definitions indicating that 
it is not due to improperly corrected extinction in the H$\alpha$ map.
}
\label{fig:vary-dustcut}
\ep

It has long been known that, although a map of H$\alpha$ is an excellent tracer 
of free-free emission from ionized gas, it is not perfect.  In particular, 
extinction of H$\alpha$ by dust makes the map a poor tracer of free-free in 
regions of high dust column density.  To account for this \citet{finkbeiner03} 
(whose map we have used in this paper) apply a dust correction which assumes 
uniform mixing between the dust and gas.  \citet{dickinson03} take a slightly 
different approach by matching slices through the galactic plane to an 
extinction model plus dust column density model and attempting to directly 
measure the extinction.

The effects of dust on the H$\alpha$ map could manifest themselves in two ways.  
First, recall that our mask is generated by avoiding regions where the 
extinction due to dust at H$\alpha$ is $E(B-V) > 1$ mag.  It is difficult to 
envision that our mask could have a sufficiently large impact on our derived 
spectrum to produce the bump, but it is instructive to consider the possibility.  
In \reffig{vary-dustcut} we show the derived H$\alpha$-correlated spectrum for a 
wide range of dust cuts (i.e., for numerous masks).  As can be seen in the 
figure, the bump feature is robust to variations in the mask.

The second potential issue is that perhaps the H$\alpha$ map has been over 
corrected for dust so that it is actually a rather poor tracer of free-free, 
potentially leading to the bump as a spurious artifact.  To test this 
hypothesis, we performed our template fit with an H$\alpha$ map that was not 
corrected for dust extinction at all.  Even in this extreme case of
\emph{under}-correction, we find that the bump persists.  We conclude
that the bump does not result from errors in the extinction correction
of the H$\alpha$ map.

\subsection{CMB Cross-Correlation Bias}

The CMB cross-correlation bias described above (and more 
extensively in DF08a) is both large and ubiquitous in CMB 
foreground analyses. Because the effect on the derived 
spectrum of each component can be substantial, it is 
important to rule out the possibility that this bias is 
producing the bump.  This possibility represents the ``null 
hypothesis'' discussed in \refsec{hacorr}.  The basic 
question is, can the derived cross-correlation spectrum be 
fit by a simple linear combination of free-free and CMB 
spectra?  As we showed in that section, this $D_0 = 0$ case 
is ruled out at \emph{very} high confidence with a $\chi^2$ 
per degree of freedom of 245 (for 3 degrees of freedom at 
$D_0 = 0$).

\subsection{Cross-Correlation Between the Templates}

If there is any chance spatial cross-correlation between the 
templates, than this will impact the shape of the derived 
spectra.  Furthermore, if one of the templates is 
contaminated by H$\alpha$ morphology or vice-versa, than the 
spectrum of each foreground will be ``contaminated'' by the 
others.  Again, a clear example of this would be poor dust 
correction of the H$\alpha$ map adding dust-like morphology 
to that map.  Another example would be reflection of 
H$\alpha$ photons created in the plane off of dust at high 
latitudes, which would also imprint a dust-like morphology 
on the H$\alpha$ map.  There is also the fact that the WIM 
emits thermal radiation as well and so there is some 
H$\alpha$ morphology present in the FDS99 map.

DF08b showed that the chance spatial cross-correlations between the templates
used in the analysis is sufficiently small that there should be relatively
little contamination of one foreground spectrum onto another.  We explore this
point in greater detail in \refapp{imperfect-templates}, where we specifically
address the fact that WIM features appear at a low level in the dust map.

\subsection{Variations of $\Tgas$ with Position}

Gas temperatures in the interstellar medium are known to 
vary with position.  For example, there is evidence that 
there is a decrease in $\Tgas$ towards the Galactic center 
\citep{quireza06}.  However, even though relatively small 
variations in $\Tgas$ lead to large changes in the H$\alpha$ 
to free-free ratio, the effect is simply an amplitude shift 
\citep{valls98}.  Thus, when we fit the nearly full sky, we 
are averaging over many regions with slightly different 
amplitudes, \emph{but all with spectra $\propto 
\nu^{-0.15}$}.  Thus, our resultant spectrum should be 
$\propto \nu^{-0.15}$ indicating that a superposition of 
many regions with varying $\Tgas$ would \emph{not} produce a 
bump in the average spectrum.

\subsection{Model Independent Identification of the Anomalous Emission}

\bpm
\centerline{
  \includegraphics[width=0.85\textwidth]{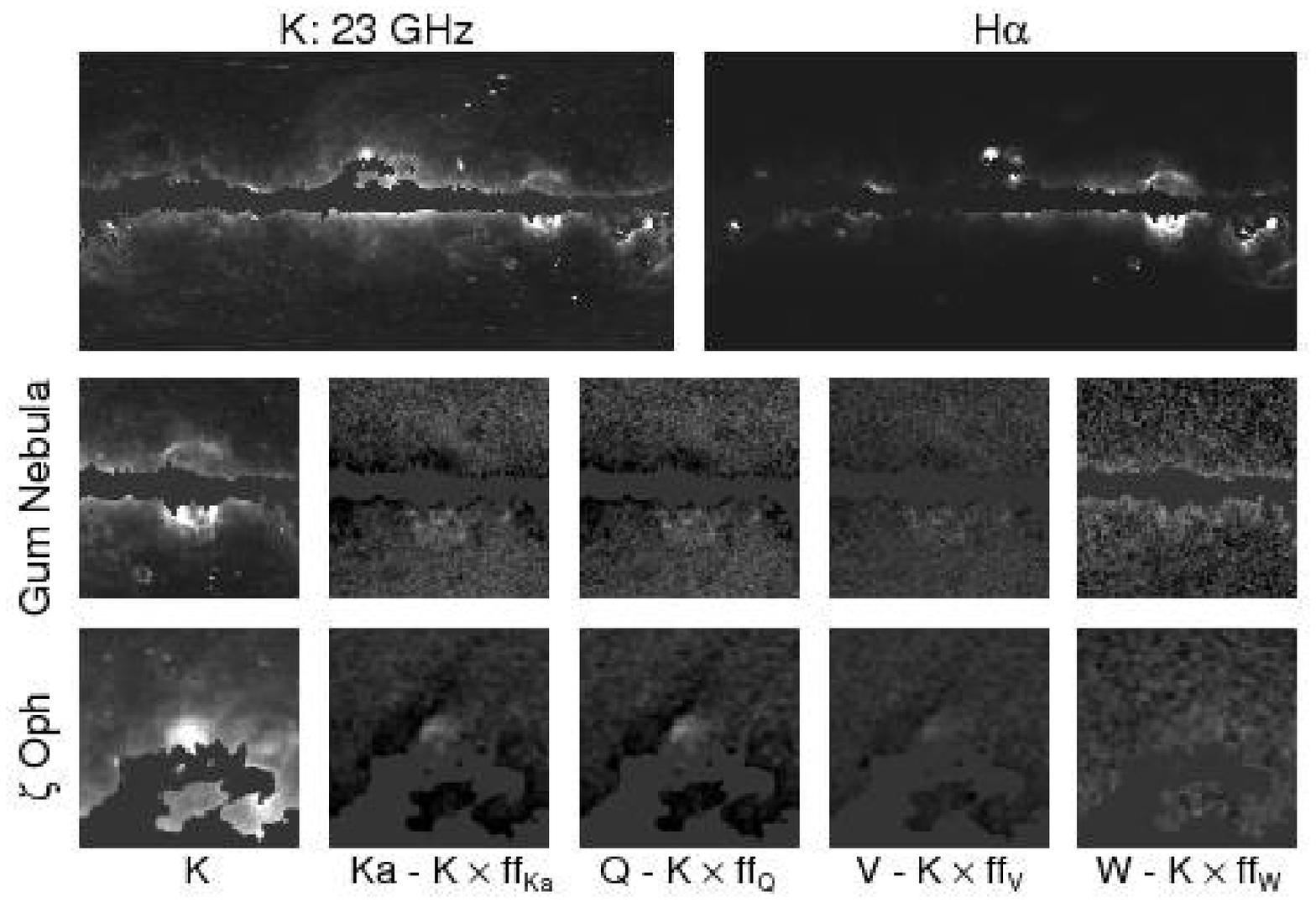}
}
\caption{
\emph{Top left:} WMAP K band minus CMB5.  \emph{Top right:} 
The H$\alpha$ map.  \emph{First column:} A cut out of K band 
around the Gum Nebula (row 1) and $\zeta$ Oph (row 2).  
\emph{Second through fifth column}: K band subtracted from 
Ka, Q, V, and W bands using a free-free $\nu^{-2.15}$ 
spectrum.  Clearly there is residual emission in both the 
southern Gum Nebula and the $\zeta$ Oph cloud using a 
free-free spectrum indicating the need for a separate 
component.  This component is harder than free-free from 23 
to 41 GHz and softer than free-free from 41 to 94 GHz, which 
is indicative of a spinning dust spectrum.  Although there 
are significant regions of over-subtraction (due to the 
substantial synchrotron component at low frequencies), the 
H$\alpha$-correlated emission is clearly under-subtracted by 
a free-free spectrum.
}
\label{fig:ffres}
\epm

Rather than attempting to address every possible 
contamination mechanism that could result in a bump in the 
H$\alpha$-correlated spectrum, it is possible to identify 
the bump \emph{without the use of any template fitting and 
only using the WMAP data}. The top two panels of 
\reffig{ffres} show the WMAP data at 23 GHz (with CMB5 
subtracted) as well as the H$\alpha$ map.  We identify two 
regions that are bright in H$\alpha$ and also bright at K 
band (23 GHz): $\zeta$ Oph near $(\ell,b) = (6,23)$ degrees 
and the Gum Nebula near $\ell = 260$ and $-22 < b < 17$ 
degrees .  If the H$\alpha$ map traces only free-free 
emission, then subtracting K band scaled by a free-free 
spectrum from Ka, Q, V, and W band should completely 
eliminate these features.

The bottom grid in \reffig{ffres} shows K band in the left 
most column and then the result of subtracting K band scaled 
by a free-free spectrum from Ka, Q, V, and W bands (columns 
2, 3, 4, and 5 respectively) for both the Gum Nebula region 
(row 1) and the $\zeta$ Oph region (row 2).  Since the 
spectrum of free-free is harder than synchrotron and softer 
than thermal dust, synchrotron emission is highly 
over-subtracted and thermal dust emission is 
under-subtracted (note the prominence of thermal dust 
emission in column 5).  What we find is that with a 
free-free spectrum, there is a clear excess in the southern 
Gum Nebula region and particularly in $\zeta$ Oph.  The 
implication is that, $\zeta$ Oph and the Gum Nebula contain 
an emission component which is \emph{harder} than free-free 
from 23 GHz to 41 GHz, but softer than free-free from 41 GHz 
to 94 GHz.  Furthermore, this spectrum is not soft enough at 
23-61 GHz to be synchrotron, yet is vastly brighter than the 
expected thermal dust emission, so it cannot be any of the 
standard foreground components.  Rather, it has a spinning 
dust type spectrum with peak frequency near $\sim 40$ GHz.

\section{Discussion}

We have repeated the multi-linear regression template fit 
outlined in \citet[][DF08a]{DF08a} on the 5-year WMAP 
data and found that the anomalous ``bump'' at $\sim 40$ GHz 
in the H$\alpha$-correlated emission spectrum persists and 
has in fact become more pronounced.  The most significant 
changes in the spectrum from 3-year to 5-year data 
come from large scale modes (particularly a large dipole at 
61 GHz V band) of amplitude $\sim 10$ $\mu$K that were 
corrected in the 5-year data.

We interpret this bump as a ``warm ionized medium'' (WIM) 
spinning dust component that is traced by the H$\alpha$ map.  
We find that for WIM parameters, the total spinning dust 
emission per grain scales roughly as the ambient ionized gas 
density squared (see \reffig{nujnu_per_em}).  Therefore, a 
map of emission measure ($\EM = \int n_e^2 d\ell$) like the 
H$\alpha$ map would be \emph{expected} to trace this 
component.

Any derived foreground spectrum is subject to the CMB 
correlation bias (see DF08a) from chance spatial 
correlations of the foreground emission with the CMB.  Thus, 
we fit a three component model consisting of free-free, WIM 
spinning dust, and CMB spectra to the derived 
H$\alpha$-correlated emission in which the amplitude of each 
component is allowed to float.  The best fit model gives a 
value for the ion density of the ambient medium of $\nH \sim 
0.15 \ \cm^{-3}$ and a characteristic dipole moment for the 
grains of $\sim 3.5$ debye referenced to 1 nm grains.  
However, there is a strong degeneracy between these two 
parameters as they have similar effects on the peak 
frequency.

The amplitude of the spinning dust spectrum must be reduced 
by a factor $\sim 0.3$ compared to the model to fit the 
data, possibly indicating that PAHs are depleted in the WIM.  
From the amplitude coefficient of the free-free component, 
we use the H$\alpha$ to free-free ratio to infer a gas 
temperature of $\sim 3000$ K.  Near the peak frequency at 41 
GHz, the total spinning dust emission is roughly 20\% of the 
free-free emission.  We also find that the bump persists in 
smaller regional fits of the Gum Nebula as well with roughly 
the same amplitude.  There is some evidence that the peak 
frequency varies with region, but this parameter is the most 
poorly constrained in our analysis.

As in numerous other studies, we also find the now familiar 
rise in the \emph{thermal dust}-correlated spectrum from 41 
to 23 GHz from anomalous dust-correlated emission.  An 
attempt to fit a three component model of thermal dust, 
``cold neutral medium'' (CNM) spinning dust, and CMB spectra 
to the dust-correlated emission yields a poor fit.  However, 
we expect that, along any line of sight, there will be 
multiple regions with different environmental properties so 
that the low frequency dust-correlated emission may 
represent a superposition of spinning dust spectra.  
Although other possibilities exist for this emission 
(magnetic dust and ``dust-correlated synchrotron'' being 
two), we point out that since the H$\alpha$ map is a density 
squared map, and since WIM spinning dust emission goes 
roughly as density squared, we believe it is very likely 
that the bump in the H$\alpha$-correlated emission 
represents a spinning dust component.  \emph{Magnetic dust 
emission would not correlate with H$\alpha$ in this way.}

We have shown that the bump cannot be explained by numerous 
possible systematic uncertainties in our fit.  For example, 
the bump cannot be the result of the CMB correlation bias 
because a fit to the spectrum with a two component free-free 
plus CMB model yields a very bad $\chi^2 = 740$ with three 
degrees of freedom.  It also cannot be due to poor 
correction for dust extinction of the H$\alpha$ map since 
the bump persists even if the template fit is performed with 
a completely uncorrected map, nor is it sensitive to our 
choice of mask for which we cut on dust extinction of 
H$\alpha$.  Variations in gas temperature with position and 
cross-correlation between the templates used in the fit are 
also incapable of producing the bump.

Perhaps the most striking visualization that the bump is not an artifact of the 
fitting procedure is by subtracting the K band WMAP data scaled with a free-free 
spectrum from Ka, Q, V, and W bands.  This reveals that in regions that are 
very bright in K band and H$\alpha$ which were thought to be completely free-free 
dominated at these frequencies (e.g., the Gum Nebula and $\zeta$ Oph), there is 
a positive residual at 33, 41, and 61 GHz.  Scaling by a free-free plus bump 
spectrum instead removes the residual.  The implication is that there is 
emission coming from these regions that has a spectrum that is harder than 
free-free from 23-41 GHz and then softer than free-free from 41-94 GHz, which 
resembles a spinning dust spectrum.

The interpretation of the dust-correlated anomalous emission as spinning dust
(or magnetic dust) has been plagued by the fact that ancillary data sets at
lower frequencies are needed to distinguish that emission from synchrotron. 
Here we have shown that a turnover in the spinning dust spectrum is completely
recoverable \emph{within} the WMAP frequency range.  The only surprise is that
it correlates with the H$\alpha$ map.  In hindsight, this is not very surprising
since the H$\alpha$ map traces density squared emission, though it does suggest
that there is no ``true'' template for the spinning dust.  Rather, it is
potentially ubiquitous and probably will not correlate precisely with any map of
the sky.

\acknowledgements
We acknowledge informative discussions with Simon Cassasus, 
Clive Dickinson, Joanna Dunkley, Ben Gold, Gary Hinshaw, Al 
Kogut, and David Spergel.  We thank Matt Haffner and Carl 
Heiles for insights on gas temperatures in the warm ionized 
medium, and Alex Lazarian, Chris Hirata, and Yacine 
Ali-Hamoud for stimulating discussions on excitation 
mechanisms in spinning dust models.  Some of the results in 
this paper were derived using HEALPix 
\citep{gorski99,calabretta07}.  This research made use of 
the IDL Astronomy User's Library at 
Goddard\footnote{Available at 
\texttt{http://idlastro.gsfc.nasa.gov}}.  GD and DPF are 
partially supported by NASA LTSA grant NAG5-12972.  BTD is 
partially supported by NSF grant AST-0406883.

\appendix

\section{Imperfect Template Morphology}
\label{app:imperfect-templates}

\bpm
\centerline{
  \includegraphics[width=0.9\textwidth]{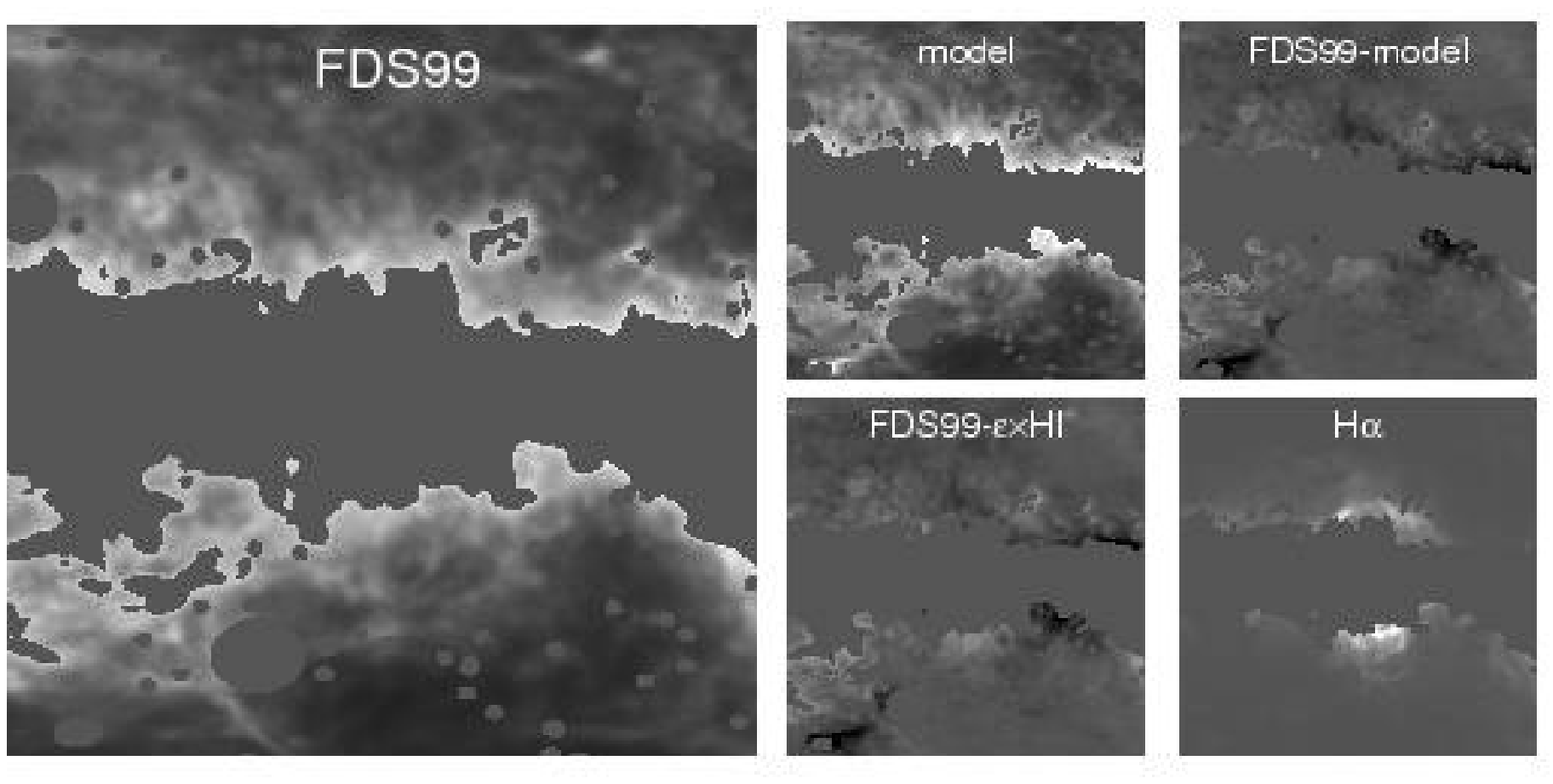}
}
\caption{
An illustration of the contamination of FDS99 by H$\alpha$ 
morphology.  The left panel shows the FDS99 dust map 
centered on the Gum Nebula ($\{\ell,b\} = \{260,0\}$, a 
strong H$\alpha$ feature).  The top middle panel shows the 
best fit FDS99 = $\epsilon_0 L + \delta_0 H$ where $L$ is a 
map of \ion{H}{1} emission and $H$ is the H$\alpha$ map.  
The top right panel shows that the difference between the 
model and the FDS99 map contains very little H$\alpha$ 
morphology (lower right panel).  Nevertheless, the effect is 
subtle as shown in the residual plus $\delta_0 H$ map in the 
lower middle panel.  Our fit of the level of contamination of 
the FDS99 map by H$\alpha$ morphology is $\delta_0 \sim 4.5 
\times 10^{-5}$ mK/R.
} \label{fig:rm_hal_fds_maps}
\epm

\bpm
\centerline{
  \includegraphics[width=0.33\textwidth]{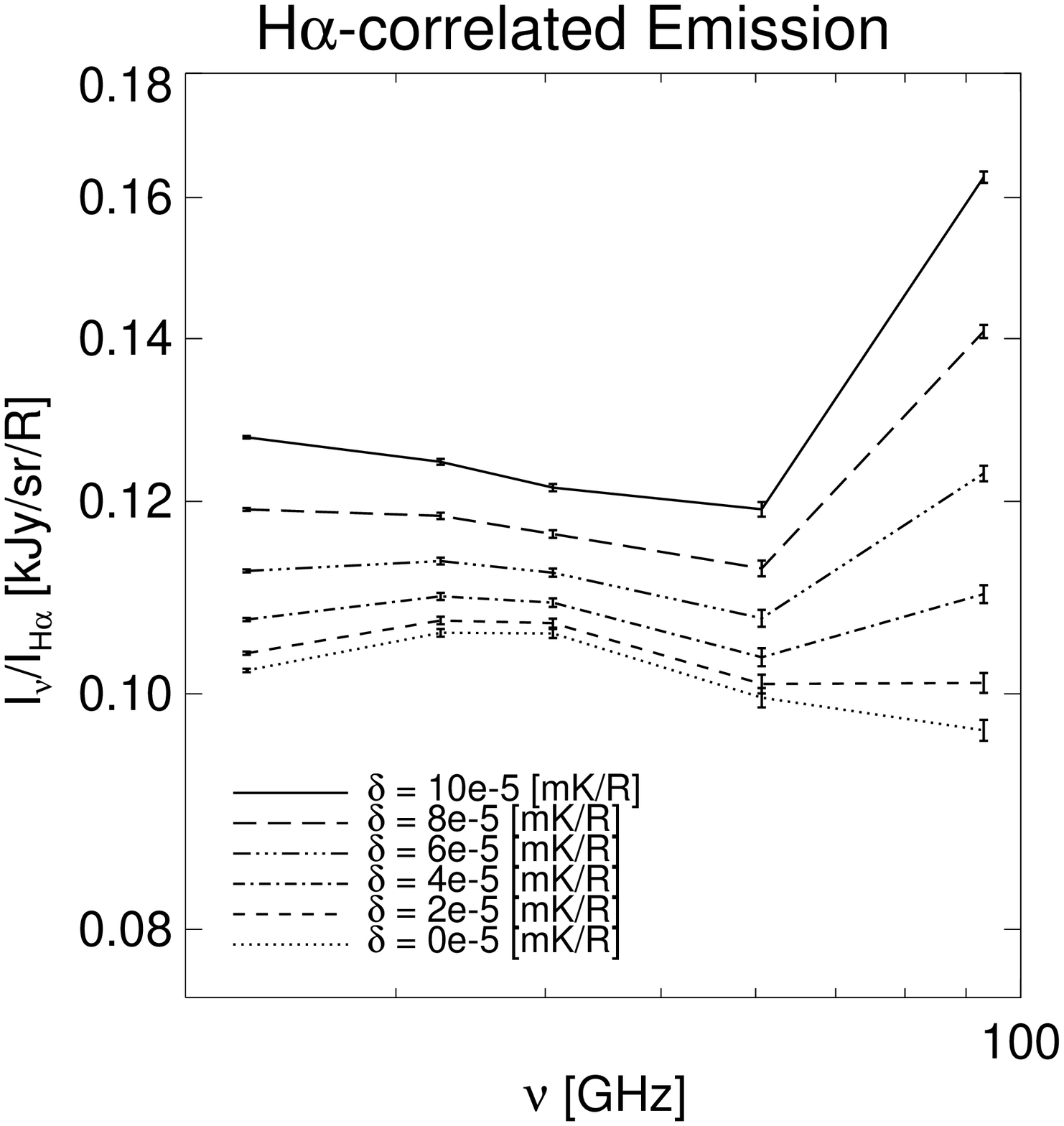}
  \includegraphics[width=0.33\textwidth]{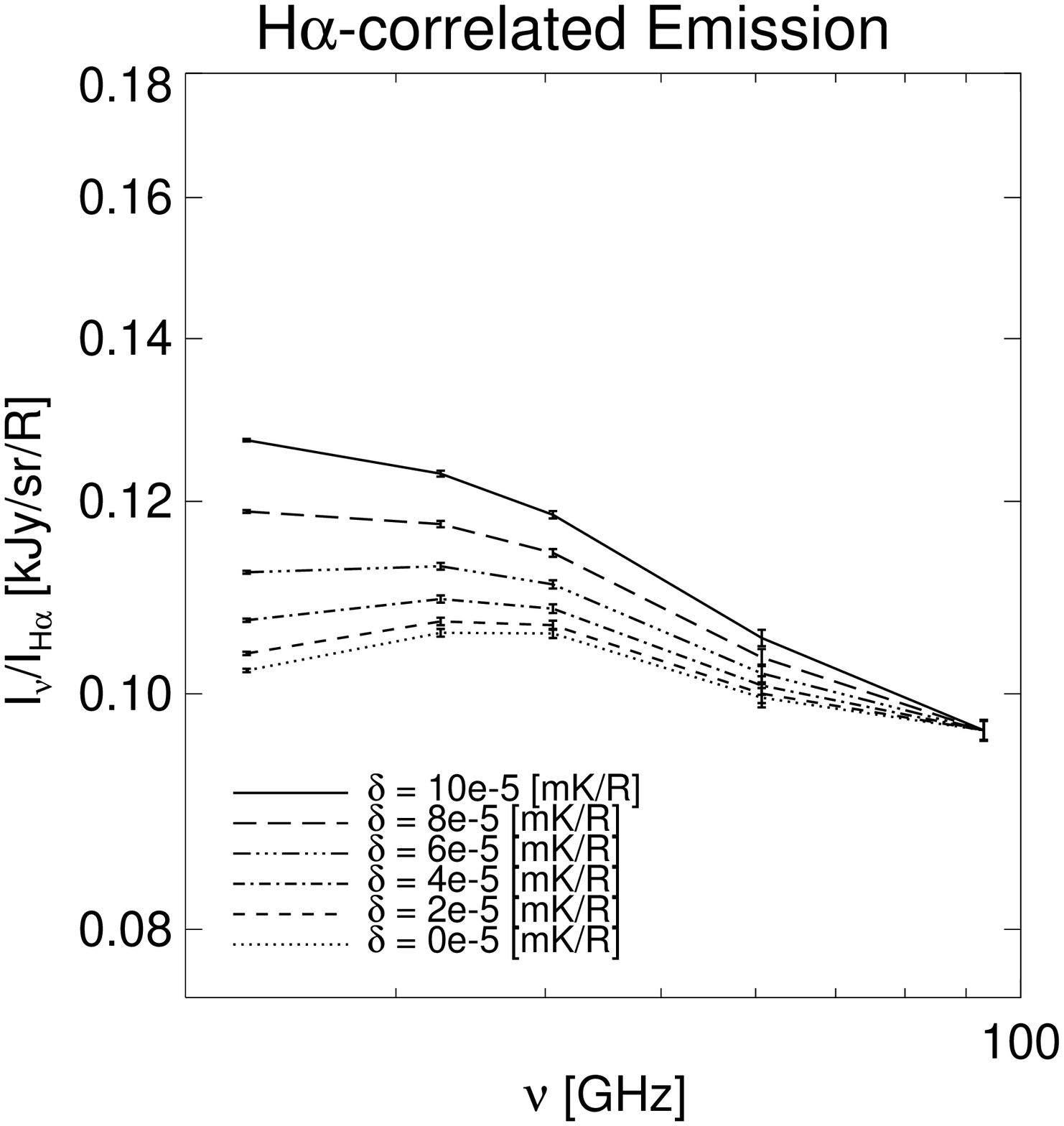}
  \includegraphics[width=0.33\textwidth]{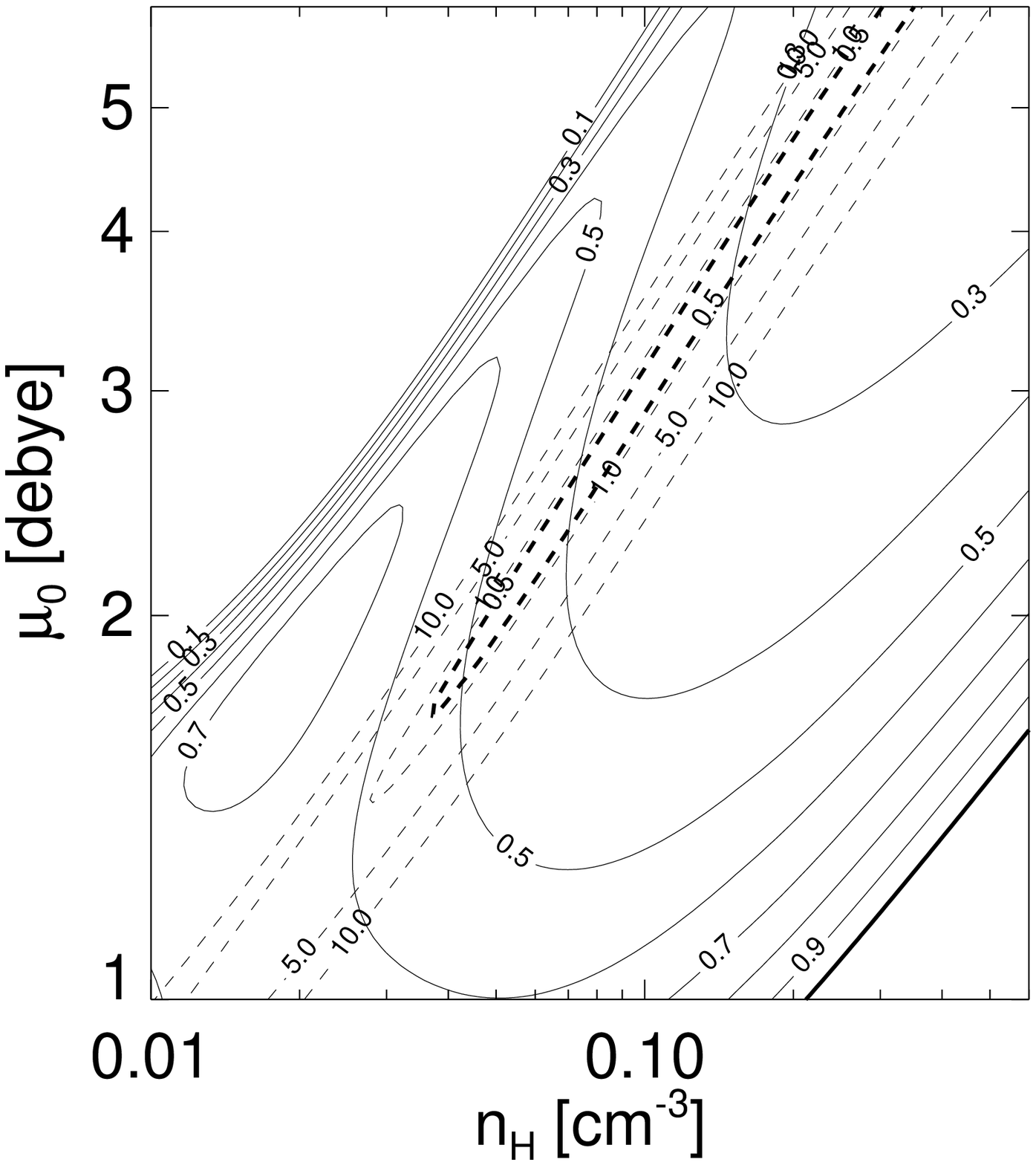}
}
\caption{
\emph{Left:} the spectrum of H$\alpha$-correlated emission 
using FDS99 $-$ $\delta$ $\times$ H$\alpha$ as a CNM 
template for various values of $\delta$. Note that $\delta 
\sim 4 \times 10^{-5}$ mK/R is roughly required to 
appropriately remove H$\alpha$ morphology from FDS99 as 
shown in \reffig{rm_hal_fds_maps}.  As expected, as $\delta$ 
is increased, the spectrum from \reffig{halpha-twopanel} is 
contaminated with FDS99-correlated spectrum in 
\reffig{tsc-fit} (see text, \refeq{hal_fds_contam}).  
\emph{Middle:} the spectrum of H$\alpha$-correlated emission 
assuming that FDS99 exactly traces thermal dust emission 
with $T_{\rm dust} = \mbox{FDS99} \times (\nu/94\mbox{ 
GHz})^{1.7}$.  For $\delta = 0$ mK/R, the bump in the 
H$\alpha$-correlated spectrum represents the difference of 
WIM and CNM spectra, for $\delta = 4\times 10^{-5}$ mK/R, 
the spectrum is almost entirely WIM, and as $\delta$ is 
increased further, the bump represents an addition of WIM 
plus increasingly more CNM spectra.  Indeed, the peak 
frequency moves to the lower CNM value as $\delta$ gets 
large. \emph{Right:} The $\chi^2$ (dashed) and $D_0$ 
contours in the $(\nH,\mmu)$ plane for the spectrum shown in 
the middle panel with $\delta = 4 \times 10^{-5}$ mK/R.  
With this small level of contamination, our results for the 
best fit $\nH$, $\mmu$, and $D_0$ values are almost 
completely unchanged (cf., \reffig{comp-contours}).
} \label{fig:rm_hal_fds_spec}
\epm

Here we consider the effects of ``contaminating'' one of the templates with
another, and determine the effects on the resultant spectra.  We phrase this in
terms of a template (denoted with primes) and a ``true'' map that perfectly
traces the emission for the component we're solving for.  We consider two cases. 
First, we consider that the H$\alpha$ map is not a perfect tracer of free-free
emission and show that this does not mix the spectrum of other foregrounds into
the H$\alpha$-correlated spectrum.  That is, the bump in the spectrum is not due
to contamination effects from other foreground spectra.  Second, we consider the
case in which the dust template, FDS99, is tracing both a CNM and WIM spinning
dust component at low frequencies.  This case does have an impact on the
spectrum, but we show that it cannot produce the observed bump and that it does
not significantly impact our main conclusions.

\subsection{H$\alpha$ as an imperfect tracer of free-free}
\label{sec:app1}

For the template fits the equation we are trying to solve, neglecting the CMB 
and the associated cross correlation bias and considering only two 
foregrounds, is
\be
        a_1 H' + a_2 D = t_1 H + t_2 D,
        \label{eq:one}
\ee
for $a_1$ and $a_2$, where $H$ is the true H$\alpha$ map which we assume 
exactly traces free-free emission, $D$ is the dust map, $t_1$ and 
$t_2$ are the true amplitudes (at an arbitrary frequency), and $H'$ is 
the observed H$\alpha$ map (which may be contaminated in some way).

By multiplying \refeq{one} by $H'$ and solving for $a_1$, we obtain
\be
        a_1 = t_1\frac{\mean{H H'}}{\mean{H'^2}} + t_2 \frac{\mean{D 
        H'}}{\mean{H'^2}} - a_2 \frac{\mean{D H'}}{\mean{H'^2}}
        \label{eq:two}
\ee
and likewise by multiplying by $D$ and solving for $a_2$ we find
\be
        a_2 = t_2 + t_1 \frac{\mean{D H}}{\mean{D^2}} - a_1 \frac{\mean{D 
        H'}}{\mean{D^2}}.
        \label{eq:three}
\ee
To show that the H$\alpha$-correlated emission spectrum is not 
affected by the \emph{spectrum} of the dust-correlated emission, 
we need to show that $a_1 \neq a_1(a_2,t_2)$.  Plugging \refeq{three} 
into \refeq{two} we get,
\be
        a_1 = t_1\frac{\mean{H H'}}{\mean{H'^2}} - t_1\frac{\mean{D 
        H}\mean{D H'}}{\mean{D^2}\mean{H'^2}} + a_1\frac{\mean{D H'}\mean{D 
        H'}}{\mean{D^2}\mean{H'^2}},
\ee
or
\be
        a_1\left(1-\frac{\mean{D H'}\mean{D 
        H'}}{\mean{D^2}\mean{H'^2}}\right) = 
        t_1\left(\frac{\mean{H H'}}{\mean{H'^2}} - \frac{\mean{D H}\mean{D 
        H'}}{\mean{D^2}\mean{H'^2}}\right).
        \label{eq:five}
\ee

\refeq{five} shows two things.  First, $a_1$ does not depend on either 
$a_2$ or $t_2$, implying that the spectrum of H$\alpha$ correlated 
emission does not depend on either the spectrum or the inferred 
spectrum of dust correlated emission.  But, there is an overall 
multiplicative normalization, that depends on cross correlations of 
$H$, $H'$, and $D$.  Note that as $H \rightarrow H'$, $a_1 \rightarrow 
t_1$.  \refeq{five} is the general case; but now suppose we assume a 
specific form for $H'$, namely we mix in a bit of the dust map so that 
$H' = H + \epsilon D$.  This can originate from imperfect dust correction of 
the H$\alpha$ map or scattering of H$\alpha$ photons off of dust for 
example.  Then the right hand side of \refeq{five} reads,
\begin{eqnarray}
        t_1\left(\frac{\mean{H H'}}{\mean{H'^2}} - \frac{\mean{D H}\mean{D
        H'}}{\mean{D^2}\mean{H'^2}}\right) & = &
        t_1\left(\frac{\mean{(H' - \epsilon D) H'}}{\mean{H'^2}} - 
        \frac{\mean{D (H' - \epsilon D)}\mean{D 
        H'}}{\mean{D^2}\mean{H'^2}}\right) \nonumber\\
        & = &
        t_1
        \left(
        1 - \epsilon\frac{\mean{D H'}}{\mean{H'^2}} - \frac{\mean{D 
        H'}\mean{D H'}}{\mean{D^2}\mean{H'^2}} + 
        \epsilon\frac{\mean{D^2}\mean{D H'}}{\mean{D^2}\mean{H'^2}}
        \right) \nonumber\\
        & = &
        t_1\left(1-\frac{\mean{D H'}\mean{D
          H'}}{\mean{D^2}\mean{H'^2}}\right).
\end{eqnarray}
Comparing the term in parentheses with the left hand side of 
\refeq{five} shows, that, in the case where $H'$ is a linear 
superposition of $H$ and $D$, $a_1 = t_1$.  That is, we recover the true 
spectrum.  The bottom line is that, if the H$\alpha$ map does not exactly trace 
the free-free morphology, the spectrum will be affected, but only at most by a 
constant multiplicative factor (i.e., it will not produce a bump in the 
spectrum).

\subsection{FDS99 as a tracer of both WIM and CNM spinning dust}

So far we have assumed that FDS99 is a tracer of exclusively 
the CNM dust, as if WIM dust only appeared as 
H$\alpha$-correlated emission.  While the WIM dust indeed 
has a hotter color temperature (e.g. in the 60/100 micron 
ratio) than CNM dust, it certainly also appears in the FDS99 
map.  Therefore, in our regression analysis, fitting the CNM 
component with the FDS99 template will inadvertently absorb 
some of the WIM emission and affect the inferred spectrum of 
the WIM. The sense of the effect is to bias the 
H$\alpha$-correlated WIM emission with some amount of the 
CNM spectrum, pushing the WIM spectrum to higher peak 
frequency and lower amplitude.  In this section we estimate 
the extent of this bias.

In order to calculate an estimate of the contamination of 
FDS99 by H$\alpha$ morphology, we model the FDS99 map as a 
linear combination of maps of \ion{H}{1} emission and 
H$\alpha$ emission,
\be
\label{eq:lab_hal_fds_fit}
  D = \epsilon_0 L + \delta_0 H, 
\ee 
where again, $H$ is the H$\alpha$ map, and $L$ is the 
\cite{LAB} \ion{H}{1} map.  With this notation, 
$\delta_0$ has the units of mK/R.  We solve the above 
equation over unmasked pixels (where the mask consists of 
the point source mask described in \refsec{templates} plus a 
masking of all pixels for which the dust extinction at 
H$\alpha$ is $2.65E(B-V) > 0.5$ mag) and find that $\delta_0 
\sim 4.5 \times 10^{-5}$ mK/R.  Although this value varies 
slightly from region to region \citep{heiles99}, 
the contamination is very small and extremely 
subtle (see \reffig{rm_hal_fds_maps}).  Nevertheless, based 
on our value of $\delta_0$, H$\alpha$ morphology does indeed 
``contaminate'' the FDS99 map since the large grains in the 
WIM emit thermal radiation at 94 GHz as well.

To calculate the affect on the inferred spectrum, we can use the same procedure
as \refsec{app1} but now taking $D' \rightarrow D + \delta_0 H$.  It is
straightforward to show that the recovered H$\alpha$-correlated spectrum
becomes,
\bel{hal_fds_contam}
  a_1 = t_1 - \delta_0 t_2.
\ee
That is, the inferred $H\alpha$-correlated spectrum is 
contaminated by the FDS99-correlated spectrum at a level 
that is given by $\delta_0$. Assuming that $H$ exactly 
traces both the free-free emission and the WIM spinning dust 
emission and that $D$ exactly traces CNM emission (thermal 
and spinning dust), the implication is that the bump in the 
inferred spectrum of the H$\alpha$-correlated emission 
represents the WIM spinning dust spectrum \emph{minus} some 
amount of the CNM spinning dust spectrum, determined by how 
much the H$\alpha$ morphology has leaked into the FDS99 map via 
large grain thermal emission in the WIM.

The left panel of \reffig{rm_hal_fds_spec} shows the level of contamination of
the H$\alpha$-correlated spectra by taking $D' = \mbox{FDS99} - \delta \times
\mbox{H}\alpha$ as a tracer of CNM dust.  As expected from
\refeq{hal_fds_contam}, for large values of $\delta$, the H$\alpha$ map is being
over-subtracted from FDS99 and the resultant spectrum is highly contaminated by
a CNM thermal plus spinning dust spectrum (cf., Figures
\ref{fig:halpha-twopanel} and \ref{fig:tsc-fit}).  For $\delta = 0$ mK/R on the
other hand, some of the WIM spinning dust bump is being absorbed in the
$D'$-correlated spectrum and so the bump in the H$\alpha$-correlated spectrum
represents a WIM spinning dust minus some CNM spinning dust spectrum.  Based on
the fit in \refeq{lab_hal_fds_fit}, the $\delta = \delta_0 \approx 4 \times
10^{-5}$ mK/R case is a nearly pure mixture of free-free plus CMB plus WIM
spinning dust spectrum.

However, since the WIM does also emit thermal dust radiation, if we assume that 
FDS99 is a perfect representation of the thermal emission at 94 GHz, then 
removing the WIM (H$\alpha$) morphology from FDS99 implies that the 
H$\alpha$-correlated spectrum also contains that thermal emission.  We can 
estimate the amplitude of this effect by assuming that the WIM thermal emission 
has a $\nu^\beta$ dependence on frequency with $\beta = 3.7$ (in intensity 
units, $\beta = 1.7$ in antenna temperature) and adding $\delta \times H \times 
(\nu/94\mbox{ GHz})^{3.7}$ back into the H$\alpha$-correlated spectrum.  This is 
equivalent to setting,
\be
  D' = \mbox{FDS99} - \left[1-\left( \frac{\nu}{94\mbox{ GHz}} 
  \right)^{1.7} \right] \times \delta \times \mbox{H}\alpha.
\ee
The resultant spectra are shown in the middle panel of \reffig{rm_hal_fds_spec}.  
In this case, the bump is ideally due to \emph{only} spinning dust emission from the 
WIM for $\delta = 4 \times 10^{-5}$ mK/R.  For larger values of $\delta$ the CNM 
spinning dust spectrum begins to leak into the fit so that by $\delta = 10 
\times 10^{-5}$ mK/R, the spinning dust bump has moved towards the lower CNM 
value, and the amplitude has increased since the WIM and CNM spectra are now 
both present in the data.

Lastly, the right panel of \reffig{rm_hal_fds_spec} shows 
$\chi^2$ and $D_0$ contours in the $(\nH,\mmu)$ plane for a 
free-free plus CMB plus WIM spinning dust fit to the $\delta 
= 4 \times 10^{-5}$ mK/R spectra in the middle panel of 
\reffig{rm_hal_fds_spec}.  These contours indicate that the 
best fit values of $\nH$ and $\mmu$ are not significantly 
affected by the presence of WIM morphology in the FDS99 map 
for our estimate of $\delta = 4 \times 10^{-5}$ mK/R.  
Nevertheless, the exact level of PAH depletion measured by 
our technique ($D_0$) will depend on precisely how one 
estimates the WIM leakage in FDS99.  We defer a detailed 
analysis of this to future work.


\begin{thebibliography}{}

  \bibitem[\protect\citeauthoryear{Bennett et al.}{2003}]{bennett03}
    Bennett C.L. et al., 2003, ApJS, 148, 97

  \bibitem[\protect\citeauthoryear{Boughn \& Pober}{2007}]{boughn07}
    Boughn S. P.\& Pober J. C. 2007, ApJ, 661, 938

  \bibitem[\protect\citeauthoryear{Calabretta \& Roukema}{2007}]{calabretta07}
    Calabretta M.R. \& Roukema B.F., 2007, MNRAS, 381, 865

  \bibitem[\protect\citeauthoryear{Cassasus et al.}{2008}]{cassasus08}
    Cassasus et al., 2008, arXiv:0809.3965

  \bibitem[\protect\citeauthoryear{DIRBE Exp. Supp.}{1995}]{dirbesupp95}
    \COBE\ Diffuse Infrared Background Experiment (DIRBE) Explanatory Supplement,
    ed. M. G. Hauser, T. Kelsall, D. Leisawitz, \& Weiland, J. 1995,
    \COBE\ Ref. Pub. No. 95-A (Greenbelt, MD: NASA/GSFC),
    available electronically from the NSSDC [DIRBE~Exp.~Supp.]
    
  \bibitem[\protect\citeauthoryear{FIRAS Exp. Supp.}{1997}]{firas_supp}
    \COBE\ Far Infrared Absolute Spectrophotometer (FIRAS)
    Explanatory Supplement, Version 4, 1997,
    ed.\ S. Brodd, D. J. Fixsen, K. A. Jensen, J. C. Mather, \& R. A. Shafer,
    \COBE\ Ref. Pub. No. 97-C (Greenbelt, MD: NASA/GSFC),
    available in electronic form from the NSSDC [FIRAS Exp. Supp.]

  \bibitem[\protect\citeauthoryear{de Oliveira-Costa et al.}{1997}]{deO97}
    de Oliveira-Costa A., Kogut A., Devlin M.J., Netterfield C.B., Page L.A., \& 
Wollack E.J., 1997, ApJ, 482, L17

  \bibitem[\protect\citeauthoryear{de Oliveira-Costa et al.}{1998}]{deO98}
    de Oliveira-Costa A., Tegmark M., Page L., \& Boughn S., 1998, ApJ, 509, L9

  \bibitem[\protect\citeauthoryear{de Oliveira-Costa et al.}{1999}]{deO99}
    de Oliveira-Costa A. et al., 1999, ApJ, 527, L9

  \bibitem[\protect\citeauthoryear{de Oliveira-Costa et al.}{2000}]{deO00}
    de Oliveira-Costa A. et al., 2000, ApJ, 542, L5

  \bibitem[\protect\citeauthoryear{de Oliveira-Costa et al.}{2002}]{deO02}
    de Oliveira-Costa A. et al., 2002, ApJ, 567, 363

  \bibitem[\protect\citeauthoryear{de Oliveira-Costa et al.}{2004}]{deO04}
    de Oliveira-Costa A. et al., 2004, ApJ, 606, L89

  \bibitem[\protect\citeauthoryear{Davies et al.}{2006}]{davies06}
    Davies R.D. et al., 2006, MNRAS, 370, 1125

  \bibitem[\protect\citeauthoryear{Dennison et al.}{1998}]{dennison98}
    Dennison B., Simonetti J.H., \& Topasna G., 1998,
    Publ. Astron. Soc. Australia, 15, 147

  \bibitem[\protect\citeauthoryear{Dickinson et al.}{2003}]{dickinson03}
    Dickinson C., Davies R. D., \& Davis R. J. 2003, MNRAS, 341, 369

  \bibitem[\protect\citeauthoryear{Dickinson et al.}{2006}]{dickinson06}
    Dickinson C., Casassus S., Pineda J. L., Pearson T. J.,
    Readhead A. C. S., \& Davies, R. D. 2006, ApJ, 643, L111

  \bibitem[\protect\citeauthoryear{Dobler \& Finkbeiner}{2008a}]{DF08a}
    Dobler G. \& Finkbeiner D.P, 2008, ApJ, 680, 1222

  \bibitem[\protect\citeauthoryear{Dobler \& Finkbeiner}{2008b}]{DF08b}
    Dobler G. \& Finkbeiner D.P, 2008, ApJ, 680, 1235

  \bibitem[\protect\citeauthoryear{Draine \& Lazarian}{1998a}]{DL98a}
    Draine B.T. \& Lazarian A., 1998a, ApJ, 494, L19

  \bibitem[\protect\citeauthoryear{Draine \& Lazarian}{1998b}]{DL98b}
    Draine B.T. \& Lazarian A., 1998b, ApJ, 508, 157

  \bibitem[\protect\citeauthoryear{Draine \& Lazarian}{1999}]{DL99}
    Draine B.T. \& Lazarian A., 1999, ApJ, 512, 740

  \bibitem[\protect\citeauthoryear{Draine \& Li}{2007}]{draine07}
    Draine B.T. \& Li A., 2007, ApJ, 657, 810

  \bibitem[\protect\citeauthoryear{Erickson}{1957}]{erickson57}
    Erickson W. C. 1957, \apj, 126, 480
    
  \bibitem[\protect\citeauthoryear{Ferrara \& Dettmar}{1994}]{ferrara94}
    Ferrara A., \& Dettmar R. -J. 1994, \apj, 427, 155
 
  \bibitem[\protect\citeauthoryear{Finkbeiner et al.}{2002}]{F02}
    Finkbeiner D.P., Schlegel D.J, Frank, C., \& Heiles C.,
    2002, ApJ, 566, 898

  \bibitem[\protect\citeauthoryear{Finkbeiner}{2003}]{finkbeiner03}
    Finkbeiner D.P., 2003, ApJS, 146, 407

  \bibitem[\protect\citeauthoryear{Finkbeiner}{2004}]{finkbeiner04}
    Finkbeiner D.P., 2004, ApJ, 614, 186

  \bibitem[\protect\citeauthoryear{Finkbeiner et al.}{1999}]{finkbeiner99}
    Finkbeiner D.P., Davis M., \& Schlegel D.J., 1999, ApJ, 524, 867

  \bibitem[\protect\citeauthoryear{Finkbeiner, Langston, \& Minter}{2004}]{gb04}
    Finkbeiner D. P., Langston G. I., \& Minter A. H. 2004, ApJ, 617, 350

  \bibitem[\protect\citeauthoryear{Gaustad et al.}{2001}]{gaustad01}
    Gaustad J.E., McCullough P.R., Rosing W., \& Van Buren D., 2001, 
    PASP, 113, 1326

  \bibitem[\protect\citeauthoryear{G\'orski et al.}{1999}]{gorski99}
    G\'{o}rski K.M., Hivon E., \& Wandelt B.D., 1999, in MPA/ESO Cosmology 
Conf., Evolution of Large-Scale Structure, ed. A. J. Banday, R. K. Sheth, \& L. 
N. da Costa (Garching: ESO), 37

  \bibitem[\protect\citeauthoryear{Haslam et al.}{1982}]{haslam82}
    Haslam C.G.T., Stoffel H., Salter C.J., \& Wilson W.E., 1982, A\&AS, 47, 1

  \bibitem[\protect\citeauthoryear{Haffner et al.}{2003}]{haffner03}
    Haffner L.M., Reynolds R.J., Tufte S.L., Madsen G.J., Jaehnig K.P., \& 
    Percival J.W., 2003, ApJS, 149, 405

  \bibitem[\protect\citeauthoryear{Heiles et al.}{1999}]{heiles99}
    Heiles C., Haffner L. M., \& Reynolds R. J., 1999, ASPC, 168, 211

  \bibitem[\protect\citeauthoryear{Heiles}{2001}]{heiles01}
    Heiles C., 2001, ApJ, 551, L105

  \bibitem[\protect\citeauthoryear{Hinshaw et al.}{2007}]{hinshaw07}
    Hinshaw G. et al., 2007, ApJS, 170, 288

  \bibitem[\protect\citeauthoryear{Hinshaw et al.}{2008}]{hinshaw08}
    Hinshaw G. et al., 2008, arXiv:0803.0732

  \bibitem[\protect\citeauthoryear{Hooper et al.}{2007}]{hooper07}
    Hooper D., Finkbeiner D.P., \& Dobler G., 2007, PRD, 76, 3012

  \bibitem[\protect\citeauthoryear{Kalberla et al.}{2005}]{LAB}
    Kalberla P.M.W. et al., 2005, A\&A, 440, 775

  \bibitem[\protect\citeauthoryear{Kogut et al.}{1996}]{kogut96}
    Kogut A. et al., 1996, ApJ, 464, L5

  \bibitem[\protect\citeauthoryear{Kogut et al.}{2007}]{kogut07}
    Kogut A. et al., 2007, ApJ, 665, 355

  \bibitem[\protect\citeauthoryear{La Porta et al.}{2008}]{laporta08}
    La Porta L., Burigana C., Reich W., \& Reich P., 2008, A\&A, 479, 641

  \bibitem[\protect\citeauthoryear{Langston et al.}{2000}]{langston00}
    Langston, G., Minter, A., D'Addario, L., Eberhardt, K., Koski, K., \&
    Zuber, J. 2000, AJ, 119, 2801
    
  \bibitem[\protect\citeauthoryear{Leitch et al.}{1997}]{leitch97}
    Leitch E. M., Readhead A. C. S., Pearson T. J., \& Myers S. T.
    1997, \apj, 486, L23

  \bibitem[\protect\citeauthoryear{Madsen et al.}{2006}]{madsen06}
    Madsen G.J., Reynolds R.J., \& Haffner L.M., 2006, ApJ, 652, 401

  \bibitem[\protect\citeauthoryear{Page et al.}{2007}]{page07}
    Page L. et al., 2007, ApJS, 170, 335

  \bibitem[\protect\citeauthoryear{Quireza et al.}{2006}]{quireza06}
    Quireza C. et al., 2006, ApJ, 653, 1226

  \bibitem[\protect\citeauthoryear{Schlegel et al.}{1998}]{schlegel98}
    Schlegel D.J., Finkbeiner D.P., \& Davis M., 1998, ApJ, 500, 525

  \bibitem[\protect\citeauthoryear{Spergel et al.}{2003}]{spergel03}
    Spergel D.N. et al., 2003, ApJS, 148, 175

  \bibitem[\protect\citeauthoryear{Spergel et al.}{2007}]{spergel07}
    Spergel D.N. et al., 2007, ApJS, 170, 377

  \bibitem[\protect\citeauthoryear{Spitzer}{1978}]{spitzer}
    Spitzer, L. 1978, \emph{Physical Processes in the Interstellar Medium},
    Wiley, New York

  \bibitem[\protect\citeauthoryear{Valls-Gabaud}{1998}]{valls98}
    Valls-Gabaud D., 1998, Publ. Aston. Soc. Australia, 15, 111

  \bibitem[\protect\citeauthoryear{Wheelock et al.}{1994}]{issa94}
    Wheelock, S. L. et al. 1994, \IRAS\ Sky Survey Atlas: Explanatory Supplement,
    JPL Publication 94-11 (Pasadena: JPL)

\end{thebibliography}
\end{document}